\documentclass[12pt]{article}
\usepackage{amsfonts,amssymb,graphicx,epsfig}
\usepackage{float}
\usepackage[usenames]{color}
\usepackage{pstcol}
\usepackage{setspace}

\makeatletter
\def\numberbysection{\@addtoreset{equation}{section}
 	\def\theequation{\thesection.\arabic{equation}}}
\makeatother

\numberbysection 

\usepackage[nosort]{cite}
\usepackage[backref=false]{hyperref}
\hypersetup{
colorlinks=true,
citecolor=red,
linkcolor=darkblue,
urlcolor=darkblue
}
\definecolor{darkblue}{rgb}{0,0,.8}
\definecolor{red}{rgb}{1,0,0}

\newcommand{\be}{\begin{eqnarray}}
\newcommand{\ee}{\end{eqnarray}}
\newcommand{\non}{\nonumber}

\newcommand{\BB}{\ensuremath{\mathcal{V}}}
\def\Re{\mathop{\mbox{Re}}\nolimits}
\def\Im{\mathop{\mbox{Im}}\nolimits}

\def\calF{{\cal F}}
\def\calU{{\cal U}}
\def\calV{{\cal V}}
\long\def\ignore#1{}

\newcommand{\RR}[5]{\renewcommand{\arraystretch}{0.5}
\mathcal{R}^{(\scriptstyle{#1},\scriptstyle{#2})}\!
\begin{array}{l}{\ \ \scriptstyle {#3}}
\\ {\scriptstyle{#4}}\\ 
{\ \ \scriptstyle{#5}}\end{array}
\renewcommand{\arraystretch}{1.0}}

\newcommand{\BP}[5]{\renewcommand{\arraystretch}{0.5}
\overline{B}^{(\scriptstyle{#1},\scriptstyle{#2})}\!
\begin{array}{l}{\ \ \scriptstyle {#3}}
\\ {\scriptstyle{#4}}\\ 
{\ \ \scriptstyle{#5}}\end{array}
\renewcommand{\arraystretch}{1.0}}

\newcommand{\R}[5]{\renewcommand{\arraystretch}{0.5}
B^{(\scriptstyle{#1},\scriptstyle{#2})}\!
\begin{array}{l}{\ \ \scriptstyle {#3}}
\\ {\scriptstyle{#4}}\\ 
{\ \ \scriptstyle{#5}}\end{array}
\renewcommand{\arraystretch}{1.0}}

\newcommand{\bpic}{\begin{picture}}
\newcommand{\epic}{\end{picture}}
\renewcommand{\ss}{\scriptstyle}
\newcommand{\ds}{\displaystyle}
\newcommand{\p}[2]{\makebox(0,0)[#1]{$#2$}}
\newcommand{\pp}[2]{\makebox(0,0)[#1]{$\ss#2$}}
\newcommand{\text}[6]{\begin{picture}(#1,#2)
\put(#3,#4){\p{#5}{\ds#6}}\end{picture}}

\begin{document}

\begin{titlepage}
\vspace{.5in}
\begin{center}

{\bf\LARGE Boundary reflection matrices of 
massive}\\[6pt]
{\bf\LARGE \boldmath$\phi_{1,3}$-perturbed unitary minimal models}\\[1.0in]
\large 
Zoltan Bajnok\,\footnote{
   Wigner Research Centre for Physics,
   1121 Budapest, Konkoly-Thege Miklós út 21-23,
   Hungary; email: bajnok.zoltan@wigner.hu},
Rafael I. Nepomechie\,\footnote{
       Department of Physics, P.O. Box 248046, University of Miami,
       Coral Gables, FL 33124 USA; e-mail: nepomechie@miami.edu},
Paul A. Pearce\,\footnote{
       School of Mathematics and Statistics, University of 
       Melbourne, Parkville, Victoria 3010, Australia; 
       School of Mathematics and Physics, University of Queensland, St Lucia, Brisbane, Queensland 4072, \mbox{Australia};
       High Energy Physics Research Unit, Faculty of Science, Chulalongkorn University, Pathumwan, Bangkok 10330, Thailand;
       e-mail: papearce@unimelb.edu.au}
 \\

\end{center}

\vspace{.5in}

\begin{abstract}
\noindent
We propose explicit expressions for the boundary reflection matrices of the ${\cal A}_m+(r,s)$ series of massive scattering theories, obtained by perturbing the ${\cal A}_m$ unitary minimal models with $(r,s)$ boundary conditions with both bulk and boundary $\phi_{1,3}$ operators.
We identify the vacua that live on the boundary with the allowed edges of the $(r,s)$
conformal boundary conditions of the $A_m$ Andrews-Baxter-Forrester  model. 
The boundary reflection matrices
are then ``direct sums'' of certain pairs of 
$A_{m-1}$ Behrend-Pearce
solutions of the boundary Yang-Baxter equation and are consistent with
the boundary bootstrap and the recently-introduced crossing, as well as the $Z_{2}$ (height-reversal), Kac table and non-invertible symmetries. 
\end{abstract}

\vspace{.5in}

\end{titlepage}

\newpage
\setstretch{.85}
\tableofcontents

\newpage
\setstretch{1.04}
\hyphenpenalty=10000

\setcounter{footnote}{0}

\section{Introduction}
\label{sec:intro}

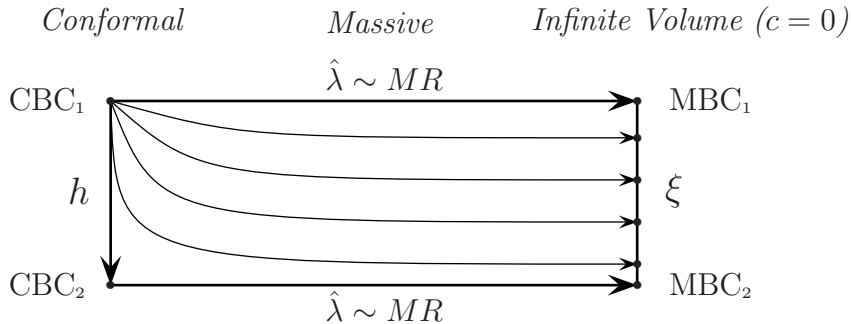
\begin{figure}[t]
\begin{center}
\psset{unit=14mm}
\setlength{\unitlength}{14mm}
\begin{pspicture}(5,2.75)
\put(0,2.5){\pp{}{\mbox{\em Conformal}}}
\put(2.5,2.5){\pp{}{\mbox{$\ $\em Massive}}}
\put(5.5,2.5){\pp{}{\mbox{\em Infinite Volume ($c=0$)}}}
\put(-.3,.9){\pp{}{\mbox{\large $h$}}}
\put(5.35,.9){\pp{}{\mbox{\large $\xi$}}}
\put(2.6,2.){\pp{}{\mbox{$\hat \lambda\sim MR$}}}
\put(2.6,-.2){\pp{}{\mbox{$\hat \lambda\sim MR$}}}
\put(-.6,1.75){\pp{}{\mbox{CBC}_1}}
\put(-.6,0){\pp{}{\mbox{CBC}_2}}
\put(5.7,1.75){\pp{}{\mbox{MBC}_1}}
\put(5.7,0){\pp{}{\mbox{MBC}_2}}
\psline[linewidth=1pt,arrowsize=8pt]{->}(0,1.75)(5,1.75)
\psline[linewidth=1pt,arrowsize=8pt]{->}(0,0)(5,0)
\psline[linewidth=1pt,arrowsize=8pt]{->}(0,1.75)(0,0)
\psline[linewidth=1pt,arrowsize=8pt]{-}(5,1.75)(5,0)
\psbezier[linewidth=.5pt,arrowsize=5pt]{->}(0,1.75)(.05,.2)(.05,.2)(5,.2)
\psbezier[linewidth=.5pt,arrowsize=5pt]{->}(0,1.75)(.45,.6)(.45,.6)(5,.6)
\psbezier[linewidth=.5pt,arrowsize=5pt]{->}(0,1.75)(.85,1.0)(.85,1.0)(5,1.0)
\psbezier[linewidth=.5pt,arrowsize=5pt]{->}(0,1.75)(1.25,1.4)(1.25,1.4)(5,1.4)
\put(0,0){\circle*{.08}}
\put(5,0){\circle*{.08}}
\put(5,1.75){\circle*{.08}}
\put(0,1.75){\circle*{.08}}
\put(5,0.2){\circle*{.08}}
\put(5,0.6){\circle*{.08}}
\put(5,1.0){\circle*{.08}}
\put(5,1.4){\circle*{.08}}
\end{pspicture}
\end{center}
\label{twoflow}
\caption{Schematic representation of the one-parameter family of flows of the ${\cal A}_m$ unitary minimal model as described by the boundary reflection matrix. These flows are parametrized by the boundary parameter $\xi$ and the source of the flows is associated with an initial conformal boundary condition $\mbox{CBC}_1=(r,s)$. The boundary RG flow between conformal fixed points $\mbox{CBC}_1$ and $\mbox{CBC}_2$ is on the left. On the right is a line of massive fixed points parametrized by $\xi$. The mass is introduced via the relevant perturbation (\ref{bulkaction}) with coupling $\hat \lambda$.}
\end{figure}

The discovery~\cite{Zamolodchikov:1987jf, Zamolodchikov:1989hfa} that certain perturbations of conformal field
theories~\cite{Belavin:1984vu, ZZ:1989, Ginsparg:1988ui,BYB:2019} are integrable was a watershed in the study
of integrable quantum field theory~\cite{Zamolodchikov:1978xm}.  
Here we consider the $A$-series~\cite{Cardy:1986ie, Cardy:1986gw, Cappelli:1986hf} of
massive scattering theories obtained from the 
unitary minimal models with central charge 
$c=1 - {6\over
m(m+1)}$ (which for brevity we shall denote by  ${\cal A}_{m}$, with $m=3 \,,
4 \,, \ldots$) by integrable perturbations with the relevant bulk 
and boundary $\phi_{1,3}$ operators 
as shown schematically in Figure~\ref{twoflow}.
These theories correspond to the massive continuum limit~\cite{Pearce:1997nv} 
of the off-critical $A_m$ Restricted-Solid-On-Solid (RSOS) models of Andrews-Baxter and Forrester~\cite{Andrews:1984af} in Regime~III. 
In this low-temperature regime these models exhibit $m-1$ coexisting phases (vacua) 
where the heights $a=1,2,\ldots,m-1$ on the two sublattices of the square lattice
alternate between $a$ and $a+1$. 
The problem of
determining bulk $S$ matrices for the $A$
scattering theories 
was solved by Zamolodchikov~\cite{Zamolodchikov:1989rd}, Bernard and
LeClair~\cite{LeClair:1989wy, Bernard:1990cw}, and Reshetikhin and Smirnov~\cite{Reshetikhin:1989qg}.  These
authors showed that the spectrum consists of massive kinks
that interpolate between neighboring RSOS vacua, 
labeled by $a=1,2,\ldots,m-1$, 
with $S$ matrices of 
``critical" RSOS type~\cite{Andrews:1984af}.  They
also showed that these models possess integrals of motion of
fractional spin $2/m$ (residual quantum group symmetries) which
commute with the $S$ matrix.  
Analysis using thermodynamic Bethe ansatz (TBA)~\cite{Zamolodchikov:1991vh} and the truncated conformal
space approach (TCSA)~\cite{Yurov:1989yu, Klassen:1992qy} supports these results.
Recently, it was emphasized that the crossing relation for the 
scatterings of kink particles has to be modified~\cite{Copetti:2024rqj, Copetti:2024dcz}, see also \cite{Smirnov:1991uw, Colomo:1991gw, Klassen:1992qy}. This modification is related to the 
normalization of the vacua and kink particles and can be calculated from non-invertible symmetries. 
The modified crossing results in a different (actually simpler) scattering matrix, which is 
proportional to the corresponding RSOS lattice Boltzman weights. 
For this new scattering matrix, the 
previously-obtained and thoroughly-tested  thermodynamics is the same.   

The problem of determining the corresponding {\it boundary} $S$ or reflection
matrices has also received considerable attention.  The foundation was
established by Cardy~\cite{Cardy:1989ir}, who identified a class of conformal
boundary conditions (CBCs) of the ${\cal A}_{m}$ models that are in
one-to-one correspondence with the $(r,s)$ primary fields; and by
Affleck and Ludwig~\cite{Affleck:1991tk, Affleck:1992ng}, who used conformal perturbation theory
(CPT) to show that boundary perturbations can cause RG flow from one
CBC to another.  Ghoshal and Zamolodchikov~\cite{Ghoshal:1993tm} then formulated
the general framework for analyzing this problem.  They argued that
the boundary conformal field theory ${\cal A}_{m} + (r,s)$ perturbed by both
bulk and boundary $\phi_{1,3}$ operators is integrable, and they found
the boundary reflection matrix for the first case, $m=3$, corresponding to
the Ising model.  The case $m=4$, corresponding to the tricritical
Ising model, was then analyzed by Chim~\cite{Chim:1995kf}.  
Further results
were subsequently obtained through a variety of approaches: direct
solution~\cite{Ahn:1996nq, Ahn:1995yn, Ahn:1998} of the boundary Yang-Baxter equation 
(BYBE)~\cite{Cherednik:1984vvp}, vertex operators~\cite{Miwa:1996ht}, TBA~\cite{Lesage:1998qf, Nepomechie:2002ak, Feverati:2003rb},
solutions on the lattice~\cite{Saleur:1988zx, Behrend:2000us,Behrend:1995zj,OBrien:1997}, 
CPT~\cite{Recknagel:2000ri, Fredenhagen:2002qn, Fredenhagen:2003xf, Graham:2001pp, Graham:2003nc}, and TCSA~\cite{Graham:2000si}.
Recently, boundary crossing unitarity for kink particles was also modified~\cite{Shimamori:2025ntq} due to the 
normalization of kink particles and vacua.
The authors also considered a very specific boundary condition, which was invariant under the non-invertible symmetries and determined the corresponding (diagonal) reflection matrices.
(Results for non-unitary minimal models have also been obtained~\cite{Ghoshal:1993tm, Dorey:1997yg, Dorey:1999cj, Dorey:1998kt, Dorey:unpub}.)  However, boundary reflection  matrices corresponding
to the $\phi_{1,3}$ bulk and boundary perturbation of ${\cal A}_{m} + (r,s)$
models for all the possible Cardy CBCs $(r,s)$ have not been proposed.

We attempt to fill this gap here.  This task is not as daunting as it
might first appear.  Indeed, Behrend and Pearce~\cite{Behrend:2000us} have already
found ``elementary'' solutions of the BYBE.
Although these elementary solutions do not satisfy the boundary
bound-state bootstrap equation~\cite{Ghoshal:1993tm}, one can obtain solutions of
both BYBE and boundary bound-state bootstrap equations
by forming ``direct sums''  of the elementary solutions.  In
order to match these solutions with the various 
``critical" solutions we make use of
\begin{itemize}
\item[(i)] a proposed identification, 
extending earlier work~\cite{Ghoshal:1993tm, Chim:1995kf}, 
of ``boundary subsets'' $U_{(r,s)} \subseteq \{ 1, 2, \ldots \,, m-1 \}$ of allowed
RSOS vacua at the boundary
with allowed edges~\cite{Cardy:1989ir, Saleur:1988zx, Behrend:2000us} of the $(r,s)$ ``critical" solutions; and also
\item[(ii)] the fact~\cite{Ghoshal:1993tm} that the boundary reflection matrix corresponding to 
the bulk and boundary $\phi_{1,3}$-perturbation of ${\cal A}_{m} + (r,s)$
necessarily describes the 
boundary renormalization group (RG) flows of the CBC
$(r,s)$ under $\phi_{1,3}$. Fredenhagen and Schomerus~\cite{Fredenhagen:2002qn, Fredenhagen:2003xf} (see also Graham and Watts~\cite{Graham:2003nc}) have made a proposal for all such 
boundary flows, based on earlier work of Lesage
{\it et al.}~\cite{Lesage:1998qf} and Recknagel {\it et al.}~\cite{Recknagel:2000ri}.
\end{itemize}
We use the term ``critical" to emphasize that we use critical trigonometric solutions of the RSOS BYBE rather than off-critical elliptic solutions.

We recall some pertinent facts about the bulk theory~\cite{Zamolodchikov:1989rd, LeClair:1989wy, Bernard:1990cw,
Reshetikhin:1989qg} in Section~\ref{sec:Bulk}.  In Section~\ref{sec:CBC}, we briefly
review Cardy's~\cite{Cardy:1989ir} classification of CBCs for the ${\cal A}_{m}$
models.
We identify ``boundary subsets'' of allowed RSOS vacua
at the boundary with the allowed edges of corresponding 
``critical" boundary conditions.  These identifications, together with
the boundary flows~\cite{Graham:2003nc}, enable us later to associate boundary
reflection matrices with CBCs.  In Section~\ref{sec:Boundary} we recall some
pertinent facts about integrable boundary scattering theory~\cite{Ghoshal:1993tm}.
In Section~\ref{sec:Elementary} we recall the ``elementary'' solutions
of the BYBE found by Behrend and Pearce~\cite{Behrend:2000us}, and work out their
further properties.  In Section~\ref{sec:Pairs} we construct 
``paired'' solutions by forming ``direct sums'' of certain pairs of 
elementary solutions. We identify these paired solutions as the boundary reflection matrices
corresponding to the $\phi_{1,3}$ bulk and boundary perturbation of
${\cal A}_{m} + (r,s)$.  We verify that these reflection matrices are
consistent with the boundary bootstrap and the $Z_{2}$ (height-reversal) symmetries
of the conformal boundary conditions.  The conformal and massive boundary flows are discussed in Section~\ref{sec:BoundaryFlows}. 
We demonstrate in Section \ref{sec:Noninvertible} that the boundary reflection matrices transform covariantly under the non-invertible symmetries of the theory. 
The boundary reflection matrices for the cases $m=3,4,5$ are worked out in
detail in Section~\ref{sec:Examples} as examples.  We conclude with a
brief discussion of our results in Section \ref{sec:Conclude}.
 
\section{Bulk Scattering Theory}
\label{sec:Bulk}

We denote by ${\cal A}_{m}$ the unitary $A$-series~\cite{Cardy:1986ie, Cardy:1986gw, Cappelli:1986hf} of Virasoro minimal models 
${\cal M}(m,m+1)$ with central
charge $c =1 - {6\over m(m+1)}$.  The chiral primary fields
$\phi_{r,s}\equiv\phi_{r',s'}$, where
\be
r' = m-r \,, \qquad s'= m+1-s \,,
\label{primes}
\ee
have conformal weights
\be
\Delta_{r,s}= {\left[ r(m+1) - s m \right]^{2} - 1\over 4 m (m+1)} 
\,,
\ee
where $r$ and $s$ are integers satisfying $1 \le r \le m-1$ and $1 \le s 
\le m$.

In this Section, we briefly review some pertinent results about the
bulk theory defined by the ``action''~\cite{Zamolodchikov:1989rd, LeClair:1989wy, Bernard:1990cw}
\be
\mathcal{S} = \mathcal{S}_{{\cal A}_{m}} + \hat \lambda \int_{-\infty}^{\infty} dy 
\int_{-\infty}^{\infty} dx\  
\Phi_{1,3}(x \,, y) \,, \qquad \hat \lambda < 0 \,, 
\label{bulkaction}
\ee
where $\mathcal{S}_{{\cal A}_{m}}$ is the action of ${\cal A}_{m}$, and $\Phi_{1,3}$ is the
spinless operator with (left, right) dimensions $( \Delta_{1,3} \,,
\Delta_{1,3})$, where $\Delta_{1,3}={m-1\over m+1}$.  Moreover,
$\hat \lambda$ is a bulk parameter with dimension length${}^{-{4\over
m+1}}$.  Since both the CFT ${\cal A}_{m}$ and the operator $\Phi_{1,3}$ are 
invariant under spin-reversal, then so is the model (\ref{bulkaction}). The operator algebra is 
invariant under this $Z_{2}$ transformation~\cite{ZZ:1989,Ginsparg:1988ui,BYB:2019}:
\be
\phi_{r,s} \mapsto (-1)^{r+1} \phi_{r,s} \,, \quad  m = 
\mbox{even};\qquad  
\phi_{r,s} \mapsto (-1)^{s+1} \phi_{r,s} \,, \quad  m = 
\mbox{odd} \,.  
\ee

We restrict our attention to the case $\hat \lambda < 0$, for which there
is an $(m-1)$-fold vacuum degeneracy, and the spectrum consists of
kinks $K_{a \,, b}(\theta)$ with nonzero mass $M \propto 
|\hat \lambda|^{m+1\over 4}$ 
and rapidity $\theta$ 
\footnote{Hence, a kink has energy 
$e= M \cosh \theta$ and momentum $p = M \sinh \theta$.}
that separate neighboring vacua 
\be
a \,, b \in \{ 1\,, 2\,, \ldots \,, m-1 \} \,, \label{heights}
\ee
with $|a - b|=1$. 

\begin{figure}
\begin{center}
\setlength{\unitlength}{9mm}
\psset{unit=9mm}
\begin{pspicture}(4,4)
\psline[linewidth=.75pt,arrowsize=5pt]{->}(0,2)(4,2)
\psline[linewidth=.75pt,arrowsize=5pt]{->}(2,0)(2,4)
\rput(-.4,2){\large$\theta_1$}
\rput(2,-.5){\large$\theta_2$}
\rput(1,1){\large$a$}
\rput(3,1){\large$b$}
\rput(3,3){\large$c$}
\rput(1,3){\large$d$}
\end{pspicture}
\end{center}
\caption{Bulk $S$ matrix $S_{a\ b}^{d\ c}(\theta_{1}-\theta_{2})$.}
\label{fig:bulkSmatrixa}
\end{figure}
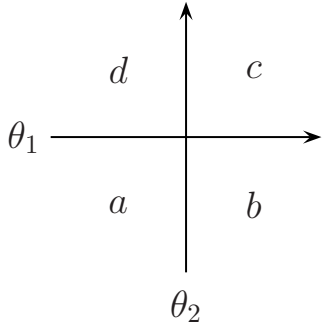

The two-kink $S$ matrix $S_{a\ b}^{d\ c}(\theta)$ is defined by the
relation (see Figure \ref{fig:bulkSmatrixa})
\be
K_{d\,, a}(\theta_{1})\ K_{a\,, b}(\theta_{2}) =
\sum_{d} S_{a\ b}^{d\ c}(\theta_{1}-\theta_{2})\
K_{d\,, c}(\theta_{2})\ K_{c\,, b}(\theta_{1}).
\label{bulkS0}
\ee

It was given by \cite{Zamolodchikov:1989rd, LeClair:1989wy, Bernard:1990cw, Klassen:1992qy} but modified due to the new crossing\footnote{The ``old'' scattering matrix contains an additional factor $\left({[a][c]\over [b][d]}\right)^{-\frac{\theta}{2\pi i}} $ on the rhs of (\ref{bulkS}).} in \cite{Copetti:2024rqj}  
\be
S_{a\ b}^{d\ c}(\theta) &=&{U(\theta)}\ 
\overline{S}_{a\ b}^{d\ c}(\theta) 
\label{bulkS}
\ee
where the ``reduced'' matrix elements $\overline{S}_{a\ b}^{d\ c}(\theta)$
are related to the Boltzmann weights of the $A_{m-1}$ lattice models 
\be 
\overline{S}_{a\ b}^{d\ c}(\theta) =
\sinh \biggl ({i \pi-\theta\over m}\biggr ) \delta_{a c} +
\left({[a][c]\over [b][d]}\right)^{1\over 2} 
\sinh \biggl ({\theta\over m}\biggr ) \delta_{b d}
\label{reducedbulkS}
\ee
with\,\footnote{In \cite{LeClair:1989wy, Bernard:1990cw, Klassen:1992qy}, the quantity $[a]$ is defined
with a further phase factor $(-1)^{a+1}$, which can be removed by
a gauge transformation. We do not include
this factor here, which conforms with \cite{Zamolodchikov:1989rd, Behrend:2000us}.}
\be
[a] 
= {\sin( {\pi a\over m})\over \sin( {\pi \over m})} \,.
\label{box}
\ee 

Observe the shift in $m$: the conformal field theory 
${\cal A}_{m}$ and its massive perturbation are related to the continuum limit of the $A_m$ lattice model, while the scattering matrix is related
to the Boltzmann weights of the $A_{m-1}$ lattice model. 
The scalar factor $U(\theta)$ satisfies 
\be
U(\theta)\ U(-\theta) = {1\over 
\sinh ({1\over m}( i \pi+\theta )) \sinh ({1\over m}(i\pi - \theta))} 
\,, \qquad
U(i \pi-\theta)=U(\theta)\,,
\label{Urelations}
\ee
and has no poles in the physical strip $0 \le \Im \theta< \pi$.  Two 
useful integral representations for this function are  
\begin{eqnarray}
U(\theta) &=& {1 \over \sinh({1\over m}(\theta - i \pi))}
\exp \left( i \int_{0}^{\infty} {dt\over t} {\sin({\theta t\over \pi}) 
\sinh({(m-1)t\over 2})\over \sinh({m t\over 2}) \cosh({t\over 2})} \right) 
 \label{Utheta}\\
&=& {i \over \sin({\pi \over m})}\exp\left(\int_{-\infty}^\infty {\cosh((m-2)\pi t)\sin(\theta t)\sin((\pi i-\theta)t)\over
t\sinh (m\pi t)\cosh( \pi t)}\,dt\right)
\label{UBaxter}
\end{eqnarray}
The first is a ``reduction'' of the well-known integral representation 
for the corresponding scalar factor of the sine-Gordon $S$ matrix~\cite{Zamolodchikov:1978xm}
with ${8 \pi\over \gamma'} = {1\over m}$. The second is related to the bulk free energy of the critical $A_{m-1}$ lattice model as given in Appendix~C after the replacements $u\mapsto {\theta\over im}$ and $t\mapsto mt$ with $\lambda={\pi\over m}$. This observation is important because it tells us that such scalar factors (which remove non-universal terms) are related to the free energies of the associated critical $A_{m-1}$ lattice models.

The $S$ matrix (\ref{bulkS}) has the following 
diagonal symmetries in its indices
\be
S_{a\ b}^{d\ c}(\theta) = S_{c\ b}^{d\ a}(\theta) = S_{a\ d}^{b\ c}(\theta) \label{symm}
\ee
and, since $[\bar a] = [a]$, it is height-reversal symmetric
\be
S_{a\ b}^{d\ c}(\theta) = S_{\bar a\ \bar b}^{\bar d\ \bar c}(\theta),\qquad \bar a = m - a
\,, \label{bulkSspinrev}
\ee
It also satisfies the recently-found  crossing relation\footnote{In the old crossing, the factor $\left({[a][c]\over [b][d]}\right)^{1\over 2}$ is absent.} \cite{Copetti:2024rqj, Copetti:2024dcz}
\be
S_{a\ b}^{d\ c}(\theta) =\left({[a][c]\over [b][d]}\right)^{1\over 2}  S_{d\ a}^{c\ b}(i\pi - \theta) 
\label{crossing}
\ee
and the unitarity relation
\be
\sum_{g} S_{a\ b}^{d\ g}(\theta)\ S_{g\ b}^{d\ c}(-\theta)
= \delta_{a\,, c}\, G_{a\,, b}\ G_{a\,, d} \,,
\label{unitarity}
\ee
where $G_{a\,, b}$ is the 
adjacency matrix 
\be
G_{a\,, b} = \delta_{a \,, b-1} + \delta_{a \,, b+1} \,.
\label{adjacency}
\ee
Moreover, it satisfies the Yang-Baxter (star-triangle) equation
\be
\lefteqn{\sum_{g} 
S_{b\ c}^{a\ g}(\theta_{1} - \theta_{2})\
S_{c\ d}^{g\ e}(\theta_{1} - \theta_{3})\ 
S_{g\ e}^{a\ f}(\theta_{2} - \theta_{3})} \non  \\
&=& \sum_{g} 
S_{c\ d}^{b\ g}(\theta_{2} - \theta_{3})\ 
S_{b\ g}^{a\ f}(\theta_{1} - \theta_{3})\
S_{g\ d}^{f\ e}(\theta_{1} - \theta_{2})
\,. \label{YBE}
\ee

\section{Conformal Boundaries and Boundary Subsets}
\label{sec:CBC}
\subsection{Cardy boundary conditions}
Cardy~\cite{Cardy:1989ir} identified a consistent class of conformal boundary
conditions (CBCs) of the ${\cal A}_{m}$ models that are in one-to-one
correspondence with the primary fields $\phi_{r,s}=\phi_{r',s'}$
and labeled by the Kac labels $(r,s)\equiv (r',s')$.  
(See also \cite{Behrend:1999bn}.)
These CBCs are characterized in part
by their boundary entropy ($g$ factor)~\cite{Affleck:1991tk, Affleck:1992ng}
\be
g_{r,s}= \left({8\over m(m+1)} \right)^{1\over 4}
{\sin({\pi r \over m}) \sin({\pi s \over m+1})\over
\sqrt{\sin({\pi\over m}) \sin({\pi \over m+1})}} \,,
\ee 
in a similar way that a bulk conformal field theory is characterized 
in part by its central charge. (CBCs are also characterized by
their boundary subsets, which we discuss in the following Section.)
A general CBC is a linear combination of Cardy CBCs with positive
integer coefficients. (See, e.g., \cite{Chim:1995kf, Recknagel:2000ri, Graham:2001pp}.)

The ${\cal A}_m$ conformal minimal models are related to the critical $A_m$ ABF lattice models \cite{Andrews:1984af} as their continuum limits. This provides an enlightening visualization of their conformal boundary conditions. The critical ABF lattice models are defined in Appendix~A.
The $\mathcal{A}_{m}$ CBC $(r,s)$ has a convenient graphical representation  \cite{Behrend:2000us, Graham:2001pp}. 
Let us consider the Dynkin diagram for the Lie
algebra $A_{m}$ with the nodes labeled by $1 \,, 2\,, \ldots \,, m
\in A_m$ and the edges labeled by $1 \,, 2 \,, \ldots \,, m-1 \in A_{m-1}$, as follows
\setlength{\unitlength}{14mm}
\be  
\raisebox{-18pt}{\bpic(8,1)
\put(0,0.5){\line(1,0){2.5}}\put(3.5,0.5){\line(1,0){2.0}}
\put(6.5,0.5){\line(1,0){1.5}}
\multiput(2.7,0.5)(0.4,0){2}{\line(1,0){0.2}}
\multiput(5.7,0.5)(0.4,0){2}{\line(1,0){0.2}}
\multiput(0,0.5)(1,0){3}{\pp{}{\bullet}}
\multiput(4,0.5)(1,0){2}{\pp{}{\bullet}}
\multiput(7,0.5)(1,0){2}{\pp{}{\bullet}}
\put(0,0.85){\pp{t}{1}}\put(1,0.85){\pp{t}{2}}\put(2,0.85){\pp{t}{3}}
\put(4,0.85){\pp{t}{j}}\put(5,0.85){\pp{t}{j+1}}
\put(7,0.85){\pp{t}{m-1}}\put(8,0.85){\pp{t}{m}}
\put(0.5,0.35){\pp{t}{1}}\put(1.5,0.35){\pp{t}{2}}
\put(4.5,0.35){\pp{t}{j}}\put(7.5,0.35){\pp{t}{m-1}}\epic}
\label{figDynkin}
\ee 
Note that the number of edges $m-1$ is the same as the number of 
vacua (\ref{heights}). 
The CBC $(r,s)$ with $r < s$ can
be represented by the \emph{subgraph} of connected nodes from $s-r$ to $r+s$
as follows\,\footnote{In both  (\ref{figid2}) and (\ref{figid1}) we 
assume $r+s \le m$.  For $r+s >m$, the maximum node of the subgraph is
$2m+1-r-s$ instead of $r+s$.} 
\be
\psset{unit=14mm}
\setlength{\unitlength}{14mm}
\raisebox{8pt}{\begin{pspicture}[shift=.1](0,.7)(10,.8)
\multiput(2.0,0.5)(2.5,0){2}{\psline[linewidth=1.5pt](0,0)(1.5,0)}
\multiput(1.25,0.5)(0.25,0){3}{\circle*{0.001}}
\multiput(3.7,0.5)(0.4,0){2}{\line(1,0){0.2}}
\multiput(6.25,0.5)(0.25,0){3}{\circle*{0.001}}
\multiput(0,0.5)(1,0){4}{\pp{}{\bullet}}
\multiput(5,0.5)(1,0){2}{\pp{}{\bullet}}
\multiput(7,0.5)(1,0){2}{\pp{}{\bullet}}
\put(0,0.85){\pp{t}{1}}\put(1,0.85){\pp{t}{2}}
\put(2,0.85){\pp{t}{s-r}}\put(6,0.85){\pp{t}{r+s}}
\put(7,0.85){\pp{t}{m-1}}\put(8,0.85){\pp{t}{m}}
\put(2.5,0.35){\pp{t}{s-r}}
\put(5.5,0.35){\pp{t}{r+s-1}}
\put(9,.43){$(r < s)$}
\end{pspicture}}
\label{figid2}
\ee 
and the CBC $(r,s)$ with $r \ge s$
can be represented by the \emph{subgraph} of connected nodes from $r-s+1$ to
$r+s$ as follows
\be  
\psset{unit=14mm}
\setlength{\unitlength}{14mm}
{\raisebox{0pt}{\begin{pspicture}[shift=.5](0,.9)(10,.9)
\multiput(2.0,0.5)(2.5,0){2}{\psline[linewidth=1.5pt](0,0)(1.5,0)}
\multiput(1.25,0.5)(0.25,0){3}{\circle*{0.001}}
\multiput(3.7,0.5)(0.4,0){2}{\line(1,0){0.2}}
\multiput(6.25,0.5)(0.25,0){3}{\circle*{0.001}}
\multiput(0,0.5)(1,0){4}{\pp{}{\bullet}}
\multiput(5,0.5)(1,0){2}{\pp{}{\bullet}}
\multiput(7,0.5)(1,0){2}{\pp{}{\bullet}}
\put(0,0.85){\pp{t}{1}}\put(1,0.85){\pp{t}{2}}
\put(2,0.85){\pp{t}{r-s+1}}\put(6,0.85){\pp{t}{r+s}}
\put(7,0.85){\pp{t}{m-1}}\put(8,0.85){\pp{t}{m}}
\put(2.5,0.35){\pp{t}{r-s+1}}
\put(5.5,0.35){\pp{t}{r+s-1}}
\put(9,.43){$(r \ge s)$}
\end{pspicture}}}
\label{figid1}
\ee 
We have used the Kac table symmetry (\ref{uuprime}) to restrict ourselves to $r+s\le m$. 

\subsection{Boundary subsets}
\label{sec:BoundarySubsets}

A key point is that, for a given $\mathcal{A}_{m}$ CBC $(r,s)$, only a subset
\begin{equation}
\mathcal{U}_{(r,s)} \subseteq \{1, 2, \ldots, m-1\}
\end{equation}
of heights (vacua, or boundary edges) are allowed on the boundary. This set of heights, which we call \emph{boundary subset}, uniquely characterizes the CBC
and will play a fundamental role in our
analysis.

The boundary subset $\mathcal{U}_{(r,s)}$ is given by the set of \emph{edges} of the $(r,s)$ connected subgraph (\ref{figid2})-(\ref{figid1}). Explicitly, 
the boundary subset $\calU_{(r,s)}$ is given by 
sets of consecutive heights
\be
\calU_{(r,s)} = \{ i^{m}_{r,s} \,, i^{m}_{r,s}+1 \,, \ldots \,, 
j^{m}_{r,s} \} \,, 
\label{symmbsubset}
\ee
where the minimum and maximum heights $i^{m}_{r,s}$ and $j^{m}_{r,s}$  
are given by
\be
i^{m}_{r,s} = \min \{ |r-s+1|+1 \,, |r-s|+1 \} \,,  \qquad 
j^{m}_{r,s} = \min \{r+s-1 \,, 2m-r-s \} \,,
\label{minmaxheight}
\ee
respectively. Equivalently,
\be
\calU_{(r,s)} = 
\left\{ \begin{array}{ll}
\{ s-r  \,, s-r+1 \,, \ldots \,,
 j^{m}_{r,s} \} & \mbox{if} \qquad r < s \\[4pt]
 \{ r-s+1 \,, r-s+2 \,, \ldots \,,  
 j^{m}_{r,s} \} & \mbox{if} \qquad r \ge s 
 \end{array} \right. \,. \label{bsubset}
\ee

As expected, the boundary subsets corresponding
to the CBCs $(r,s)$ and $(r',s')$ are the same,
\be
\calU_{(r,s)} = \calU_{(r',s')} \,,
\label{uuprime}
\ee
where $r'$ and $s'$ are defined in (\ref{primes}).
Further checks are that 
$\calU_{(1,s)} =\{s-1,s\}$ and $\calU_{(r,1)} =\{ r \}$,
in agreement with \cite{Cardy:1989ir} and \cite{Saleur:1988zx}, respectively.
The CBCs $(r,s)$ and corresponding
boundary subsets $\calU_{(r,s)}$ 
for the cases $m=3\,, 4\,, 5$ are given in Tables~\ref{figm3},
\ref{figm4}, \ref{figm5}, respectively.  
The table for $m=3$ agrees
with Cardy~\cite{Cardy:1989ir}; his designations of the CBCs are given
in the third column of Table~\ref{figm3}.  The 
table for $m=4$ agrees
with Chim~\cite{Chim:1995kf}; his designations\,\footnote{Both
Zamolodchikov~\cite{Zamolodchikov:1989rd} and Chim~\cite{Chim:1995kf} label the vacua (edges)
for the tricritical Ising model ($m=4$) by $\{-1 \,, 0 \,, 1\}$
rather than $b\in\{ 1 \,, 2 \,, 3 \}$.}
$\!$of the CBCs are given in the third column of Table~\ref{figm4}.
Table~\ref{figm5} shows the boundary subsets for $m=5$.

\setlength{\unitlength}{1.mm}
\def\bx#1{\makebox[.6in]{#1}}

\begin{table}[p] 
\begin{center}
\begin{tabular}{r|c|c|l} 
\multicolumn{4}{l}{$s$}\\
\cline{2-3}
3&\bx{$\{2\}$} & \bx{$\{1\}$} &\raisebox{-1\unitlength}{\rule{0pt}{0\unitlength}} \\ \cline{2-3}
2&\bx{$\{1,2\}$} & \bx{$\{1,2\}$} &\raisebox{-1\unitlength}{\rule{0pt}{0\unitlength}} \\ \cline{2-3}
1&\bx{$\{1\}$} &\bx{$\{2\}$} &\raisebox{-1\unitlength}{\rule{0pt}{0\unitlength}}\\ \cline{2-3}
\multicolumn{4}{l}{\hspace{.8\unitlength}
\begin{tabular}{ccc}
$\;\,\,$\bx{1}&\bx{2}&$\;r$
\end{tabular}}
\end{tabular}
\hspace{.75in}
\begin{tabular}{r|c|c|l} 
\multicolumn{4}{l}{$s$}\\
\cline{2-3}
3&\bx{$(-)$} & \bx{$(+)$} &\raisebox{-1\unitlength}{\rule{0pt}{0\unitlength}} \\ \cline{2-3}
2&\bx{$(f)$} & \bx{$(f)$} &\raisebox{-1\unitlength}{\rule{0pt}{0\unitlength}} \\ \cline{2-3}
1&\bx{$(+)$} &\bx{$(-)$} &\raisebox{-1\unitlength}{\rule{0pt}{0\unitlength}}\\ \cline{2-3}
\multicolumn{4}{l}{\hspace{.8\unitlength}
\begin{tabular}{ccc}
$\;\,\,$\bx{1}&\bx{2}&$\;r$
\end{tabular}}
\end{tabular}
   \caption{Boundary subsets $\calU_{(r,s)}$ for $m=3$. Cardy's designation of the CBCs are shown on the right.}
  \label{figm3}
   \end{center}
\end{table}

\begin{table}[p] 
\begin{center}
\begin{tabular}{r|c|c|c|l} 
\multicolumn{5}{l}{$s$}\\
\cline{2-4}
4&\bx{$\{3\}$} & \bx{$\{2\}$} & \bx{$\{1\}$}&\raisebox{-1\unitlength}{\rule{0pt}{0\unitlength}} \\ \cline{2-4}
3&\bx{$\{2,3\}$} & \bx{$\{1,2,3\}$} &\bx{$\{1,2\}$}&\raisebox{-1\unitlength}{\rule{0pt}{0\unitlength}} \\ \cline{2-4}
2&\bx{$\{1,2\}$} & \bx{$\{1,2,3\}$} & \bx{$\{2,3\}$}&\raisebox{-1\unitlength}{\rule{0pt}{0\unitlength}} \\ \cline{2-4}
1&\bx{$\{1\}$} &\bx{$\{2\}$} & \bx{$\{3\}$}&\raisebox{-1\unitlength}{\rule{0pt}{0\unitlength}}\\ \cline{2-4}
\multicolumn{5}{l}{\hspace{.8\unitlength}
\begin{tabular}{cccc}
$\;\,\,$\bx{1}&\bx{2}&\bx{3}&$\;r$
\end{tabular}}
\end{tabular}
\hspace{.2in}
\begin{tabular}{r|c|c|c|l} 
\multicolumn{5}{l}{$s$}\\
\cline{2-4}
4&\bx{$(+)$} & \bx{$(0)$} & \bx{$(-)$}&\raisebox{-1\unitlength}{\rule{0pt}{0\unitlength}} \\ \cline{2-4}
3&\bx{$(0+)$} & \bx{$(d)$} &\bx{$(-0)$}&\raisebox{-1\unitlength}{\rule{0pt}{0\unitlength}} \\ \cline{2-4}
2&\bx{$(-0)$} & \bx{$(d)$} & \bx{$(0+)$}&\raisebox{-1\unitlength}{\rule{0pt}{0\unitlength}} \\ \cline{2-4}
1&\bx{$(-)$} &\bx{$(0)$} & \bx{$(+)$}&\raisebox{-1\unitlength}{\rule{0pt}{0\unitlength}}\\ \cline{2-4}
\multicolumn{5}{l}{\hspace{.8\unitlength}
\begin{tabular}{cccc}
$\;\,\,$\bx{1}&\bx{2}&\bx{3}&$\;r$
\end{tabular}}
\end{tabular}
   \caption{Boundary subsets $\calU_{(r,s)}$ for $m=4$. Chim's designation of the CBCs are shown on the right.}
  \label{figm4}
   \end{center}
\end{table}

\def\bx#1{\makebox[.9in]{#1}}
\begin{table}[p] 
\begin{center}
\begin{tabular}{r|c|c|c|c|l} 
\multicolumn{6}{l}{$s$}\\
\cline{2-5}
5&\bx{$\{4\}$} & \bx{$\{3\}$} & \bx{$\{2\}$}& \bx{$\{1\}$}&
\raisebox{-1\unitlength}{\rule{0pt}{0\unitlength}} \\ \cline{2-5}
4&\bx{$\{3,4\}$} & \bx{$\{2,3,4\}$} & \bx{$\{1,2,3\}$}& \bx{$\{1,2\}$}&
\raisebox{-1\unitlength}{\rule{0pt}{0\unitlength}} \\ \cline{2-5}
3&\bx{$\{2,3\}$} & \bx{$\{1,2,3,4\}$} &\bx{$\{1,2,3,4\}$}& \bx{$\{2,3\}$}&
\raisebox{-1\unitlength}{\rule{0pt}{0\unitlength}} \\ \cline{2-5}
2&\bx{$\{1,2\}$} & \bx{$\{1,2,3\}$} & \bx{$\{2,3,4\}$}& \bx{$\{3,4\}$}&
\raisebox{-1\unitlength}{\rule{0pt}{0\unitlength}} \\ \cline{2-5}
1&\bx{$\{1\}$} &\bx{$\{2\}$} & \bx{$\{3\}$}& \bx{$\{4\}$}&
\raisebox{-1\unitlength}{\rule{0pt}{0\unitlength}}\\ \cline{2-5}
\multicolumn{6}{l}{\hspace{.8\unitlength}
\begin{tabular}{cccccc}
$\;\,\,$\bx{1}&\bx{2}&\bx{3}&\bx{4}&$\;r$
\end{tabular}}
\end{tabular}
   \caption{Boundary subsets $\calU_{(r,s)}$ for $m=5$.}
  \label{figm5}
   \end{center}
\end{table}

Let us determine how the CBCs transform under 
height-reversal. 
The Dynkin diagram (\ref{figDynkin}) is evidently invariant under the
$Z_{2}$ transformation which takes node $n$ to node $m+1-n$ and
link $j$ to link $\bar j = m - j$.  By acting with this
transformation on the subgraphs (\ref{figid2}) and (\ref{figid1}), we
see that $j \mapsto m - j$ implies
\be
\calU_{(r,s)} &\mapsto& \calU_{(r,s')} = \calU_{(r',s)} \,,
\ee
where $r'$ and $s'$ are defined by (\ref{primes}), and we have 
made use of (\ref{uuprime}).
It follows that the transformation of CBCs under  
height-reversal is given by
\be
(r,s) &\mapsto& (r,s') =  (r',s) \,.  
\label{spinrev}
\ee
In particular, the CBCs $(r, {m+1\over 2})$ with $m$ odd, and the CBCs
$({m\over 2},s)$ with $m$ even, are invariant under 
height-reversal, corresponding to the cases $s=s'$ and $r=r'$, respectively.

\subsection{Elementary boundary subsets}

For later reference, it is convenient to define here
the \emph{elementary boundary subset}
\be
\calV_{(r,s)}
=\{b\in A_{m-1}\,|\,F_{bs}^r>0\}
=\{b\in A_m\,|\, \tilde{F}_{bs}^r \tilde{F}_{b+1\,s}^{r+1}>0\}\subseteq A_{m-1} \,,
\label{elemboundsubset}
\ee
where here $A_m=\{1, 2, ..., m\}$, and
$F^r$, $\tilde{F}^r$ are the fused adjacency matrices of $A_{m-1}$, $A_m$ as defined in Appendix~A. Note that the parity of $b\in\calV_{(r,s)}$ is the same as the parity of $r+s+1$. 
The elementary subset $\calV_{(r,s)}$ is given explicitly by
\be
\calV_{(r,s)} = \{ |s-r|+1 \,, |s-r|+3 \,, \ldots \,, k^{m}_{r,s} \} \,,
\label{elembsubset}
\ee
where $k^{m}_{r,s}$ is given by
\be
k^{m}_{r,s} = \min \{r+s-1 \,, 2m-1-r-s \} \,.
\label{maxheightelem}
\ee

We note that the boundary subset $\calU_{(r,s)}$ can be expressed as the disjoint union of two elementary boundary subsets
\be
\calU_{(r,s)}\!=\!\calV_{(r,s-1)}\cup\calV_{(r,s)}
\!=\!\calV_{(s-1,r)}\cup\calV_{(m-s,m-r)},\quad r\!=\!1,2,\ldots,m-1;\ \ s\!=\!1,2,\ldots,m
\label{Fsubset}
\ee
where ${\cal V}_{r,s}$ is the empty set whenever $r$ or $s$ is zero,
see (3.41) in \cite{Behrend:2000us}.
This disjoint union partitions the boundary subset $\calU_{(r,s)}$ into odd and even heights. 

\section{Boundary Scattering Theory}
\label{sec:Boundary}
 
Consider now ${\cal A}_{m} + (r,s)$ with both bulk and boundary $\phi_{1,3}$ 
perturbations~\cite{Ghoshal:1993tm},
\be
{\cal S} = {\cal S}_{{\cal A}_{m} + (r,s)} +\hat \lambda \int_{-\infty}^{\infty} dy 
\int_{-\infty}^{0} dx\  \Phi_{1,3}(x \,, y) 
+ h \int_{-\infty}^{\infty} dy\  \phi_{1,3}(y) 
\,, \qquad \hat\lambda < 0 
\,, \label{boundaction}
\ee
The boundary perturbation is present only when the fusion coefficient $N^{(r,s)}_{(1,3)\ (r,s)} \ne 0$ (cf., Eqs.  (\ref{bulkaction}),
(\ref{boundmasslessaction})). 
Ghoshal and Zamolodchikov have argued~\cite{Ghoshal:1993tm} that this 
two-parameter perturbation is integrable. The bulk parameter $\lambda$ sets the mass-scale, while  $h$ parametrizes integrable boundary conditions.

Following~\cite{Ghoshal:1993tm, Chim:1995kf}, we introduce the boundary operator
$B_{a}$, where $a$ labels the vacuum at the boundary.  
As explained in the previous section, we make the
important assumption that $a \in \calU_{(r,s)}$, where the boundary subset
$\calU_{(r,s)}$ is given by (\ref{symmbsubset}).  Multi-kink states have
the form
\be
K_{a_{1} \,, a_{2}}(\theta_{1})\ 
K_{a_{2} \,, a_{3}}(\theta_{2}) \ldots 
K_{a_{N} \,, a}(\theta_{N})\ B_{a} \,. \non 
\ee

\begin{figure}[htb]
	\centering
	\includegraphics[width=0.15\textwidth]{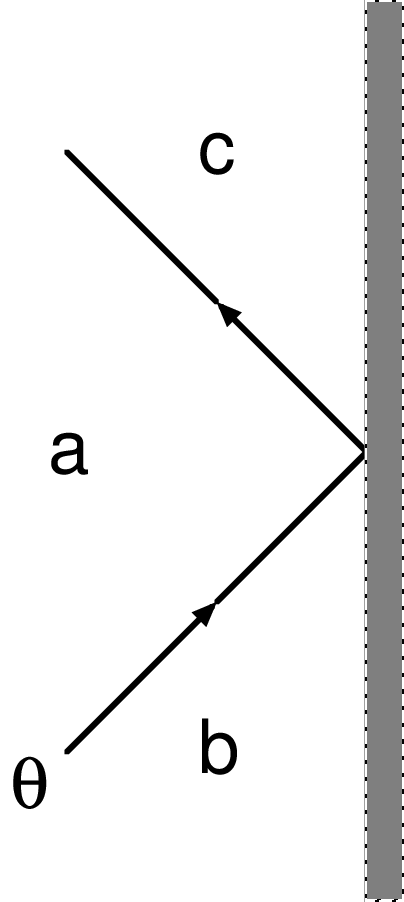}
	\caption[xxx]{\parbox[t]{0.6\textwidth}{
	Boundary reflection matrix $\RR{r}{s}{c}{a}{b}(\theta\,, \xi)$}
	}
	\label{fig:boundSmatrix}
\end{figure}

{$\!\!$}The kink boundary reflection matrix $\RR{r}{s}{c}{a}{b}(\theta \,, \xi)$ is
defined by the relation (see Figure~\ref{fig:boundSmatrix})
\be
K_{a \,, b}(\theta)\ B_{b} = \sum_{c} \RR{r}{s}{c}{a}{b}(\theta\,, \xi)\ 
K_{a\,, c}(-\theta)\ B_{c} \,.
\label{boundS0}
\ee
Evidently, 
\be
\RR{r}{s}{c}{a}{b}(\theta\,, \xi) = 0 \qquad 
\mbox{  unless both } b \,, c \in \calU_{(r,s)} \,. \label{bsubsetSmatrix}
\ee 
The parameter $\xi$ is related to the parameters $h$ and $\hat \lambda$ in
the action (\ref{boundaction}) in some way which at present is not known.  However, as
discussed for the case $m=4$ by Chim~\cite{Chim:1995kf}, $\xi={\pi\over 2}$
corresponds to $h=0$; and $h \mapsto -h$ corresponds to $\xi \mapsto
\pi - \xi$. Following Ghoshal and Zamolodchikov, and Chim,
we will assume that $h$ remains real as $\xi$ moves (i) from ${\pi\over 2}$ to $-{\Lambda_-}$ along the real axis and then to $-{\Lambda_-}\pm i\infty$ with $\Re\xi=-{\Lambda_-}$ and (ii) from ${\pi\over 2}$ to $\Lambda_+=\Lambda_-+\pi $ along the real axis and then to $\Lambda_+\pm i\infty$ with $\Re\xi=\Lambda_+$. The parameters $\Lambda_{\pm}$ have to be determined from the unknown $h-\xi$ relation, at a later stage. 
Varying $\xi$ along these paths will induce the boundary Renormalization Group (RG) flows and we will refer the two flows as $\xi < \frac{\pi}{2}$ and  $\xi > \frac{\pi}{2}$, respectively. 

\begin{figure}[H]
	\centering
	\includegraphics[width=0.4\textwidth]{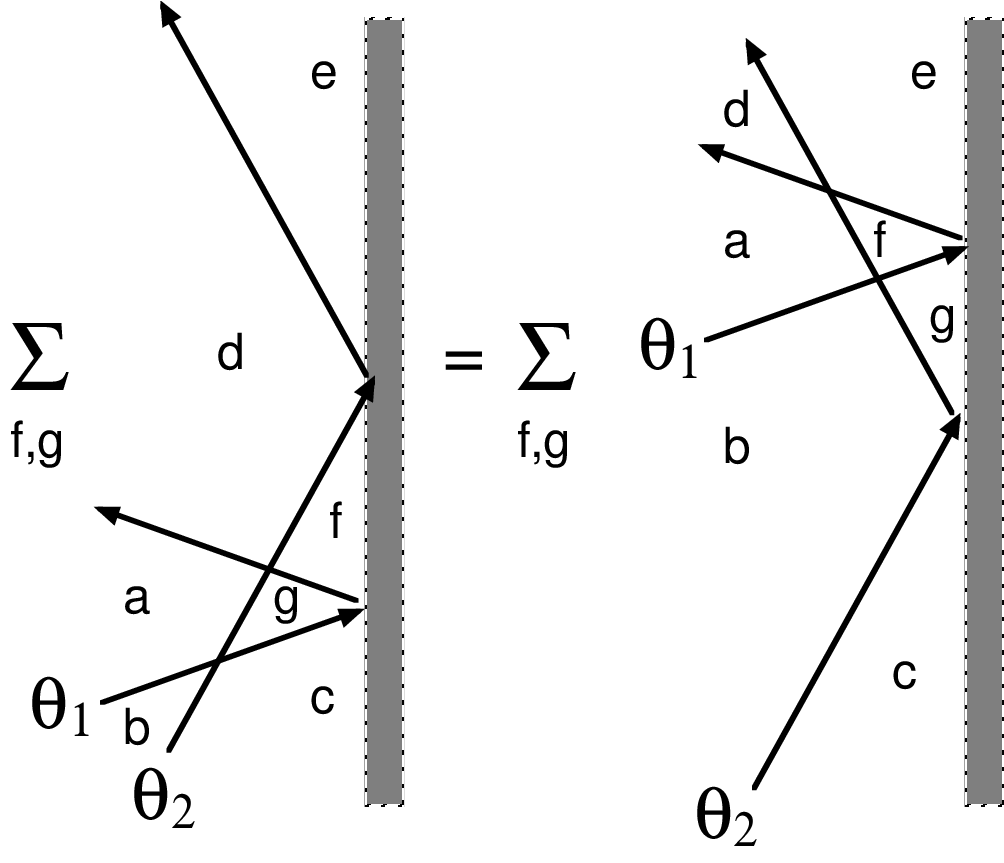}
\caption{The boundary Yang-Baxter equation.}
	\label{fig:BYBE}
\end{figure}

The boundary reflection matrix must obey the 
boundary unitarity relation
\be
\sum_{c}
\RR{r}{s}{c}{a}{b}(\theta\,, \xi)\ \RR{r}{s}{d}{a}{c}(-\theta\,, \xi)
= \delta_{b \,, d}\ G_{a\,, b}\ {\cal U}_d\ \,,
\label{boundunitarity}
\ee
where $G_{a\,, b}$ is the adjacency matrix (\ref{adjacency}), 
and ${\cal U}_{d}$ equals $1$ if $d$ is an allowed
boundary height (i.e., $d \in {\cal U}_{(r,s)}$) and equals zero otherwise. It must also obey
the recently-introduced boundary crossing-unitarity relation\footnote{In the old crossing the factor  $\left (\frac{[d]}{[b]}\right )^\frac{1}{2}$ is absent.} \cite{Shimamori:2025ntq}
\be
\RR{r}{s}{a}{b}{c}({i \pi\over 2}-\theta\,, \xi) 
= \sum_{d} \left (\frac{[d]}{[b]}\right )^\frac{1}{2}
S_{b\ c}^{a\ d}(2\theta)\  
\RR{r}{s}{a}{d}{c}({i \pi\over 2}+\theta\,, \xi) 
\,; \label{boundcrossunit}
\ee
and the boundary Yang-Baxter equation (BYBE)~\cite{Ghoshal:1993tm, Chim:1995kf, Ahn:1996nq, Cherednik:1984vvp} (see Figure \ref{fig:BYBE})
\be
\lefteqn{\sum_{f \,, g}
S_{b\ c}^{a\ g}(\theta_{1} - \theta_{2})\
\RR{r}{s}{f}{g}{c}(\theta_{1}\,, \xi)\
S_{g\ f}^{a\ d}(\theta_{1} + \theta_{2})\
\RR{r}{s}{e}{d}{f}(\theta_{2}\,, \xi)} \non  \\
&=& \sum_{f \,, g}\
\RR{r}{s}{g}{b}{c}(\theta_{2}\,, \xi)\
S_{b\ g}^{a\ f}(\theta_{1} + \theta_{2})\
\RR{r}{s}{e}{f}{g}(\theta_{1}\,, \xi)\
S_{f\ e}^{a\ d}(\theta_{1} - \theta_{2})
\,. \label{BYBE}
\ee

\begin{figure}[H]
	\centering
	\includegraphics[width=0.15\textwidth]{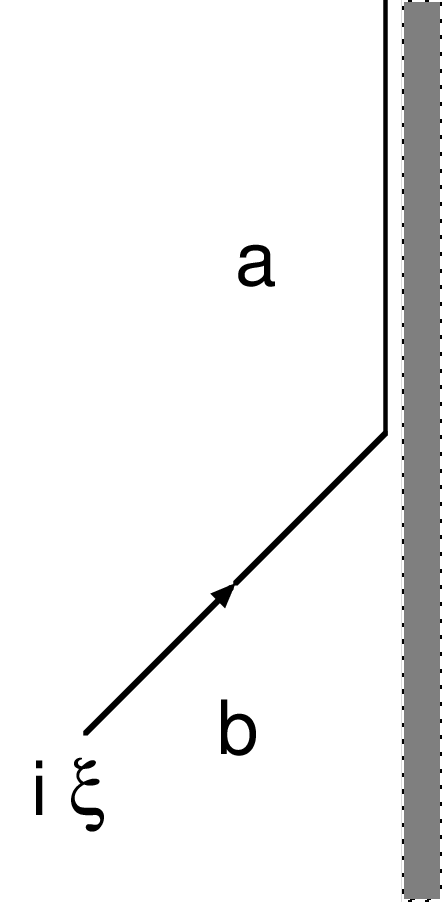}
\caption{A boundary bound state.}
	\label{fig:boundstate}
\end{figure}

\begin{figure}[H]
	\centering
	\includegraphics[width=0.155\textwidth]{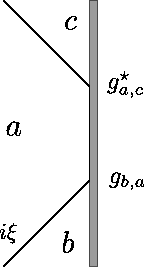} \hspace{2cm} \includegraphics[width=0.13\textwidth]{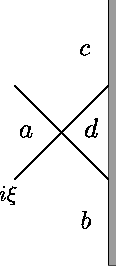} 
\caption{Simple poles of the reflection factors are related to the excitation or decay of boundary bound-states.}
	\label{fig:Bdrypole}
\end{figure}

\begin{figure}[H]
	\centering
	\includegraphics[width=0.4\textwidth]{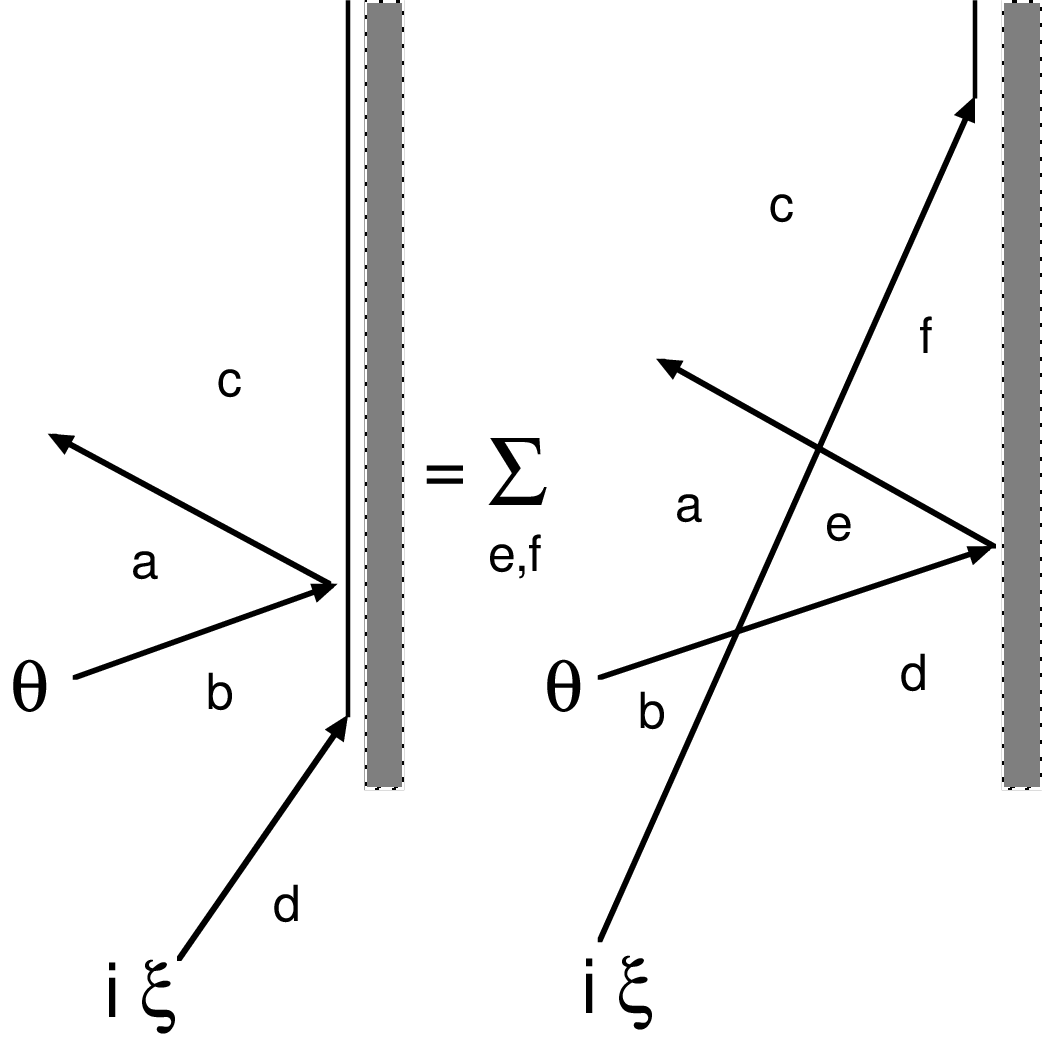}
\caption{The boundary bound-state bootstrap equation.}
	\label{fig:bootstrap}
\end{figure}

Following \cite{Ghoshal:1993tm, Chim:1995kf}, suppose a boundary state of height $a$ and
energy $e_{a}$ can be interpreted as a boundary bound state; i.e., as
a bound state of a kink and a boundary state of height $b$ and energy
$e_{b}$, such that (see Figure \ref{fig:boundstate})
\be
e_{a} = e_{b} + M \cos \xi \,,
\label{energyboundstate}
\ee 
where $0 < \xi < {\pi\over 2}$ and $M$ is the kink mass.  This can be
expressed formally as \cite{Bajnok:2002vp}
\be
B_{a} = {1\over g_{b, a}(\xi)} K_{a\,, b}(i \xi)\ B_{b} \,,
\label{boundstate}
\ee
where $g_{b, a}(\xi)$ is a particle-boundary coupling constant~\cite{Ghoshal:1993tm}, 
which is related to the residue of the  pole of the boundary reflection
matrix at $\theta=i\xi$ as  
\be 
-2i\, {\rm res} _{\theta=i\xi} \RR{r}{s}{c}{a}{b}(\theta\,, \xi)= g_{b,a}(\xi)g^\star_{c,a} (\xi) \,. 
\ee 

Simple poles of the reflection matrix are related to the 
excitation or the decay  of bound states, see Figure \ref{fig:Bdrypole}. 
For ground-state boundaries, a simple pole at $\theta =i\xi$ in the physical strip $0<\xi\leq \pi/2$ 
implies a boundary bound state. On the left, a particle with rapidity $i \xi$ 
can stick to the ground-state boundary, excite it and then decay, leading to a pole in the reflection process at $i\xi$. On the right, the excited boundary first decays and is then re-excited by the particle with rapidity $i\xi$. This leads to a pole in the reflection process at the same position $i\xi$, but for the excited boundary. Thus, the excited boundary always has the same pole as the 
ground-state boundary and so is related to the ground state and not to a new excited state. Depending on the parameter range, the roles of the ground state and the excited state can be interchanged. 

Integrability
implies that the boundary reflection matrix must also satisfy the boundary
bound-state bootstrap equation~\cite{Ghoshal:1993tm, Chim:1995kf}
\be
g_{d, b}(\xi)\ \RR{r}{s}{c}{a}{b}(\theta\,, \xi) = \sum_{e, f} g_{f, c}(\xi)\ 
S^{a\ e}_{b\ d}(\theta-i\xi)\ \RR{r}{s}{f}{e}{d}(\theta\,, \xi)\
S^{a\ c}_{e\ f}(\theta+i\xi) \,,
\label{bootstrap}
\ee 
as shown in Figure \ref{fig:bootstrap}.

Our objective
in the remainder of this paper is 
to determine the boundary reflection matrices $\RR{r}{s}{c}{a}{b}(\theta\,,
\xi)$ for all the possible 
initial Cardy CBCs $(r,s)$.

\section{Elementary Solutions}
\label{sec:Elementary}

A 
complete set of independent solutions of the BYBE (\ref{BYBE}) has been found by Behrend
and Pearce~\cite{Behrend:2000us}.  We 
call these ``elementary'' solutions,
since these are the building blocks with which we 
construct the boundary reflection matrices. Recall that the bulk scattering matrix corresponds to 
the bulk weights of the $A_{m-1}$ lattice model, so we need to use the boundary weights of this theory.

\subsection{Basic properties}

The (reduced) elementary solutions $\BP{p}{q}{c}{a}{b}(\theta, \xi,\mu)$ with $p,q=1,2,\ldots,m-1$
are given for $\xi\ne 0$ by (see   
(4.50) in \cite{Behrend:2000us}, with the replacements 
$u\mapsto -i \theta/m \,,\  \xi \mapsto \xi/m$, $r\mapsto p,\ s\mapsto q$  and $\mu=\pm 1$)
\be
\BP{p}{q}{c\pm 1}{c}{c\pm 1}\!\!(\theta, \xi,\mu) \!\!&\!\!=\!\!&\!\! 
{1\over [p] 
\sqrt{[c][c\pm 1]}}
\Big\{
[(p\mp c +q)/2] [(c\pm q \mp p)/2] 
s(i\xi + \theta)\ s(i(p \pi +\xi) - \theta) 
\non \\
&&\hspace{-.65in}\mbox{}+ [(p\pm c +q)/2] [(c \mp q \pm p)/2] 
s(i\xi - \theta)\ s(i(p \pi +\xi) + \theta) 
\Big\} F^{p}_{c\pm 1, q} \label{elem1}
\ee
and by
\be
\BP{p}{q}{c\mp 1}{c}{c\pm 1}\!\!(\theta, \xi,\mu) \!\!&\!\!=\!\!&\!\!
{\sqrt{[(p-c+q)/2] [(p+c-q)/2] [(c+q-p)/2] [(c+q+p)/2]}\over
\sqrt[4]{[c-1] [c+1]} \sqrt{[c]}} \non \\
\!\!&\!\!\times\!\!&\!\! \mu\,s(2\theta)\,F^{p}_{c+1, q} F^{p}_{c-1, q}\,, \label{elem2}
\ee
where the bracket notation $[*]$ is defined in (\ref{box}), and 
\be
s(x) = [-i x/\pi] = {\sinh ( {x\over m})
\over i\sin ({\pi\over m})} \,.
\ee
Here $F^{p}$ is the level-$p$ fused adjacency matrix
of $A_{m-1}$ as defined in Appendix \ref{app:Adjacency}.  
Its matrix elements $F^{p}_{a, b}$ are either 0 or 1 and are completely symmetric in the three indices. The scalar $\mu=\pm 1$ corresponds to a gauge freedom associated to the non-diagonal boundary amplitudes.

Although in the first instance we will be concerned with the region $\Re\xi>0$, we note that the reduced elementary solutions have the symmetry property
\be
\BP{p}{q}{c}{a}{b}(\theta, \xi,\mu) =  
-\BP{m-p}{m-q}{c}{a}{b}(\theta, -\xi,-\mu) \,.
\label{reducedsymmetry}
\ee 
They also have the ${\Bbb Z}_2$ symmetry
\be
\BP{p}{q}{c}{a}{b}(\theta, \xi,\mu) = 
\BP{p}{m-q}{\bar c}{\bar a}{\bar b}(\theta, \xi,\mu) \,,\qquad p,q=1,2,\ldots,m-1
\label{BPelemz2}
\ee

Recall from (\ref{elemboundsubset}) that the \emph{elementary boundary subset} $\calV_{(p,q)}$ is
\be
\calV_{(p,q)}=\{b\,|\,F^p_{bq}>0\}\subseteq A_{m-1} \,.
\ee
It follows from (\ref{elem1}), (\ref{elem2}) that
\be
\BP{p}{q}{c}{a}{b}(\theta\,, \xi,\mu) = 0 \qquad 
\mbox{  unless both } b \,, c \in \calV_{(p,q)} \,. 
\label{elembsubsetSmatrix}
\ee 
That is, the relation of $\calV_{(p,q)}$ to the elementary
solution $\BP{p}{q}{c}{a}{b}$ is the same as the relation of the
boundary subset $\calU_{(r,s)}$ to the boundary reflection matrix
$\RR{r}{s}{c}{a}{b}$ (\ref{bsubsetSmatrix}). This accounts 
for our choice of name for $\calV_{(p,q)}$. 
As is evident from (\ref{elembsubset}), the heights in the
elementary boundary subset $\calV_{(p,q)}$ are either all even (if
$|q-p|+1$ is even) or all odd (if $|q-p|+1$ is odd).  Correspondingly,
we assign to the elementary solutions $\BP{p}{q}{c}{a}{b}$ a 
\emph{parity} 
\be
\Pi_{(p,q)} = (-1)^{|p-q|+1} \,.
\label{elemparity}
\ee 
 
The quantities $\BP{p}{q}{c}{a}{b}(\theta \,, \xi,\mu)$ are solutions of the BYBE  (\ref{BYBE}). 
In order to normalize them correctly 
we define $\R{p}{q}{c}{a}{b}(\theta \,, \xi,\mu)$ by\,\footnote{With the old crossing an extra   $\left({[b][c]\over [a]^{2}}\right)^{-{\theta \over {2 \pi i} }}$ factor would be included here.}
\be
\R{p}{q}{c}{a}{b}(\theta \,, \xi,\mu) 
= {V^{(p,q)}(\theta \,, \xi,\mu)\over s(i\xi)\, s(i(p\pi+\xi))}
\left({[b][c]\over [a]^{2}}\right)^{{1\over 4}} 
\BP{p}{q}{c}{a}{b}(\theta \,, \xi,\mu) \,,
\label{elemgauged}
\ee
where $\BP{p}{q}{c}{a}{b}(\theta \,, \xi,\mu)$ is given by
(\ref{elem1}), (\ref{elem2}).
For later convenience, we have introduced a
$\theta$-independent gauge factor to help to satisfy boundary unitarity
(\ref{elemboundunitarity}) and boundary crossing  (\ref{boundunitarity}), as well as a scalar factor 
$V^{(p,q)}(\theta \,, \xi,\mu)$ that we determine below 
(\ref{boundaryscalarfactor}). Just as the bulk scalar factor was related to the bulk free energy of the associated critical $A_{m-1}$ lattice model, so these boundary scalar factors will be related to the boundary free energies of the associated critical $A_{m-1}$ lattice model as in Appendix~C. 
In the first instance, we restrict $\xi$ to the region ${\pi\over 2}\le\Re\xi\le {3\pi\over 2}$ corresponding to the physical region of the lattice models. The analytic continuation to the region $0<\Re\xi<{\pi\over 2}$ will then yield the ground state in the field theory.

Despite our notation, it will turn out that the scalar factors are independent of $q$ and $\mu$.
It follows that the elementary solutions 
enjoy the $Z_{2}$ symmetry
\be
\R{p}{q}{c}{a}{b}(\theta, \xi,\mu) = 
\R{p}{m-q}{\bar c}{\bar a}{\bar b}(\theta, \xi,\mu) \,,\qquad p,q=1,2,\ldots,m-1
\label{elemz2}
\ee
where $\bar a = m-a$, since the scalar factors satisfy
\be
V^{(p,q)}(\theta, \xi,\mu) =
V^{(p,m-q)}(\theta, \xi,\mu)
\label{scalarconstraint1}
\ee
being independent of $q$.

\subsection{Boundary unitarity and boundary crossing}

The elementary solution (\ref{elemgauged}) also satisfies a boundary
unitarity relation (cf,  (\ref{boundunitarity}))
\be
\lefteqn{\sum_{c}
\R{p}{q}{c}{a}{b}(\theta\,, \xi,\mu)\ \R{p}{q}{d}{a}{c}(-\theta\,, \xi,\mu)
= \delta_{b \,, d}\ G_{a\,, b}\ \BB_{d}}  
\label{elemboundunitarity}\\
&\times& V^{(p,q)}(\theta \,, \xi,\mu)\ V^{(p,q)}(-\theta \,, \xi,\mu)\ 
{s(i(p \pi +\xi) + \theta)\ s(i(p \pi +\xi) - \theta)\;s(i \xi + \theta)\ s(i \xi - \theta)\over
s(i\xi)^2s(i(p\pi+\xi))^2} \,,\non
\ee
where $G_{a\,, b}$ is the adjacency matrix (\ref{adjacency}), 
and $\BB_{d}$ equals $1$ if $d$ is an allowed
boundary height (i.e., $d \in \calV_{(p,q)}$) and equals zero otherwise.
The boundary crossing-unitarity relation (cf,  
(\ref{boundcrossunit})) is
\be
\lefteqn{V^{(p,q)}({i \pi\over 2}-\theta\,, \xi,\mu) \BP{p}{q}{a}{b}{c}({i \pi\over 2}-\theta\,, \xi,\mu)} \non \\*
&=&  U(2\theta)\, V^{(p,q)}({i \pi\over 2}+\theta\,, \xi,\mu)\sum_{d}
{\bar S}_{b\ c}^{a\ d}(2\theta)\  
\BP{p}{q}{a}{d}{c}({i \pi\over 2}+\theta\,, \xi,\mu) \,.
\label{elemboundcrossunit}
\ee
Note that the $\theta$-independent gauge factors just cancels with the new factor in the crossing unitarity relation  (\ref{boundcrossunit}).

The unitarity and crossing relations for the elementary solutions
(\ref{elemboundunitarity}) and (\ref{elemboundcrossunit}) can be made 
to coincide with those for the boundary reflection matrix 
(\ref{boundunitarity}) and (\ref{boundcrossunit}) by demanding
\be
V^{(p,q)}(\theta \,, \xi,\mu)\,V^{(p,q)}(-\theta, \xi,\mu)\,
{s(i(p \pi +\xi) + \theta)\, s(i(p \pi +\xi) - \theta)\,s(i \xi + \theta)\, s(i \xi - \theta)\over
s(i\xi)^2s(i(p\pi+\xi))^2} = 1 \quad
\label{Vrelation1}
\ee
and
\be 
V^{(p,q)}(\frac{i \pi}{2}-\theta, \xi,\mu)=
 U(2\theta)\, \sinh\Big(\frac{i\pi+2\theta}{m}\Big)\,V^{(p,q)}(\frac{ i\pi}{2} +\theta\,, \xi,\mu)  \,,
\label{Vrelation2}
\ee
respectively. Note that iteration of (\ref{Vrelation2}) gives the 
unitarity relation for $U(\theta)$ (\ref{Urelations}), as required for 
consistency.

\subsection{Boundary bound-state bootstrap} 

We have shown that the elementary solutions (\ref{elemgauged}) satisfy
the boundary unitarity (\ref{boundunitarity}), boundary crossing
(\ref{boundcrossunit}) and boundary Yang-Baxter 
equations (\ref{BYBE}).  However, these are not the sought-after boundary reflection
matrices, since they do not satisfy the boundary bound-state bootstrap
equation (\ref{bootstrap}).  Indeed, we find that the elementary solutions satisfy instead
\be
\lefteqn{-{s(i(\pi-\xi))\,s(i(\xi+p\pi))\, s(i\xi+\theta)\,s( i (\xi +\pi)-\theta)\over 
 V^{(m-p-1,m-q)}(\theta, \pi -\xi,\mu)}\,
 \bar g_{d, b}(\xi)\,\R{m-p-1}{m-q}{c}{a}{b}\!(\theta, \pi - \xi,\mu)}
\label{elembootstrap0} \\
\mbox{}\!\!\!&=&\!\!\!{s(i\xi)\,s(i(\xi+p\pi))\over \sin^{2}({\pi\over m}) 
U(\theta\! -\! i \xi) U(\theta \!+\! i \xi) V^{(p,q)}(\theta, \xi,\mu)}
\sum_{e, f} \bar g_{f, c}(\xi)
S^{a\ e}_{b\ d}(\theta-i\xi) \R{p}{q}{f}{e}{d}\!\!(\theta, \xi,\mu) 
S^{a\ c}_{e\ f}(\theta+i\xi) \non
\ee
for $p,q  = 1\,, 2\,, \ldots \,, m-1$. 
This follows by viewing  the fusion construction of Behrend and Pearce~\cite{Behrend:2000us} iteratively as a map from 
$\R{p}{q}{f}{e}{d}\!\!(\theta, \xi,\mu)$ to $-\R{p+1}{q}{c}{a}{b}\!\!(\theta,\xi-\pi,-\mu)$ and 
then applying (\ref{reducedsymmetry}). 
The particle-boundary coupling constants are given by 
\be
\bar g_{a\pm 1,a}(\xi)=\sqrt{\frac{[(\pm p \mp q-a)/2][(a\mp p\mp q)/2][2\xi/\pi]}{[a]}} \,,
\label{particle-boundary}
\ee
where the $\xi$-independent factors are related to the lattice fusion vectors of Behrend and Pearce. They can be calculated from
 $\BP{p}{q}{c}{a}{b}\!\!(i\xi,\xi,\mu)\left({[b][c]\over [a]^{2}}\right)^{{1\over 4}}=  \bar g_{b,a}(\xi)\bar g^\star_{c,a}(\xi)$. The quantities  
$\bar g_{b,a}(\xi)$ are real and  can be extracted both from the 
 diagonal and the off-diagonal reflections, which give the same result. 
Imposing the constraint
\begin{eqnarray}
V^{(m-p-1,m-q)}(\theta, \pi\!-\!\xi,\mu)=\hspace{10cm}&&\nonumber\\
-\sin^{2}({\pi\over m})\,
{s(i(\pi\!-\!\xi))s(i\xi\!+\!\theta)s(i(\xi\! +\!\pi)\!-\!\theta) 
\over s(i\xi)}\, 
U(\theta \!-\! i \xi) U(\theta \!+\! i \xi) V^{(p,q)}(\theta, \xi,\mu) \,,\quad&&
\label{bootstrapconstraint}
\end{eqnarray}
we conclude that the elementary solutions obey 
the boundary bound-state bootstrap relation
\be
{\!\!\!\!}g_{d, b}(\xi)\R{m-p-1}{m-q}{c}{a}{b}\!(\theta, \pi\!-\!\xi,\mu) \!=\! \sum_{e, f} g_{f, c}(\xi)
S^{a\ e}_{b\ d}(\theta\!-\!i\xi)\R{p}{q}{f}{e}{d}\!(\theta, \xi,\mu)
S^{a\ c}_{e\ f}(\theta\!+\!i\xi)
\label{elembootstrap}
\ee 
where $g_{a,b}(\xi)=-i\,{\rm res}_{\theta =i\xi}\,\frac{V^{(p,q)}(\theta,\xi,\mu)}{s(i\xi)s(i(p\pi +\xi))}\,\bar g_{a,b}(\xi)$.

In contrast to (\ref{bootstrap}),
the $(p,q)$ elementary solution is mapped to the $(m\!-\!p\!-\!1,m\!-\!q)$
elementary solution under bootstrap. Clearly, this is an involution so that the bootstrap can close immediately. For $0 \le\Re\xi\le {\pi\over 2}$ this bootstrap builds the amplitudes of the excited boundary boundstate $\R{m-p-1}{m-q}{c}{a}{b}\!(\theta, \pi\!-\!\xi,\mu)$ from the amplitudes of the groundstate $\R{p}{q}{f}{e}{d}\!(\theta, \xi,\mu)$. Under the involution $m\mapsto m\!-\!1\!-\!p$, $q\mapsto m\!-\!q$, $\xi\mapsto \pi\!-\!\xi$, the roles of the groundstate and boundary boundstate are interchanged and are related to changing the sign of $h$ in the action.

\subsection{QFT scalar factors}

We now solve the  boundary unitarity  (\ref{Vrelation1}) and crossing relations
(\ref{Vrelation2}) for the scalar factors $V^{(r,s)}(\theta \,, \xi,\mu)$. Writing 
\be
V^{(r,s)}(\theta,\xi,\mu)={V_0(\theta)\,V_{CDD}(\theta,\xi)\over 
V_r(\theta,\xi)P_0(\theta)} \,,
\label{boundaryscalarfactor}
\ee
we see that (\ref{Vrelation1}) and (\ref{Vrelation2}) are satisfied if
the individual factors obey the following relations 
\be
V_p(\theta, \xi)\ V_p(-\theta, \xi)
={s(i\xi+\theta)s(i\xi-\theta)s(i(\xi+p\pi)+\theta)s(i(\xi+p\pi)-\theta)\over s(i\xi)^2s(i(\xi+p\pi))^2} \,, 
\ee
\vspace{-24pt}
\be
&\quad&V_p(\theta,\xi)=V_p(\pi i-\theta,\xi) \,,
\label{Vrreltn}\\[0pt]
P_{0}(\theta)\ P_{0}(-\theta)=1 \,, 
 &\quad&
P_0(\theta)=-{1\over \sinh({2i \pi-2\theta \over m})\,U(2\theta)}\,P_0(\pi i-\theta) \,,
\label{P0reltn}\\[0pt]
V_0(\theta)V_0(-\theta)=1 \,,&\quad& 
V_0(\theta)=-V_0(\pi i-\theta)\label{V0reltn} \,, \\[0pt]
V_{\mathrm {CDD}}(\theta, \xi)\ V_{\mathrm {CDD}}(-\theta, \xi) =1 \,, 
&\quad&
V_{\mathrm {CDD}}(\theta, \xi)=V_{\mathrm {CDD}}(\pi i-\theta, \xi) \,.
\label{Vcddreltn}
\ee
Substituting the crossing relation for $P_0(\theta)$ into its inversion relation and using the bulk inversion relation gives
\be
P_0(\pi i-\theta)P_0(\pi i+\theta)={s(2\pi i-2\theta)s(2\pi i+2\theta)\over s(\pi i-2\theta)s(\pi i+2\theta)} \,.
\ee

Making the replacements $\theta\mapsto imu$ and $\xi\mapsto m\xi$ with $\lambda=\pi/m$, the above inversion and crossing relations become precisely the lattice equations solved in Appendix~C in the context of the boundary free energies of the critical $A_{m-1}$ lattice model. Taking the solutions for 
$\kappa_r(u,\xi)$, $p_0(u)$, $v_0(u)$ and $v_{CDD}(u,\xi)$ from Appendix~C and transforming back to the QFT notation via $u\mapsto {\theta\over im}$, $\xi\mapsto {\xi\over m}$ with 
$\lambda={\pi\over m}$ gives scalar factors of the form (\ref{boundaryscalarfactor})
which are thus independent of $s$ and $\mu$. After replacing $t\mapsto mt$ in the integrals we find
\be
V_p(\theta,\xi)
&\!\!=\!\!&\exp\big(\!-2\,{\cal I}_p(\theta,\xi)\big) \,,
\qquad {\pi\over 2}\le\Re\,\xi\le{3\pi\over 2} \,,
\ee
\be
{\cal I}_p(\theta,\xi)= 
\int_{-\infty}^\infty 
{\cosh{(((m\!-\!p)\pi\!-\!2\xi)t)}\cosh{(p\pi t)}\sinh (i \theta t)\sinh((\pi\!+\!i\theta)t)
\over t\,\sinh({m\pi t})\cosh(\pi t)}\,dt \,,
\ee
\be
P_0(\theta)=\exp\!\Big(\!\!-2\!\int_0^\infty {\sinh({(m-3)\pi t\over 2})\sinh({\pi t\over 2})\sinh (2i\theta t)\over
t\,\sinh({m\pi t\over 2})\sinh(2\pi t)}\,dt\Big) \,,
\label{P0}
\ee
\be
V_0(\theta)= i\tanh({i\pi\over 4}-{\theta\over 2}) \,,
\qquad 
V_{\mathrm {CDD}}(\theta, \xi) = {\sin \xi - i \sinh \theta\over \sin \xi + i \sinh \theta} \,,
\label{VCDD}\label{V0}
\ee 
where this solution can be extended by analytic continuation into the region 
$0<\Re\,\xi<{\pi\over 2}$. Under this continuation the integral representation for ${\cal I}_p(\theta,\xi)$ remains valid and $V_{\mathrm {CDD}}(\theta, \xi)$ stays the same.

These QFT scalar factors $V^{(p,q)}(\theta, \xi,\mu)$ also satisfy the bootstrap constraint (\ref{bootstrapconstraint}). Again, after changing notation, this follows from identity (\ref{Identity2}) of Appendix~C after use of (\ref{Identity1}).

\section{Paired Solutions}
\label{sec:Pairs}

We have seen that the elementary solutions $\R{p}{q}{c}{a}{b}(\theta, \xi,\mu)$ (\ref{elemgauged}) satisfy the requirements of boundary
unitarity (\ref{boundunitarity}), boundary crossing
(\ref{boundcrossunit}) and boundary Yang-Baxter (\ref{BYBE}).
However, they do not satisfy the boundary bound-state
bootstrap equation (\ref{bootstrap}).  Indeed, we have seen
(\ref{elembootstrap}) that the bootstrap relates the $(p,q)$ and
$(m-p-1,m-q)$ solutions.
In this section we construct ``paired solutions'' that satisfy all of
these requirements by forming ``direct sums'' of certain
pairs of elementary solutions.  
\goodbreak

\subsection{Direct sums}

Recall from (\ref{Fsubset}) that 
the boundary subset $\calU_{(r,s)}$ is the \emph{disjoint} union of the elementary boundary subsets
$\{ \calV_{(s-1,r)} \,, \calV_{(m-s,m-r)}\}$ 
\be
\calU_{(r,s)} = \calV_{(s-1,r)} \cup \calV_{(m-s,m-r)} 
\,, \qquad r = 1 \,, 2\,, \ldots \,, m-1 \,, \qquad
s = 1 \,, 2 \,, \ldots \,, m \,.
\label{setpartition}
\ee
corresponding to the partition of $\calU_{(r,s)}$ into odd and even heights.
This result also follows from the formulas for the boundary subset
$\calU_{(r,s)}$ (\ref{symmbsubset}), (\ref{minmaxheight})
and for the elementary boundary subset $\calV_{(r,s)}$
(\ref{elembsubset}), (\ref{maxheightelem}).
Note that $\calV_{(0,r)}$ and $\calV_{(m,r)}$ are null sets; thus, the result
(\ref{setpartition}) for the special cases $s=1$ and $s=m$ reduces to
\be
\calU_{(r,1)} = \calV_{(m-1,m-r)} = \{ r \}  \qquad \mbox{ and  } \qquad 
\calU_{(r,m)} = \calV_{(m-1,r)} = \{ m-r \}  \,.
\label{s1case}
\ee 

We are now ready to define the notion of a direct sum of elementary
solutions.
 
 {\bf Definition}: Let  $\R{r_{1}}{s_{1}}{c}{a}{b}$, 
 $\R{r_{2}}{s_{2}}{c}{a}{b}$
 be a pair of elementary solutions 
and let $\calU_{(r,s)} = \calV_{(r_{1},s_{1})} \cup \calV_{(r_{2},s_{2})}$ be the disjoint union of
the elementary boundary subsets.
The {\bf direct sum} of these
 elementary solutions, whose boundary subset is $\calU_{(r,s)}$
 and which we denote by
 \be
 \RR{r}{s}{c}{a}{b} = \R{r_{1}}{s_{1}}{c}{a}{b} \oplus
 \R{r_{2}}{s_{2}}{c}{a}{b} \,, \non 
 \ee
 is given by $\RR{r}{s}{c}{a}{b} = \R{r_{j}}{s_{j}}{c}{a}{b}$
 for  $b \,, c \in \calV_{(r_{j},s_{j})}$.
 
This definition is unambiguous, since $\RR{r}{s}{c}{a}{b} =0$ unless
both $b, c \in \calU_{(r,s)}$; and if $b, c \in \calU_{(r,s)}$, then either
$b, c \in \calV_{(r_{1},s_{1})}$ or $b, c \in \calV_{(r_{2},s_{2})}$, but not
both, since $\{ \calV_{(r_{1},s_{1})} \,, \calV_{(r_{2},s_{2})} \}$ 
are disjoint.

In particular, 
we define the paired solutions $\RR{r}{s}{c}{a}{b}(\theta, \xi,\mu)$ by the 
direct sum 
\be
\RR{r}{s}{c}{a}{b}(\theta, \xi,\mu) &=& 
\R{s-1}{r}{c}{a}{b}(\theta, \xi,\mu) \oplus
\R{m-s}{m-r}{c}{a}{b}(\theta, \pi-\xi,\mu) \,, \label{calR} \\
& & \qquad r = 1 \,, 2\,, \ldots \,, m-1 \,, \qquad
s = 2 \,, 3 \,, \ldots \,, m-1\,, \non
\ee
where the elementary solutions are given by (\ref{elemgauged}).

There are special cases when the solution has only one single reflection factor and not two paired ones. This happens for the boundary-parameter-free cases $s=1$ and $s=m$. In this case we define
\be
\RR{r}{1}{c}{a}{b}(\theta)=\RR{m-r}{m}{c}{a}{b}(\theta)
=\lim_{\xi\to\pm i\infty} \R{r}{1}{c}{a}{b}(\theta,\xi,\mu),\quad r=1,2,\ldots,m-1 \,.
\label{calR1}
\ee
In this limit the reflection matrix simplifies considerably. First, all non-diagonal elements vanish and the rapidity dependence disappears from $\BP{r}{1}{c}{a}{b}(\theta \,, \pm i \infty ,\mu)$. The prefactor also simplifies and the $r$ dependence is only through the RSOS condition
\be
\RR{r}{1}{c\pm 1}{c}{c\pm 1}(\theta )=\RR{m-r}{m}{c\pm 1}{c}{c\pm 1}(\theta )\equiv \R{r}{1}{c\pm 1}{c}{c\pm 1}(\theta)=\frac{V_0 (\theta)}{P_0(\theta)} F^{r}_{c\pm 1, 1} 
\label{xiinfty}
\ee 
Although these boundary reflections are diagonal, the non-invertible symmetries do not leave the Cardy boundary conditions $(r,1)$ and $(m-r,m)$ invariant (see Section \ref{sec:Noninvertible}), so the solution is not related to the one in \cite{Shimamori:2025ntq}, which was obtained for invariant boundaries.

For general boundary conditions $(r,s)$, the paired solutions (\ref{calR}) are not diagonal. 
We claim that the paired solution (\ref{calR}) satisfies BYBE
(\ref{BYBE}), boundary unitarity (\ref{boundunitarity}), boundary
crossing (\ref{boundcrossunit}), and boundary bound-state bootstrap
(\ref{bootstrap}).
Indeed, the individual elementary solutions $\R{s-1}{r}{c}{a}{b}$ and
$\R{m-s}{m-r}{c}{a}{b}$ separately satisfy the BYBE, and they involve boundary
heights of definite (even or odd) parity (\ref{elemparity}).  Since
the BYBEs do not mix amplitudes with different parities, the direct sum
of these solutions also satisfies the BYBE. Moreover, since
$\R{s-1}{r}{c}{a}{b}$ and $\R{m-s}{m-r}{c}{a}{b}$ separately satisfy boundary
unitarity and boundary crossing, and the corresponding elementary
boundary subsets $\calV_{(s-1,r)}$ and $\calV_{(m-s,m-r)}$ are disjoint, it
immediately follows that the direct sum of these solutions also
satisfies boundary unitarity and boundary crossing.  The individual
solutions $\R{s-1}{r}{c}{a}{b}$ and $\R{m-s}{m-r}{c}{a}{b}$ do {\it not} 
separately satisfy boundary bound-state bootstrap.  However, 
$\R{s-1}{r}{c}{a}{b}(\theta, \xi,\mu)$ is
mapped to $\R{m-s}{m-r}{c}{a}{b}(\theta, \pi-\xi,\mu)$ and vice versa by the bootstrap 
(\ref{elembootstrap}).  Hence, the direct sum (\ref{calR}) satisfies
boundary bound-state bootstrap.

We remark that the  
disjoint union (\ref{setpartition}) is not the only
possible partition of boundary subsets $\calU_{(r,s)}$ into disjoint pairs 
of elementary boundary subsets $\calV_{(r_{j},s_{j})}$. Indeed, 
$\{\calV_{(r,s-1)} \,, \calV_{(r,s)}\}$ is another such 
disjoint union. 
It is the bootstrap relation (\ref{elembootstrap})
that singles out the particular boundary subsets (\ref{setpartition}) that we use to construct paired solutions.

\subsection{Symmetries of paired solutions}

The paired solutions (\ref{calR}) have the Kac table symmetry
\be
\RR{r}{s}{c}{a}{b}(\theta, \xi,\mu) =  
\RR{r'}{s'}{c}{a}{b}(\theta, \pi-\xi,\mu) \,,
\label{calRsymm}
\ee
where $r'$ and $s'$ are defined by (\ref{primes}). Indeed, 
\be
\RR{r'}{s'}{c}{a}{b}(\theta, \pi-\xi,\mu) &=&
\R{s'-1}{r'}{c}{a}{b}(\theta,\pi-\xi,\mu) \oplus
\R{m-s'}{m-r'}{c}{a}{b}(\theta, \xi,\mu) \non \\
&=&\R{m-s}{m-r}{c}{a}{b}(\theta, \pi-\xi,\mu) \oplus
\R{s-1}{r}{c}{a}{b}(\theta,\xi,\mu) \non \\
&=& \RR{r}{s}{c}{a}{b}(\theta, \xi,\mu)
\ee
where the conventional order of the pair has been reversed, that is, the role of the groundstates and bound states has been interchanged.

One can also prove that the paired solutions have the $Z_{2}$ 
symmetries
\be
\RR{r}{s}{c}{a}{b}(\theta, \xi,\mu) =
\RR{r'}{s}{\bar c}{\bar a}{\bar b}(\theta, \xi,\mu)
=
\RR{r}{s'}{\bar c}{\bar a}{\bar b}(\theta,\pi-\xi,\mu) \,,
\label{calRZ2symmetry}\label{mainZ2}
\ee
where $\bar a = m - a$. Indeed, using the $Z_{2}$ symmetry 
(\ref{elemz2}) of the elementary solutions,
\be
\RR{r}{s}{c}{a}{b}(\theta, \xi,\mu) &=&
\R{s-1}{r}{c}{a}{b}(\theta, \xi,\mu) \oplus
\R{m-s}{m-r}{c}{a}{b}(\theta, \pi-\xi,\mu) \non \\
&=& \R{s-1}{r'}{\bar c}{\bar a}{\bar b}(\theta, \xi,\mu)
\oplus \R{m-s}{m-r'}{\bar c}{\bar a}{\bar b}(\theta, \pi-\xi,\mu) \non \\
&=& \RR{r'}{s}{\bar c}{\bar a}{\bar b}(\theta, \xi,\mu) \,,
\label{rZ2}
\ee 
and the last relation in (\ref{calRZ2symmetry}) then follows from
(\ref{calRsymm}).

 \subsection{CDD factors and pole structure}
\label{subsec:CDD}

Consider a paired solution $\mathcal{R}^{(r,s)}$ with $s \ne 1 \,, m$.
Following \cite{Ghoshal:1993tm, Chim:1995kf}, we choose the CDD factor
$V^{(r,s)}_{\mathrm{CDD}}(\theta, \xi)$ in (\ref{boundaryscalarfactor}) so
that the pole structure of the paired solution is consistent with the
boundary bound-state bootstrap.  In particular, a nonvanishing
amplitude $\R{s-1}{r}{c}{a}{b}(\theta, \xi,\mu)$ should have a simple
pole in the physical strip $0<\Im\theta<{\pi\over 2}$ at $\theta = i \xi$ for $0 < \xi <
{\pi\over 2}$ if and only if $a \in \calV_{(m-s,m-r)}=\calV_{(s,r)}$.  
This is consistent with the interpretation of the analytic continuation of
$\R{s-1}{r}{c}{a}{b}(\theta, \xi,\mu)$ into the region $0<\Re\xi<{\pi\over 2}$
as the ground state reflection matrix. 
Moreover, a
nonvanishing amplitude $\R{m-s}{m-r}{c}{a}{b}(\theta, \pi-\xi,\mu)$ should
have a simple pole in the physical strip at $\theta = i (\pi - \xi)$ if and only if $a \in\calV_{(s-1,r)}$. 
This is consistent with the interpretation of the analytic continuation of
$\R{m-s}{m-r}{c}{a}{b}(\theta,\pi-\xi,\mu)$ into the region ${\pi\over 2}<\Re\xi<\pi$
as the ground state reflection matrix.
To this end, we propose the specific CDD factor 
\be
V_{\mathrm{CDD}}(\theta, \xi) =
{\sin \xi - i \sinh \theta\over 
\sin \xi + i \sinh \theta}
={\tan({\xi-i\theta\over 2})\over \tan({\xi+i\theta\over 2})}
\,. \label{cdd}
\ee 
which is a solution of (\ref{Vcddreltn}). 
It is also verified that the ``reduced'' elementary solution
$\BP{s-1}{r}{c}{a}{b}(\theta, \xi,\mu)$ (see (\ref{elem1}),
(\ref{elem2})) has a simple zero at $\theta = i \xi$ (which cancels
the CDD pole) only if $a \notin \calV_{(s,r)}$ and, similarly,
$\BP{s}{r}{c}{a}{b}(\theta, \pi-\xi,\mu)$ has a simple zero at $\theta = i
(\pi - \xi)$ (which cancels the CDD pole) only if $a \notin
\calV_{(s-1,r)}$.

Let us note that the CDD factor at $h=0$, i.e. at $\xi=\frac{\pi}{2}$, has a double pole. At this kinematical point, all vacua from ${\cal U}_{(r,s)}$ are degenerate and can be connected by the boundary bootstrap. In order to ensure this, the prefactor $V_0(\theta)$ renders the double pole to a single one. 

We also note that the CDD factor $V_{\mathrm{CDD}}$ appears also in the excited state reflection matrices: in $\R{m-s}{m-r}{c}{a}{b}(\theta, \pi-\xi,\mu)$ for $0 < \xi <
{\pi\over 2}$ and in $\R{s-1}{r}{c}{a}{b}(\theta, \xi,\mu)$ for ${\pi\over 2}<\Re\xi<\pi$. The corresponding pole in the excited-state reflection factor can be explained by the decay process, see Figure \ref{fig:Bdrypole}.

\newcommand{\spos}[2]{\makebox(0,0)[#1]{$\sm{#2}$}}
\newcommand{\sm}[1]{{\scriptstyle #1}}

\def\mytri#1#2#3{\setlength{\unitlength}{12mm}\raisebox{-0.7\unitlength}[0.8\unitlength][0.7\unitlength]
{\righttri{#1}{#2}{#3}{}}}
\newcommand{\righttri}[4]{\begin{picture}(1.1,1.6)
\put(0.8,0.3){\line(0,1){1}}\put(0.8,0.3){\line(-1,1){0.5}}
\put(0.8,1.3){\line(-1,-1){0.5}}
\put(0.84,0.26){\spos{tl}{#1}}\put(0.24,0.8){\spos{r}{#2}}
\put(0.84,1.34){\spos{bl}{#3}}
\put(0.72,0.8){\spos{r}{#4}}\end{picture}}
\setlength{\unitlength}{8mm}
\psset{unit=8.5mm}
\def\bc#1#2{
\hspace{.25\unitlength}
\begin{picture}(3.5,.5)
\put(0,0){\line(1,0){3}}
\put(#1,-.05){\rule{#2 \unitlength}{.1 \unitlength}}
\multiput(0,-.1)(1,0){4}{\line(0,1){.2}}
\end{picture}}
\def\bcdn#1#2{
\hspace{.25\unitlength}
\begin{picture}(3.5,.5)
\put(0,0){\line(1,0){3}}
\psline[linewidth=2pt,linecolor=black,arrowsize=0pt]{->}(#1,0)(#2,0)
\multiput(0,-.1)(1,0){4}{\line(0,1){.2}}
\end{picture}}
\def\bcd#1#2{
\hspace{.25\unitlength}
\begin{picture}(3.5,.5)
\put(0,0){\line(1,0){3}}
\psline[linewidth=2pt,linecolor=black,arrowsize=7pt]{->}(#1,0)(#2,0)
\multiput(0,-.1)(1,0){4}{\line(0,1){.2}}
\end{picture}}
\def\bcdd#1#2#3#4{
\hspace{.25\unitlength}
\begin{picture}(3.5,.5)
\put(0,0){\line(1,0){3}}
\psline[linewidth=2pt,linecolor=black,arrowsize=7pt]{->}(#1,0)(#2,0)
\psline[linewidth=2pt,linecolor=black,arrowsize=7pt]{->}(#3,0)(#4,0)
\multiput(0,-.1)(1,0){4}{\line(0,1){.2}}
\end{picture}}
\def\bcddd#1#2#3#4#5#6{
\hspace{.25\unitlength}
\begin{picture}(3.5,.5)
\put(0,0){\line(1,0){3}}
\psline[linewidth=2pt,linecolor=black,arrowsize=7pt]{->}(#1,0)(#2,0)
\psline[linewidth=2pt,linecolor=black,arrowsize=7pt]{->}(#3,0)(#4,0)
\psline[linewidth=2pt,linecolor=black,arrowsize=7pt]{->}(#5,0)(#6,0)
\multiput(0,-.1)(1,0){4}{\line(0,1){.2}}
\end{picture}}
\def\bcdddd#1#2#3#4#5#6#7#8{
\hspace{.25\unitlength}
\begin{picture}(3.5,.5)
\put(0,0){\line(1,0){3}}
\psline[linewidth=2pt,linecolor=black,arrowsize=7pt]{->}(#1,0)(#2,0)
\psline[linewidth=2pt,linecolor=black,arrowsize=7pt]{->}(#3,0)(#4,0)
\psline[linewidth=2pt,linecolor=black,arrowsize=7pt]{->}(#5,0)(#6,0)
\psline[linewidth=2pt,linecolor=black,arrowsize=7pt]{->}(#7,0)(#8,0)
\multiput(0,-.1)(1,0){4}{\line(0,1){.2}}
\end{picture}}
\def\bcx#1#2#3#4#5#6{
\hspace{.25\unitlength}
\begin{picture}(3.5,.5)
\put(0,0){\line(1,0){3}}
\psline[linewidth=2pt,linecolor=black,arrowsize=7pt]{<->}(#1,0)(#2,0)
\psline[linewidth=2pt,linecolor=black,arrowsize=7pt]{<->}(#3,0)(#4,0)
\psline[linewidth=2pt,linecolor=black,arrowsize=7pt]{->}(#5,0)(#6,0)
\multiput(0,-.1)(1,0){4}{\line(0,1){.2}}
\end{picture}}

\section{Boundary Flows}
\label{sec:BoundaryFlows}

\subsection{Conformal boundary flows}
\label{ConformalFlows}

\begin{figure}[p]
\begin{center}
\begin{pspicture}(14,19)
\psline[linewidth=1.5pt,linecolor=blue,arrowsize=10pt]{->}(2.4,6.3)(2.4,2.2)
\psline[linewidth=1.5pt,linecolor=blue,arrowsize=10pt]{->}(7.1,6.3)(7.1,2.2)
\psline[linewidth=1.5pt,linecolor=red,arrowsize=10pt]{->}(2.4,11.2)(6.95,2.3)
\psline[linewidth=1.5pt,linecolor=red,arrowsize=10pt]{->}(3,6.6)(6.8,2.1)
\psline[linewidth=1.5pt,linecolor=blue,arrowsize=10pt]{->}(3,11.7)(11.7,2.1)
\psline[linewidth=1.5pt,linestyle=dashed,linecolor=red,arrowsize=10pt]{->}(6.9,6.3)(3,1.6)
\psline[linewidth=1.5pt,linestyle=dashed,linecolor=red,arrowsize=10pt]{->}(7.65,6.2)(11.5,1.95)
\rput(7,9.5){
\begin{tabular}{r|c|c|c|l} 
\multicolumn{5}{l}{$s$}\\
\cline{2-4}
4&$\begin{array}{c}\bar B^{(1,4)}\\[10pt] \mytri 434\\ \bcd32\\[10pt] \end{array}$ 
& $\begin{array}{c}\bar B^{(2,4)}\\[10pt] \mytri 323\\ \bcd21\\[10pt] \end{array}$ 
& $\begin{array}{c}\bar B^{(3,4)}\\[10pt] \mytri 212\\ \bcd10\\[10pt] \end{array}$
&\raisebox{-1\unitlength}{\rule{0pt}{4\unitlength}} \\ \cline{2-4}
3&$\begin{array}{c}\bar B^{(1,3)}\\[10pt] \mytri {3}{2,4}{3}\\ \bcdd2123\\[10pt] \end{array}$
&$\begin{array}{c}\bar B^{(2,3)}\\[10pt] \mytri {2,4}{1,3}{2,4}\\ \bcddd101232\\[10pt] \end{array}$
& $\begin{array}{c}\bar B^{(3,3)}\\[10pt] \mytri {1,3}2{1,3}\\ \bcdd0121\\[10pt] \end{array}$
&\raisebox{-1\unitlength}{\rule{0pt}{4\unitlength}} \\ \cline{2-4}
2&$\begin{array}{c}\bar B^{(1,2)}\\[10pt]\mytri 2{1,3}2\\ \bcdd1012\\[10pt]\end{array}$ 
&$\begin{array}{c}\bar B^{(2,2)}\\[10pt]\mytri {1,3}{2,4}{1,3}\\ \bcddd012123\\[10pt]\end{array}$ 
& $\begin{array}{c}\bar B^{(3,2)}\\[10pt]\mytri {2,4}{3}{2,4}\\ \bcdd1232\\[10pt]\end{array}$&
\raisebox{-1\unitlength}{\rule{0pt}{4\unitlength}} \\ \cline{2-4}
1&$\begin{array}{c}\bar B^{(1,1)}\\[10pt] \mytri 121\\ \bcd01\\[10pt] \end{array}$ 
& $\begin{array}{c}\bar B^{(2,1)}\\[10pt]\mytri 2{3}2\\ \bcd12\\[10pt]\end{array}$
& $\begin{array}{c}\bar B^{(3,1)}\\[10pt]\mytri 3{4}3\\ \bcd23\\[10pt]\end{array}$&
\raisebox{-1\unitlength}{\rule{0pt}{4\unitlength}}\\ \cline{2-4}
\multicolumn{5}{l}{\hspace{.8\unitlength}
\begin{tabular}{ccccc}
&\hspace{.85\unitlength}1\hspace{3.8\unitlength}&\hspace{3.8\unitlength}2\hspace{3.8\unitlength}&\hspace{3.8\unitlength}3\hspace{3.8\unitlength}&\hspace{2\unitlength}$r$
\end{tabular}}
\end{tabular}}
\end{pspicture}
\vspace{.4in}
\caption{Schematic representation of the six boundary flows between conformal boundary conditions of the tricritical Ising model ($m=4$) in the language of the boundary weights $\bar B$ of the $A_4$ lattice model. The triangle represents the allowed heights, which are encoded by directed edges in the boundary subset $\calU_{(r,s)}$ of the $A_4$ diagram such that the arrow points to the middle height. We restrict to $r+s\le4$ using the Kac table symmetry.  The non-flowing boundary conditions are $(r,s)$ with $r=1,2,3$ and $s=1,4$. The $h<0$ flows
(\ref{hminus}) are shown in blue, and the $h>0$ flows (\ref{hplus})
in red. The dashed lines indicate that the flow is to the direct sum of boundary conditions.}
\label{fig:ConformalFlows}
\end{center}
\end{figure}
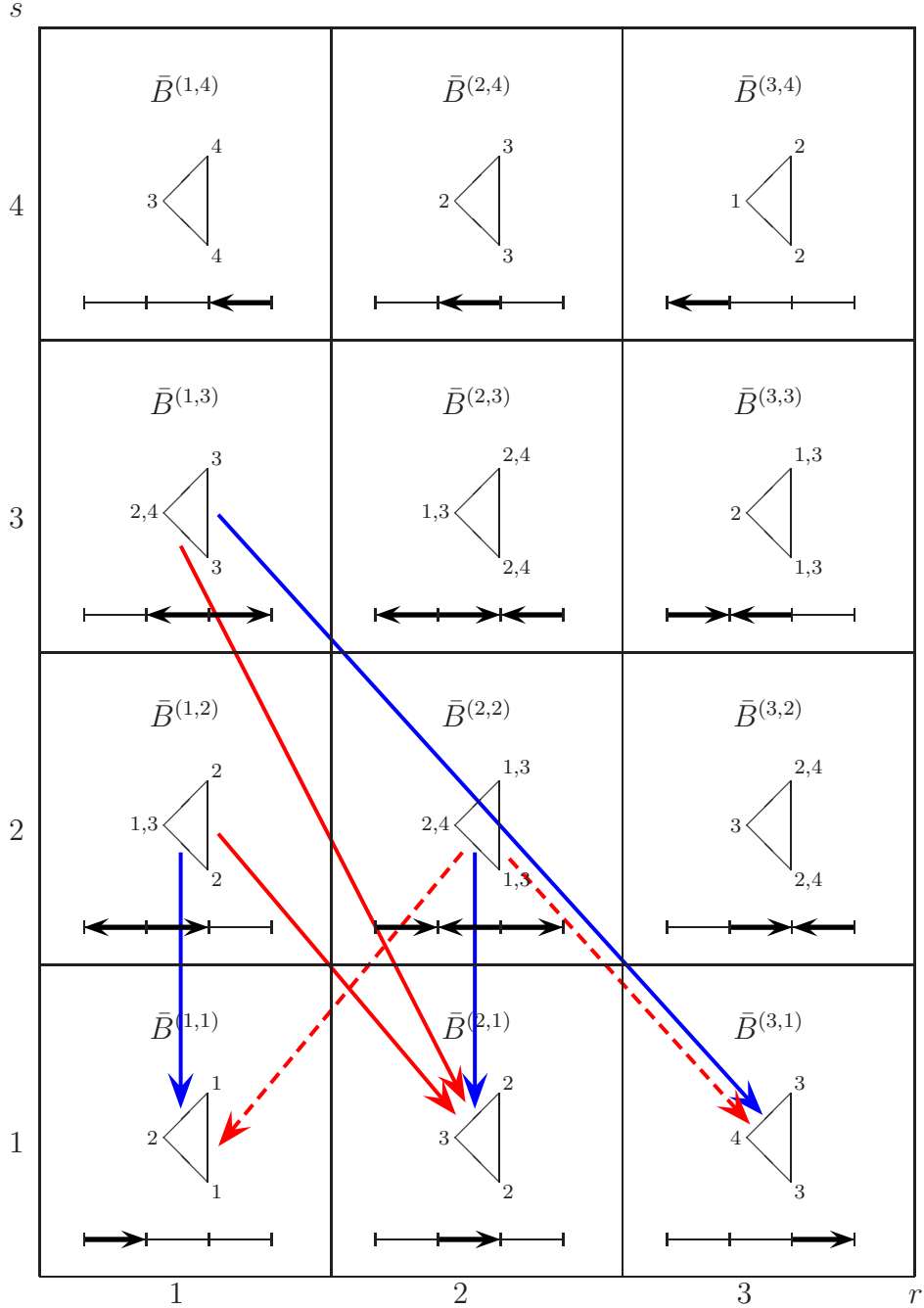

Graham and Watts~\cite{Graham:2003nc} considered the $\phi_{1,3}$
boundary perturbation of the CBC $(r,s)$ of the 
${\cal A}_m$ model with action
\be
\mathcal{S} = \mathcal{S}_{{\cal A}_{m} + (r,s)}
+ h \int_{-\infty}^{\infty} dy\  \phi_{1,3}(y) 
\,, \label{boundmasslessaction}
\ee
where $h$ is a boundary parameter that has dimensions
length${}^{-{2\over m+1}}$, and the fusion rule coefficient 
$N^{(r,s)}_{(1,3)\ (r,s)}$ must be nonzero.  They proposed, based on 
earlier work~\cite{Lesage:1998qf, Recknagel:2000ri}, the following 
RG flows between conformal boundary fixed points:
\be
(r,s) &{h<0\atop \rightarrow}& 
 \left\{ \begin{array}{ll}
 \oplus_{i=1}^{\mbox{\scriptsize min}\{r, s-1, m-r, m+1-s \}} (|r-s+1|+2i-1, 1)   
 & s \le  {m+1\over 2} \\[4pt]
 \oplus_{i=1}^{\mbox{\scriptsize min}\{r, s, m-r, m-s \}} (|r-s|+2i-1, 1)  
 & s > {m+1\over 2}
\end{array} \right. \,, \label{hminus} \\[4pt]
&{h>0\atop \rightarrow}& 
 \left\{ \begin{array}{ll}
  \oplus_{i=1}^{\mbox{\scriptsize min}\{r, s, m-r, m-s \}} (|r-s|+2i-1, 1)  
 & s \le {m+1\over 2} \\[4pt]
 \oplus_{i=1}^{\mbox{\scriptsize min}\{r, s-1, m-r, m+1-s \}} (|r-s+1|+2i-1, 1)   
 & s >  {m+1\over 2}
\end{array} \right. \,, \label{hplus}
\ee 
corresponding to $h<0$ and $h>0$, respectively,
where again we have assumed $r+s\le m$. 

As explained in \cite{Graham:2001pp}, 
the subgraphs of $A_m$ provide a simple
graphical representation of the $\phi_{1,3}$ flows (\ref{hminus}), (\ref{hplus})
of the CBC $(r,s)$.  Namely, flows correspond to projections onto
(linear superpositions of) subgraphs.  Assign a 
sign to each edge in the graph whereby
the edge connected to the
rightmost node $r+s\le m$ is negative, and contiguous links
have alternating signs.  The flows (\ref{hminus})
correspond to deleting all the negative links, while the 
flows (\ref{hplus}) correspond to deleting a subset of positive links.  

These flow patterns are also compactly encoded in the fusion rules of Fredenhagen and Schomerus~\cite{Fredenhagen:2002qn,Fredenhagen:2003xf}. Specifically, for $r+s\le m$ 
\be
(r,s)=(r,1)\times(1,s)&{h<0\atop \rightarrow}& 
\cases{(r,1)\times (s-1,1),&$s\le{m+1\over 2}$\cr 
\rule{0pt}{12pt}(r,1)\times (s,1),&$s>{m+1\over 2}$}\\[4pt]
(r,s)=(r,1)\times(1,s)&{h>0\atop \rightarrow}& 
\cases{(r,1)\times (s,1),&$s\le{m+1\over 2}$\cr 
\rule{0pt}{12pt}(r,1)\times (s-1,1),&$s>{m+1\over 2}$}
\label{FredSflows}
\ee
where the fusion products are taken in the $A_m$ coset model with  $(r,s)\in(A_{m-1},A_m)$. The pattern of flows for the tricritical Ising model ($m=4$) is shown in Figure~\ref{fig:ConformalFlows}.

\begin{figure}[p]
\begin{center}
\begin{pspicture}(14,18.5)
\psline[linewidth=1.5pt,linecolor=blue,arrowsize=10pt]{->}(.8,6.6)(2.,3.5)
\psline[linewidth=1.5pt,linecolor=blue,arrowsize=10pt]{->}(5.5,6.6)(6.9,3.7)
\psline[linewidth=1.5pt,linecolor=red,arrowsize=10pt]{->}(1.3,11.1)(6.7,3.5)
\psline[linewidth=1.5pt,linecolor=red,arrowsize=10pt]{->}(3.6,7.1)(6.4,3.3)
\psline[linewidth=1.5pt,linecolor=blue,arrowsize=10pt]{->}(3.6,11.1)(11.3,3.3)
\psline[linewidth=1.5pt,linestyle=dashed,linecolor=red,arrowsize=10pt]{->}(7.8,6.6)(2.5,3.)
\psline[linewidth=1.5pt,linestyle=dashed,linecolor=red,arrowsize=10pt]{->}(8.4,7)(11.6,3.6)
\rput(7,9.5){
\begin{tabular}{r|c|c|c|l} 
\multicolumn{5}{l}{$s$}\\
\cline{2-4}
\raisebox{20pt}{4}&\raisebox{18pt}{$\begin{array}{c}B^{(3,1)}\\[4pt] \mytri 323  \end{array}$} 
& \raisebox{18pt}{$\begin{array}{c}B^{(2,1)}\\[4pt]\mytri 2{1,3}2\end{array}$}
& \raisebox{18pt}{$\begin{array}{c}B^{(1,1)}\\[4pt]\mytri 121\end{array}$}&
\raisebox{-1\unitlength}{\rule{0pt}{4\unitlength}}\\ \cline{2-4}
\raisebox{20pt}{3}&\raisebox{18pt}{$\begin{array}{c}B^{(2,1)}\oplus B^{(1,3)}\\[4pt]\mytri 2{1,3}2\oplus\;\mytri 323\end{array}$}
&\raisebox{18pt}{$\begin{array}{c}B^{(2,2)}\oplus B^{(1,2)}\\[4pt]\mytri {1,3}{2}{1,3}\oplus\ \mytri 2{1,3}2\end{array}$}
&\raisebox{18pt}{$\begin{array}{c}B^{(2,3)}\oplus B^{(1,1)}\\[4pt]\mytri {2}{1,3}{2}\oplus\ \mytri 121\end{array}$}
&\raisebox{-1\unitlength}{\rule{0pt}{4\unitlength}} \\ \cline{2-4}
\raisebox{20pt}{2}&\raisebox{18pt}{$\begin{array}{c}B^{(1,1)}\oplus B^{(2,3)}\\[4pt]\mytri 121\oplus\;\mytri 2{1,3}2\end{array}$}
&\raisebox{18pt}{$\begin{array}{c}B^{(1,2)}\oplus B^{(2,2)}\\[4pt]\mytri 2{1,3}2\oplus\;\mytri {1,3}{2}{1,3}\end{array}$}
& \raisebox{18pt}{$\begin{array}{c}B^{(1,3)}\oplus B^{(2,1)}\\[4pt]\mytri 3{2}3\oplus\ \;\mytri 2{1,3}2\end{array}$}&
\raisebox{-1\unitlength}{\rule{0pt}{4\unitlength}} \\ \cline{2-4}
\raisebox{20pt}{1}&\raisebox{18pt}{$\begin{array}{c}B^{(1,1)}\\[4pt] \mytri 121 \end{array}$} 
& \raisebox{18pt}{$\begin{array}{c}B^{(2,1)}\\[4pt]\mytri 2{1,3}2\end{array}$}
& \raisebox{18pt}{$\begin{array}{c}B^{(3,1)}\\[4pt]\mytri 323\end{array}$}&
\raisebox{-1\unitlength}{\rule{0pt}{4\unitlength}}\\ \cline{2-4}
\multicolumn{5}{l}{\hspace{.8\unitlength}
\begin{tabular}{ccccc}
&\hspace{1.1\unitlength}1\hspace{4.1\unitlength}&\hspace{4.1\unitlength}2\hspace{4.1\unitlength}&\hspace{4.1\unitlength}3\hspace{4.1\unitlength}&\hspace{2.2\unitlength}$r$
\end{tabular}}
\end{tabular}}
\end{pspicture}
\vspace{.4in}
\caption{ Schematic representation of the six massive boundary flows 
of the tricritical Ising model ($m=4)$ in the language of the reflection matrices. We restrict to $r+s\le4$ using the Kac table symmetry.
The non-flowing boundary reflection matrices $(r,s)$ with $s=1,4$, $r=1,2,3$ are obtained as the limits $\xi=\pm i\infty$ of the elementary boundary solutions (\ref{xiinfty}). The remaining boundary reflection matrices are given by paired boundary reflection matrices.  
The $\xi>{\pi\over 2}$ flows are shown in blue and the $\xi<{\pi\over 2}$ flows are shown in red. The boundary boundstates only exists for $0<\xi<\pi$ and do not flow through. 
This pattern of flows exactly matches the pattern of conformal flows in Figure~\ref{fig:ConformalFlows}.}
\label{fig:MassiveFlows}
\end{center}
\end{figure}
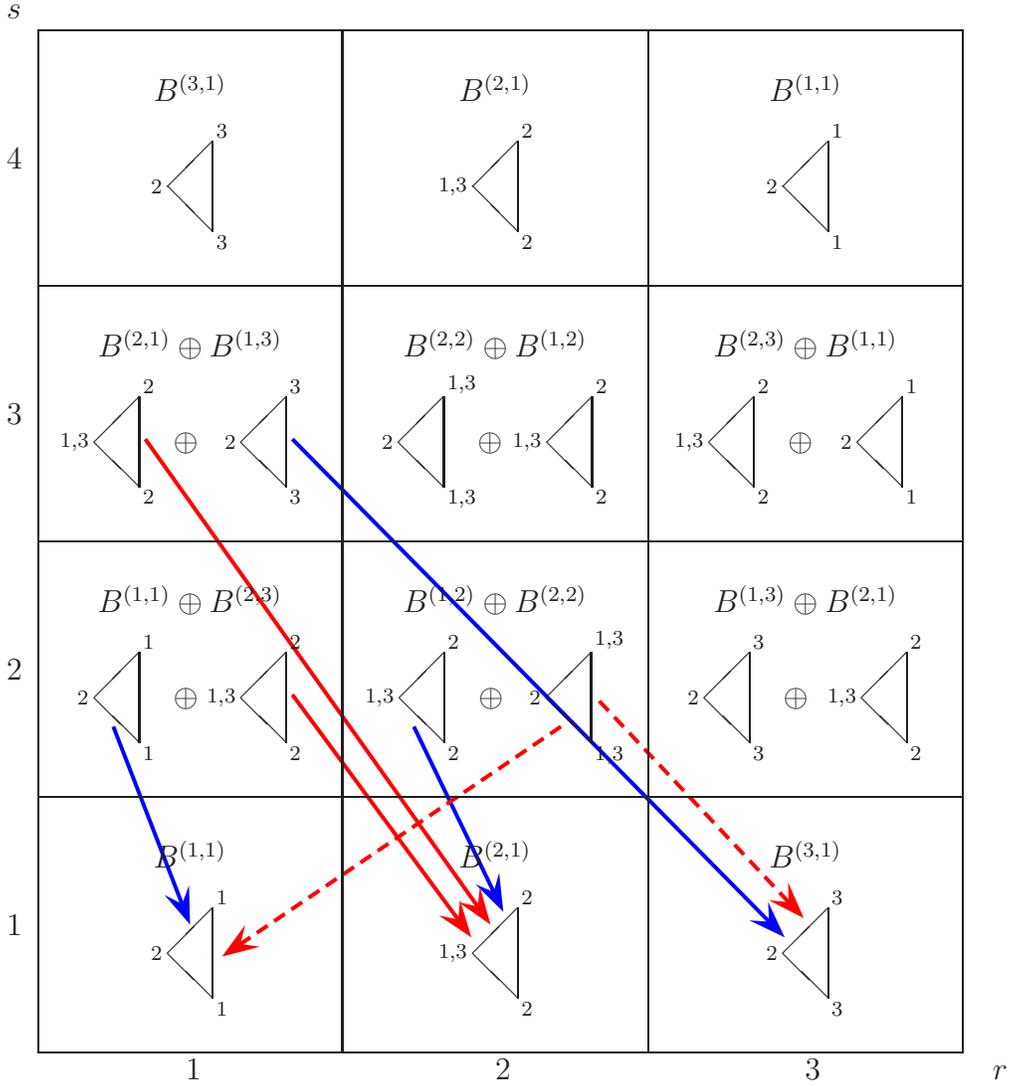

\subsection{Massive boundary flows}

Let us consider the case $s \le {m+1\over 2}$.
The massive renormalization group boundary flows corresponding 
to the paired solutions (\ref{calR})
\be
\RR{r}{s}{c}{a}{b}(\theta, \xi,\mu) &=& 
\R{s-1}{r}{c}{a}{b}(\theta, \xi,\mu) \oplus
\R{m-s}{m-r}{c}{a}{b}(\theta, \pi-\xi,\mu)
\label{pairedbdySmatrices}
\ee
with $r+s\le m$ can be deduced 
by generalizing the analysis found in \cite{Ghoshal:1993tm, Chim:1995kf}. 
Indeed, as $\xi$ varies from 
${\pi\over 2}$ to $-{\Lambda_-}$ and finally to $-{\Lambda_-} \pm i
\infty$, there is a boundary flow from the ultraviolet CBC $(r,s)$
(for which the boundary heights are in the boundary subset
$\calU_{(r,s)} = \calV_{(s-1,r)} \cup\ \calV_{(m-s,m-r)}$) to an infrared CBC for which the boundary heights are in
the elementary boundary subset $\calV_{(s-1,r)}$.  Indeed, at the
ultraviolet fixed point $\xi = {\pi\over 2}$, the boundary states with
heights $\calV_{(s-1,r)}$ and $\calV_{(m-s,m-r)}=\calV_{(s,r)}$ are degenerate.  
Away from this
fixed point with $0 < \xi < {\pi\over 2}$, the {\it ground state} boundary
reflection matrix is given by $\R{s-1}{r}{c}{a}{b}(\theta, \xi,\mu)$. Hence,
\begin{itemize}
 \item[(i)] the boundary states with heights $\calV_{(s-1,r)}$ are degenerate
ground states; and 
 \item[(ii)] the states with heights $\calV_{(m-s,m-r)}$ are
degenerate excited states, which are interpreted as boundary bound 
states of energy $M \cos \xi$ (\ref{energyboundstate}).
\end{itemize}
Point (ii) implies that (as already noted in Section~\ref{subsec:CDD})
the amplitude $\R{s-1}{r}{c}{a}{b}(\theta \,, \xi)$ should have a
simple pole at $\theta = i \xi$ if and only if $a \in \calV_{(m-s,m-r)}$,
corresponding to these boundary bound states.  Point (i) implies that,
at the infrared fixed point  $\xi \rightarrow -{\Lambda_-} \pm i
\infty$, all the boundary heights should be in $\calV_{(s-1,r)}$.

The infrared CBC is a ``superposition'' of pure Cardy CBCs.  Indeed,
using the fact $\calU_{(r,1)} = \{ r \}$ (\ref{s1case}), it
follows from (\ref{elembsubset}) that
\be 
\calV_{(s-1,r)} = \bigcup_{i=1}^{\mbox{\scriptsize min}\{r, s-1, m-r, m+1-s \}} 
\calU_{(|r-s+1|+2i-1, 1)} \,.
\ee
The infrared CBC is thus given by the superposition 
$\oplus_{i=1}^{\mbox{\scriptsize min}\{r, s-1, m-r, m+1-s \}} (|r-s+1|+2i-1, 1)$.

As $\xi$ varies from ${\pi\over 2}$ to ${\Lambda_+}$ and finally to
${\Lambda_+} \pm i \infty$, the boundary flow is instead from the
ultraviolet CBC $(r,s)$ to an infrared CBC for which the boundary
heights are in the elementary boundary subset $\calV_{(m-s,m-r)}$.  Indeed,
for ${\pi\over 2} < \xi < \pi$, the situation is reversed: the
{\it ground state} boundary reflection matrix is given by $\R{m-s}{m-r}{c}{a}{b}(\theta,
\pi-\xi,\mu)$; the boundary states with heights $\calV_{(m-s,m-r)}$ are degenerate ground states,
and the states with heights $\calV_{(s-1,r)}$ are degenerate excited
states, which are interpreted as boundary bound states.  Hence, as
already noted, the amplitude $\R{m-s}{m-r}{c}{a}{b}(\theta \,, \pi-\xi)$
should have a simple pole at $\theta = i (\pi - \xi)$ if and only if $a \in
\calV_{(s-1,r)}$, corresponding to these boundary bound states.  At the
infrared fixed point $\xi \rightarrow {\Lambda_+} \pm i \infty$, all
the boundary heights should be in $\calV_{(m-s,m-r)}$.  Since
\be 
\calV_{(m-s,m-r)} = \calV_{(s,r)} =\bigcup_{i=1}^{\mbox{\scriptsize min}\{r, s, m-r, m-s \}} \calU_{(|r-s|+2i-1, 1)} 
\,,
\ee
the infrared CBC is given by the superposition 
$\oplus_{i=1}^{\mbox{\scriptsize min}\{r, s, m-r, m-s \}} (|r-s|+2i-1, 1)$.

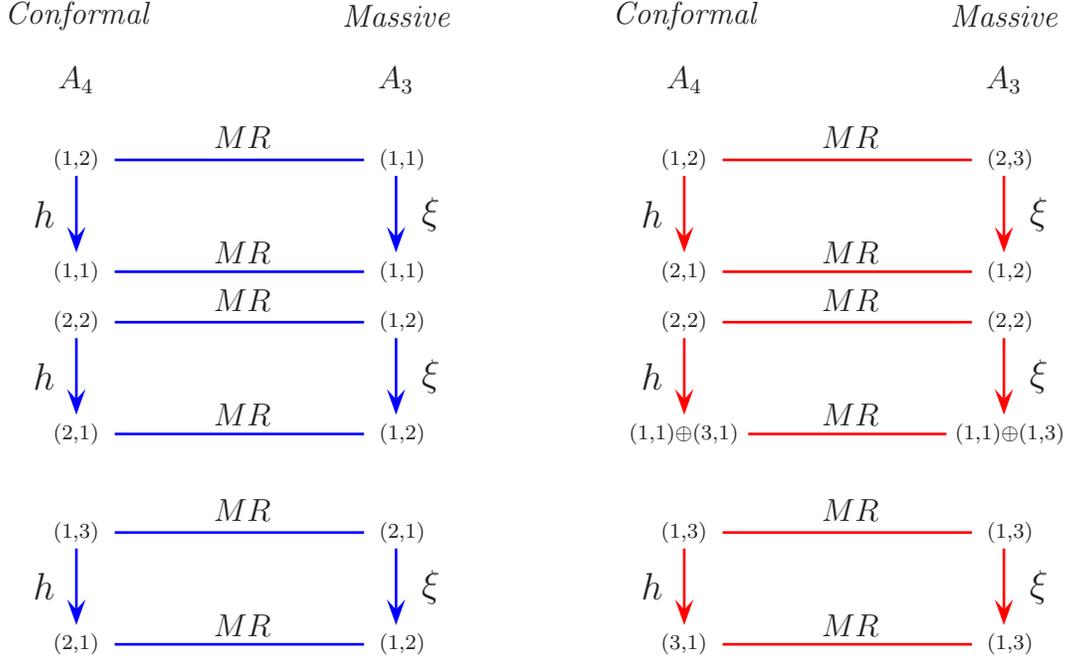
\begin{figure}[t]
\vspace{.65in}
\begin{center}
\begin{pspicture}(5,2.5)
\put(0,3){\pp{}{\mbox{$A_4$}}}
\put(5.0,3){\pp{}{\mbox{$A_3$}}}
\put(0,4){\pp{}{\mbox{\em Conformal}}}
\put(5.0,4){\pp{}{\mbox{\em Massive}}}
\put(-.5,.9){\pp{}{\mbox{\large $h$}}}
\put(5.55,.9){\pp{}{\mbox{\large $\xi$}}}
\put(2.6,2.1){\pp{}{\mbox{$MR$}}}
\put(2.6,.3){\pp{}{\mbox{$MR$}}}
\put(0,1.75){\pp{}{(1,2)}}
\put(0,0){\pp{}{(1,1)}}
\put(5.1,1.75){\pp{}{(1,1)}}
\put(5.1,0){\pp{}{(1,1)}}
\psline[linewidth=1.pt,linecolor=blue,arrowsize=7pt]{-}(.6,1.75)(4.5,1.75)
\psline[linewidth=1.pt,linecolor=blue,arrowsize=7pt]{-}(.6,0)(4.5,0)
\psline[linewidth=1.pt,linecolor=blue,arrowsize=7pt]{->}(0,1.5)(0,.3)
\psline[linewidth=1.pt,linecolor=blue,arrowsize=7pt]{->}(5,1.5)(5,.3)
\end{pspicture}
\hspace{1.4in}
\begin{pspicture}(5,2.5)
\put(0,3){\pp{}{\mbox{$A_4$}}}
\put(5.0,3){\pp{}{\mbox{$A_3$}}}
\put(0,4){\pp{}{\mbox{\em Conformal}}}
\put(5.0,4){\pp{}{\mbox{\em Massive}}}
\put(-.5,.9){\pp{}{\mbox{\large $h$}}}
\put(5.55,.9){\pp{}{\mbox{\large $\xi$}}}
\put(2.6,2.1){\pp{}{\mbox{$MR$}}}
\put(2.6,.3){\pp{}{\mbox{$MR$}}}
\put(0,1.75){\pp{}{(1,2)}}
\put(0,0){\pp{}{(2,1)}}
\put(5.1,1.75){\pp{}{(2,3)}}
\put(5.1,0){\pp{}{(1,2)}}
\psline[linewidth=1.pt,linecolor=red,arrowsize=7pt]{-}(.6,1.75)(4.5,1.75)
\psline[linewidth=1.pt,linecolor=red,arrowsize=7pt]{-}(.6,0)(4.5,0)
\psline[linewidth=1.pt,linecolor=red,arrowsize=7pt]{->}(0,1.5)(0,.3)
\psline[linewidth=1.pt,linecolor=red,arrowsize=7pt]{->}(5,1.5)(5,.3)
\end{pspicture}
\vspace{.25in}
\begin{pspicture}(5,2.5)
\put(-.5,.9){\pp{}{\mbox{\large $h$}}}
\put(5.55,.9){\pp{}{\mbox{\large $\xi$}}}
\put(2.6,2.1){\pp{}{\mbox{$MR$}}}
\put(2.6,.3){\pp{}{\mbox{$MR$}}}
\put(0,1.75){\pp{}{(2,2)}}
\put(0,0){\pp{}{(2,1)}}
\put(5.1,1.75){\pp{}{(1,2)}}
\put(5.1,0){\pp{}{(1,2)}}
\psline[linewidth=1.pt,linecolor=blue,arrowsize=7pt]{-}(.6,1.75)(4.5,1.75)
\psline[linewidth=1.pt,linecolor=blue,arrowsize=7pt]{-}(.6,0)(4.5,0)
\psline[linewidth=1.pt,linecolor=blue,arrowsize=7pt]{->}(0,1.5)(0,.3)
\psline[linewidth=1.pt,linecolor=blue,arrowsize=7pt]{->}(5,1.5)(5,.3)
\end{pspicture}
\hspace{1.4in}
\begin{pspicture}(5,2.5)
\put(-.5,.9){\pp{}{\mbox{\large $h$}}}
\put(5.55,.9){\pp{}{\mbox{\large $\xi$}}}
\put(2.6,2.1){\pp{}{\mbox{$MR$}}}
\put(2.6,.3){\pp{}{\mbox{$MR$}}}
\put(0,1.75){\pp{}{(2,2)}}
\put(0,0){\pp{}{(1,1)\oplus(3,1)}}
\put(5.1,1.75){\pp{}{(2,2)}}
\put(5.1,0){\pp{}{(1,1)\oplus(1,3)}}
\psline[linewidth=1.pt,linecolor=red,arrowsize=7pt]{-}(.6,1.75)(4.5,1.75)
\psline[linewidth=1.pt,linecolor=red,arrowsize=7pt]{-}(1.,0)(4.1,0)
\psline[linewidth=1.pt,linecolor=red,arrowsize=7pt]{->}(0,1.5)(0,.3)
\psline[linewidth=1.pt,linecolor=red,arrowsize=7pt]{->}(5,1.5)(5,.3)
\end{pspicture}
\vspace{.25in}
\begin{pspicture}(5,2.5)
\put(-.5,.9){\pp{}{\mbox{\large $h$}}}
\put(5.55,.9){\pp{}{\mbox{\large $\xi$}}}
\put(2.6,2.1){\pp{}{\mbox{$MR$}}}
\put(2.6,.3){\pp{}{\mbox{$MR$}}}
\put(0,1.75){\pp{}{(1,3)}}
\put(0,0){\pp{}{(2,1)}}
\put(5.1,1.75){\pp{}{(2,1)}}
\put(5.1,0){\pp{}{(1,2)}}
\psline[linewidth=1.pt,linecolor=blue,arrowsize=7pt]{-}(.6,1.75)(4.5,1.75)
\psline[linewidth=1.pt,linecolor=blue,arrowsize=7pt]{-}(.6,0)(4.5,0)
\psline[linewidth=1.pt,linecolor=blue,arrowsize=7pt]{->}(0,1.5)(0,.3)
\psline[linewidth=1.pt,linecolor=blue,arrowsize=7pt]{->}(5,1.5)(5,.3)
\end{pspicture}
\hspace{1.4in}
\begin{pspicture}(5,2.5)
\put(-.5,.9){\pp{}{\mbox{\large $h$}}}
\put(5.55,.9){\pp{}{\mbox{\large $\xi$}}}
\put(2.6,2.1){\pp{}{\mbox{$MR$}}}
\put(2.6,.3){\pp{}{\mbox{$MR$}}}
\put(0,1.75){\pp{}{(1,3)}}
\put(0,0){\pp{}{(3,1)}}
\put(5.1,1.75){\pp{}{(1,3)}}
\put(5.1,0){\pp{}{(1,3)}}
\psline[linewidth=1.pt,linecolor=red,arrowsize=7pt]{-}(.6,1.75)(4.5,1.75)
\psline[linewidth=1.pt,linecolor=red,arrowsize=7pt]{-}(.6,0)(4.5,0)
\psline[linewidth=1.pt,linecolor=red,arrowsize=7pt]{->}(0,1.5)(0,.3)
\psline[linewidth=1.pt,linecolor=red,arrowsize=7pt]{->}(5,1.5)(5,.3)
\end{pspicture}
\end{center}
\caption{Schematic representation of the six two-parameter RG boundary flows for the tricritical Ising model. The $h<0$ ($\xi>{\pi\over 2}$) flows are in blue on the left and the $h>0$ ($\xi<{\pi\over 2}$) flows are in red on the right. In the conformal columns $(r,s)\in(A_3,A_4)$ refers to conformal boundary condition. In the massive columns $(r,s)\in(A_2,A_3)$ refers to $B^{(r,s)}$.
\label{RGbdyflows}}
\end{figure}

To summarize, the paired solution (\ref{pairedbdySmatrices}) 
for the case $s \le {m+1\over 2}$ with $r+s\le m$ implies 
the following boundary flows:
 \be
 (r,s) \rightarrow \left\{ \begin{array}{ll}
 \oplus_{i=1}^{\mbox{\scriptsize min}\{r, s-1, m-r, m+1-s \}} (|r-s+1|+2i-1, 1) & 
 \mbox{for  } ``\xi < {\pi\over 2}" \\[4pt]
 \oplus_{i=1}^{\mbox{\scriptsize min}\{r, s, m-r, m-s \}} (|r-s|+2i-1, 1)  & 
 \mbox{for  } ``\xi > {\pi\over 2}"
\end{array} \right. \,, \label{flowsummary}
\ee
or 
\be
(r,s)=(r,1)\times(1,s)\rightarrow\left\{\begin{array}{ll}
(r,1)\times(s-1,1)&\quad
 \mbox{for  } ``\xi < {\pi\over 2}" \\[4pt]
(r,1)\times(s,1)&\quad
 \mbox{for  } ``\xi > {\pi\over 2}"
\end{array} \right. \,,
\ee
where we have introduced the expressions  $``\xi < {\pi\over 2}"$ and
$``\xi > {\pi\over 2}"$ as shorthand notations for the flows $\xi=
{\pi\over 2} \rightarrow \xi=-{\Lambda_-} \pm i \infty$ and $\xi=
{\pi\over 2} \rightarrow \xi={\Lambda_+} \pm i \infty$,
respectively. The pattern of massive flows exactly matches the pattern 
of conformal flows (\ref{hminus}), (\ref{hplus}) for $s \le {m+1\over 2}$
corresponding to $h<0$ and $h>0$, respectively. 

For the case $s > {m+1\over 2}$ with $r+s\le m$, 
the boundary reflection matrix is given instead by
\be
\RR{r}{s}{c}{a}{b}(\theta, \pi-\xi,\mu) = 
\R{s-1}{r}{c}{a}{b}(\theta, \pi-\xi,\mu) \oplus
\R{m-s}{m-r}{c}{a}{b}(\theta, \xi,\mu) \,.
\ee
Hence, the corresponding massive boundary flows are the same
as for the case $s \le {m+1\over 2}$ (\ref{flowsummary}),
except that $``\xi < {\pi\over 2}"$ and $``\xi > {\pi\over 2}"$ are
interchanged:
\be
(r,s) \rightarrow \left\{ \begin{array}{ll}
\oplus_{i=1}^{\mbox{\scriptsize min}\{r, s, m-r, m-s \}} (|r-s|+2i-1, 1)  & 
\mbox{for  } ``\xi < {\pi\over 2}" \\[4pt]
\oplus_{i=1}^{\mbox{\scriptsize min}\{r, s-1, m-r, m+1-s \}} (|r-s+1|+2i-1, 1) & 
\mbox{for  } ``\xi > {\pi\over 2}" 
\end{array} \right. \,, \label{flowsummary2}
\ee
which matches the pattern 
of conformal flows (\ref{hminus}), (\ref{hplus}) for $s > {m+1\over 2}$
corresponding to $h<0$ and $h>0$, respectively. 

The pattern of massive flows for the tricritical Ising model ($m=4$) is shown in Figure~\ref{fig:MassiveFlows}. 
The six RG boundary flows in the two-parameter space spanned by $MR$ and $h$ (or $\xi$) for the tricritical Ising model are shown in Figure~\ref{RGbdyflows}.

\section{Non-invertible symmetries}
\label{sec:Noninvertible}

\hyphenpenalty=1000

The conformal field theory ${\cal A}_m$ has categorical/non-invertible symmetries, which are implemented by topological defect lines. These defect lines  ${\cal L}_{(r,s)}$ are labeled by the Kac labels $(r,s)$ and can be continuously deformed. When two defect lines are pushed together they can be decomposed in terms of certain
elementary defect lines, to be described below. 
This ``multiplication'' follows the fusion rules. Defect lines  can also be fused to the boundaries~\cite{Graham:2003nc}. In particular, the $(r,s)$ Cardy boundary condition can be thought of as an $(r,s)$ defect fused to the identity boundary with label $(1,1)$.

\hyphenpenalty=10000

Defect lines can be represented in two different ways. When they are located at a given time slice, they act as operators on the Hilbert space. Alternatively, when they are located at a given space point, they change the Hilbert space and impose non-trivial defect conditions for the fields approaching the defect from the two sides. Modular transformation of the two descriptions leads to the classification of elementary defect lines~\cite{Petkova:2000ip}. 

\begin{figure}[tb]
	\centering
	\includegraphics[width=0.13\textwidth]{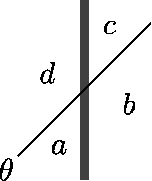}\hspace{4cm} 	\includegraphics[width=0.13\textwidth]{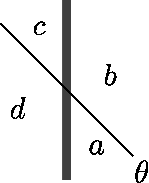}
	\caption[xxx]{
	Left $ T_{a\ b}^{d\ c}(\theta)$ and right  $\bar T_{a\ b}^{d\ c}(\theta)$ transmission factors, respectively. }
	\label{fig:transmission}
\end{figure}

Introducing a massive bulk and boundary perturbation by $\phi_{(1,3)}$ breaks not only the conformal symmetries but also some of the non-invertible symmetries of the theory. The non-invertible symmetries that commute with the perturbing fields remain symmetries after the perturbation. These symmetries are generated by the defect lines ${\cal L}_{(r,1)}$ and form the ${\cal A}_{m-1}$ fusion category. Every defect line here can be generated by fusion from the elementary defect ${\cal L}_{(2,1)}$.  When these defect lines are placed at specific time points, they act on the Hilbert space and transform the vacua according to the fusion rules following from ${\cal L}_{(a,1)}:1\to a$; that is, the vacuum labeled by $1$ will be mapped to $a$. All other transformation can be calculated from associativity, by fusing the defects first. In particular, the elementary defect ${\cal L}_{(2,1)}$ maps the vacuum $a$ to $a\pm 1$, whenever it is a valid label. 
When  a defect is placed at a given space point, it introduces non-trivial boundary conditions to the fields on the two sides. It  also introduces non-trivial transmissions, whenever particles cross defects, see Figure~\ref{fig:transmission}. These transmission factors are different when particles come from the left or from the right, but they are related by unitarity. These transmission factors must also satisfy crossing symmetry~\cite{Bajnok:2004jd} and the Yang Baxter equation. In the presence of kink particles, however, the crossing relation has to be modified as discussed in \cite{Copetti:2024rqj}.
Since the elementary defect ${\cal L}_{(2,1)}$ changes the vacua in the same way 
as the particles themselves, its transmission factor can be extracted from the $\theta\to \infty$ limit of the scattering matrices. This means that the defect behaves as an infinitely-massive particle with rapidity $ \theta_0$ sent to infinity: 
\be
T_{a\ b}^{d\ c}(\theta)=\lim_{\theta_0\to \infty}S_{a\ b}^{d\ c}(\theta-\theta_0) \,, \qquad \bar T_{a\ b}^{d\ c}(\theta)=\lim_{\theta_0\to \infty}S_{a\ b}^{d\ c}(\theta_0-\theta) \,.
\ee
As a consequence, these transmission factors do not depend on the rapidity\footnote{Here we slightly renormalized the transmission factors, by removing the overall $e^{\pm i\frac{m-1}{2m}}$ phases.}
\be
 T_{a\ b}^{d\ c} =
\delta_{a c} + e^{-i\frac{\pi}{m}}
\left({[a][c]\over [b][d]}\right)^{1\over 2} 
 \delta_{b d} \quad \,, \qquad \bar T_{a\ b}^{d\ c}=
\delta_{a c} + e^{i\frac{\pi}{m}}
\left({[a][c]\over [b][d]}\right)^{1\over 2} 
 \delta_{b d} \,,
\label{transmission}
\ee
where $T$ describes the process when the particle comes from the left and $\bar T$ when it comes from the right. Since we took the $\theta_0 \to \infty$ limit in the scattering matrix, these transmission factors are rapidity independent. 
Note that the old crossing relation has an extra S-matrix factor $ \bigl ( \frac{[a][c]}{[b][d]} \bigr )^\frac{i\theta}{2\pi} $ that does not have a well-defined $\theta \to \infty$ limit. Hence, it is crucial to modify the crossing relation, as well as the defect crossing relation in \cite{Bajnok:2004jd}, according to \cite{Copetti:2024rqj}. 

\begin{figure}[tb]
	\centering
	\includegraphics[width=0.2\textwidth]{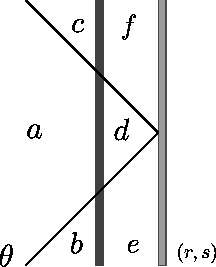}\hspace{4cm} 	\includegraphics[width=0.23\textwidth]{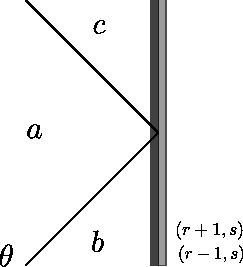}
	\caption[xxx]{
	Fusing the elementary defect to the  $(r,s)$ boundary results in the direct sum of boundary conditions $(r-1,s)$ and $(r+1,s)$.  }
	\label{fig:defectbdry}
\end{figure}

When the elementary defect line ${\cal L}_{(2,1)}$ acts on the $(r,s)$ Cardy boundary condition, the latter is changed to the sum of the boundary conditions  $(r\pm 1,s)$,  whenever $r\pm1$ is allowed. Since the defect lines commute with the perturbation, they also act in the same way on the off-critical boundary conditions.  Consistency of fusing the elementary defect to the boundary~\cite{Bajnok:2007jg} requires that the following relation between different boundary reflections be satisfied
(see also Figure \ref{fig:defectbdry})

\be
 \RR{r-1}{s}{c}{a}{b}(\theta\,, \xi)\oplus \RR{r+1}{s}{c}{a}{b}(\theta\,, \xi) = \sum_{d,e, f}
T^{a\ d}_{b\ e}\ \RR{r}{s}{f}{d}{e}(\theta\,, \xi)\
\bar T^{c\ f}_{a\ d} \,.
\label{Noninvsym}
\ee 

In checking this equation, recall from eq. (\ref{calR}) that the reflection factors are composed of paired elementary solutions ${\cal R}^{(r,s)}=B^{(s-1,r)}\oplus B^{(m-s,m-r)}$. In the direct sum, the allowed boundary weights have different parities; thus, checking (\ref{Noninvsym}) can be done for the two elementary reflections separately. Let us focus on the first term $B^{(s-1,r)}$, and notice that the prefactor 
$V^{(s-1,r)}(\theta,\xi,\mu)/(s(i\xi)s(i(s\!-\!1)\pi\!+\!\xi)))$ in the definition of the elementary solution (\ref{elemgauged}) does not depend on $r$. Thus, checking  (\ref{Noninvsym}) boils down to checking the same statement for the gauge-transformed lattice boundary weights $\bar B^{(s-1,r)}$ of Behrend and Pearce. But the higher-$r$ fused weights were constructed exactly by this fusion procedure, so they satisfy the required relation. 

This concludes the check that our proposal for the boundary reflection matrices is compatible with the unbroken non-invertible symmetries.

\section{Examples}
\label{sec:Examples}

The proposal (\ref{calR}) is the main result of this paper. 
In this Section we work out in detail the boundary reflection matrices for
the cases $m= 3, 4, 5$.

\subsection{$m=3$ (Ising model)}
\label{sec:m3}

For the case $m=3$, the  
vacua (\ref{heights}) are $\{ 1 \,, 2 \}$,
and the allowed boundary 
vacua for each 
$(r,s)$ boundary condition are given in Table
\ref{figm3}.  There is only one CBC which admits a relevant
perturbation, namely $(1,2)$.  The corresponding boundary reflection matrix
is given by 
(\ref{calR})
\be
\RR{1}{2}{c}{a}{b}(\theta,\xi)
= \R{1}{1}{c}{a}{b}(\theta,\xi,1) \oplus
\R{1}{2}{c}{a}{b}(\theta,\pi-\xi,1) \,.
\label{m3main}
\ee
According to the formula (\ref{elembsubset}) for the elementary
boundary subsets, $\calV_{(1,1)}= \{ 1 \}$; hence, the corresponding
elementary solution has only one nonvanishing amplitude, namely,
$\R{1}{1}{1}{2}{1}$.  Similarly, $\calV_{(1,2)}= \{ 2 \}$ implies that the
corresponding elementary solution has only the one nonvanishing
amplitude $\R{1}{2}{2}{1}{2}$. These amplitudes are given by 
(\ref{elemgauged}), (\ref{elem1}), (\ref{boundaryscalarfactor}),
\be
\R{1}{1}{1}{2}{1}(\theta,\xi,1) = 
\R{1}{2}{2}{1}{2}(\theta,\pi-\xi,1) =
V^{(1,1)}(\theta,\xi,1)\
{s(i\xi+ \theta) s(i(\xi+ \pi) - \theta) 
\over s(i\xi) s(i(\xi+\pi))}\,.
\ee 
Evaluating the scalar factor $V^{(1,1)}(\theta,\xi,1)$
using (\ref{boundaryscalarfactor}) 
and the results (\ref{Vrelementaryodd}), 
(\ref{Pelementary}) for $m=3$, we conclude that
\be
\R{1}{1}{1}{2}{1}(\theta,\xi,1) = 
\R{1}{2}{2}{1}{2}(\theta,\pi-\xi,1) = 
V_0(\theta)\,V_{\mathrm{CDD}}(\theta,\xi) \,.
\ee 
It follows from (\ref{m3main}) that the boundary reflection matrix
is given by
\be
\RR{1}{2}{1}{2}{1}(\theta,\xi) = \RR{1}{2}{2}{1}{2}(\theta,\xi) = V_0(\theta)\,
V_{\mathrm{CDD}}(\theta,\xi) \,,
\label{ising}
\ee 
where $V_{\mathrm{CDD}}(\theta,\xi)$ is given by (\ref{VCDD}).
The equality\,\footnote{As previously noted (\ref{spinrev}), 
the CBC $(r,{m+1\over 2})$ with $m$ odd is invariant under $Z_{2}$.} of the amplitudes $\RR{1}{2}{1}{2}{1}$ and
$\RR{1}{2}{2}{1}{2}$ is a manifestation of the $Z_{2}$ invariance
(\ref{calRZ2symmetry}).

This result coincides with
that of Ghoshal and Zamolodchikov.  Indeed, comparing (\ref{ising})
with the boundary scattering amplitude given by  (4.27) in
\cite{Ghoshal:1993tm}, we see that the two expressions coincide upon making the identification $\sin \xi = \kappa$. 
The boundary reflection matrices for the
boundary conditions $(r,s)=(1,1)$ and $(r,s)=(2,1)$ can be obtained in the limiting cases (\ref{calR1}), which is also 
compatible with the boundary flows.

\subsection{$m=4$ (tricritical Ising model)}
\label{sec:m4}

For the case $m=4$, the vacua (\ref{heights}) are $\{ 1 \,, 2 \,,
3\}$, and the allowed boundary vacua for each CBC are given in Table
\ref{figm4}.  There are three CBCs which admit a relevant
perturbation: $(1,2) \,, (1,3) \,, (2,2)$.

\subsubsection*{CBCs $(1,2)$ and $(1,3)$}
\label{subsubsec:12}

For the CBC $(1,2)$, the boundary reflection matrix is given by (\ref{calR})
\be
\RR{1}{2}{c}{a}{b}(\theta,\xi)
= \R{1}{1}{c}{a}{b}(\theta,\xi,1) \oplus
\R{2}{3}{c}{a}{b}(\theta,\pi-\xi,1) \,.
\label{m4main12}
\ee
According to the formula (\ref{elembsubset}) for the elementary
boundary subsets, $\calV_{(1,1)}= \{ 1 \}$; hence, the corresponding
elementary solution has only one nonvanishing amplitude, 
$\R{1}{1}{1}{2}{1}$, given by 
(\ref{elemgauged}), (\ref{elem1}), (\ref{boundaryscalarfactor}),
\be 
\R{1}{1}{1}{2}{1}(\theta,\xi,1)
&=& V^{(1,1)}(\theta,\xi,1)\
{ s(i\xi+ \theta) s(i(\xi+ \pi) - \theta) 
\over s(i\xi) s(i(\xi+\pi))} \,, \non  \\
&=&V_0(\theta)\,V_{\mathrm{CDD}}(\theta,\xi)\  P_{0}(\theta)^{-1} \,,
\label{m4pminus}
\ee 
In passing to the second line,
we have used the identity (for $m=4$)
\be
V_{1}(\theta,\xi) = {s(i\xi+ \theta) s(i(\xi+ \pi) - \theta) 
\over s(i\xi) s(i(\xi+\pi))} \,.
\ee
This agrees with Chim's result for $P_{-}(\theta)$ (see Eqs.  (20--23) in \cite{Chim:1995kf}).
Our factor $ P_{0}(\theta)^{-1}$, given by  (\ref{P0}), corresponds to
Chim's $P_{\mbox{\scriptsize min}}(\theta)$), except for the factors $V_0(\theta)=i \tanh( {i \pi\over
4} -{\theta\over 2})$ in (\ref{VCDD}),  
which he is missing~\cite{Miwa:1996ht}. This could be related to the fact that his S-matrix is not normalized properly. 
This prefactor is crucial, however, to decrease the second-order pole of $V_{\mathrm{CDD}}(\theta,\pi/2)$ to first order, which is required by the presence of the degenerate bound state.

Similarly, $\calV_{(2,3)}= \{ 2 \}$ implies that the
corresponding elementary solution has only two nonvanishing
amplitudes, $\R{2}{3}{2}{1}{2}$ and $\R{2}{3}{2}{3}{2}$. Hence,
\be
&&\RR{1}{2}{2}{3}{2}\!(\theta,\xi) =
\R{2}{3}{2}{3}{2}\!(\theta,\pi-\xi,1) \non\\
&=& -V^{(2,3)} (\theta,\pi-\xi,1)\
{ s(i(\xi-\pi)+ \theta) s(i(\xi+ \pi) - \theta)
\over s(i(\pi-\xi) s(i(3\pi-\xi))}\\
&=&V_0(\theta)\,V_{\mathrm{CDD}}(\theta,\xi)\ 
P_{0}(\theta)^{-1} \calF(\theta+i\xi)\calF(\theta-i\xi)
(\cos{\xi\over 2}+ i\sinh{\theta\over2}) \,, \non
\ee 
where we have used the identity (for $m=4$)
\be
V_{2}(\theta,\pi-\xi)^{-1} = \calF(\theta+i\xi) \calF(\theta-i\xi)
s(i(\pi-\xi) s(i(3\pi-\xi)) \,,
\ee
with
\be
\calF(\theta)=(\cosh{\scriptsize\theta\over 2})^{-1/2}
\exp\Big[{i\over 4}\int_0^\infty {dt\over t}\,
{\sin{\scriptsize\theta t\over\pi}\over \cosh^2{\scriptsize t\over 2}}\Big]
\,.
\ee
Moreover,
\be
{\RR{1}{2}{2}{3}{2}(\theta,\xi)\over
\RR{1}{2}{2}{1}{2}(\theta,\xi)} =
{s(i(\xi-\pi)+ \theta)\ s(i(\xi+ \pi) - \theta)\over
s(i(\xi-\pi)- \theta)\ s(i(\xi+ \pi) + \theta)}
={\cos {\xi\over 2} + i \sinh{\theta\over 2}\over
\cos {\xi\over 2} - i \sinh{\theta\over 2}} \,.
\ee 
The amplitudes $\RR{1}{2}{2}{3}{2}$ and $\RR{1}{2}{2}{1}{2}$
correspond to Chim's amplitudes $R_{+}(\theta)$ and $R_{-}(\theta)$
given in \cite{Chim:1995kf} by  (26a) and (26b), respectively. (See also
\cite{Miwa:1996ht}.)

For the CBC $(1,3)$ (which was not explicitly treated by Chim), the
boundary reflection matrix $\RR{1}{3}{c}{a}{b}(\theta,\xi)$ is related to
the one for the CBC $(1,2)$ described in Section \ref{subsubsec:12}
above by the $Z_{2}$ symmetry (\ref{calRZ2symmetry}).

\subsubsection*{CBC $(2,2)$}

For the CBC $(2,2)$, the boundary reflection matrix is given by (\ref{calR})
\be
\RR{2}{2}{c}{a}{b}(\theta,\xi)
= \R{1}{2}{c}{a}{b}(\theta,\xi,1) \oplus
\R{2}{2}{c}{a}{b}(\theta,\pi-\xi,1) \,,
\label{m4main22}
\ee
and the corresponding elementary boundary subsets are $\calV_{(1,2)}= \{ 2
\}$ and $\calV_{(2,2)}= \{ 1 \,, 3\}$.
By the $Z_{2}$ symmetry (\ref{rZ2}), \footnote{As 
previously noted (\ref{spinrev}), the CBC $({m\over 2},s)$ 
with $m$ even is invariant under $Z_{2}$.}
\be
\RR{2}{2}{2}{1}{2}(\theta,\xi) = 
\RR{2}{2}{2}{3}{2}(\theta,\xi) \,.
\ee
This amplitude is equal to 
$\ \RR{1}{2}{1}{2}{1}(\theta,\xi)$ (see  
(\ref{m4pminus})). This result is also in agreement 
(up to the missing factor $i \tanh({i \pi\over 4} - {\theta\over 2})$) 
with Chim's result $R_{-}(\theta) = R_{+}(\theta) = R(\theta)$, with 
$R(\theta)$ given by (31) in \cite{Chim:1995kf}.

The amplitudes 
\be
\RR{2}{2}{1}{2}{1}(\theta,\xi) &\!\!\!=\!\!\!&
\RR{2}{2}{3}{2}{3}(\theta,\xi) 
= V^{(2,2)} (\theta,\pi-\xi,1)  { \sin
({1\over2}(\pi-\xi))\over s(i(\pi-\xi) s(i(3\pi-\xi))}\\
&=&V_0(\theta)\,V_{\mathrm{CDD}}(\theta,\xi)\  
P_{0}(\theta)^{-1} \calF(\theta+i\xi)\calF(\theta-i\xi)
\cos{\scriptsize\xi\over 2}\non
\ee
agree with Chim's result $P_{-}(\theta) =
P_{+}(\theta) = P(\theta)$, with $P(\theta)$ given by  (32a) in
\cite{Chim:1995kf}.

Finally, the ``nondiagonal'' amplitudes are
\be
&&\RR{2}{2}{1}{2}{3}(\theta,\xi) = 
\RR{2}{2}{3}{2}{1}(\theta,\xi) \\
&=&V_0(\theta)\,V_{\mathrm{CDD}}(\theta,\xi)\
P_{0}(\theta)^{-1} \calF(\theta+i\xi)\calF(\theta-i\xi)
(-i\sinh{\scriptsize\theta\over 2})\non
\ee
in agreement with Chim's result for $V(\theta)$ (32b).

Again the amplitudes $\RR{3}{1}{b}{a}{c}(\theta,\xi)$ 
are related to the amplitudes $\RR{1}{1}{b}{a}{c}(\theta,\xi)$ by the $Z_{2}$ symmetry (\ref{rZ2}).
The special cases $(r,1)$ without the boundary perturbations,
which have so far not been analyzed,
can be obtained from the limits (\ref{calR1}). 

\subsection{$m=5$ (tetracritical Ising model)}
\label{sec:m5}

For the case $m=5$, the vacua (\ref{heights}) are $\{ 1, 2 , 3,
4\}$, and the allowed boundary vacua for each CBC are given
in Table \ref{figm5}.  There are six CBCs which admit a relevant
perturbation: $(1,2)$, $(1,3)$, $(1,4)$, $(2,2)$, $(2,3)$,
$(2,4)$. 

\subsubsection*{CBCs $(1,2)$ and $(1,4)$}

For the CBC $(1,2)$, the boundary reflection matrix is given by (\ref{calR})
\be
\RR{1}{2}{c}{a}{b}(\theta,\xi)
= \R{1}{1}{c}{a}{b}(\theta,\xi,1) \oplus
\R{3}{4}{c}{a}{b}(\theta,\pi-\xi,1) \,,
\label{m5main12}
\ee
and the corresponding elementary boundary subsets are 
$\calV_{(1,1)}= \{ 1 \}$ and $\calV_{(3,4)}= \{ 2 \}$. The 
nonvanishing amplitudes are 
\be
\R{1}{1}{1}{2}{1}(\theta,\xi,1)  
&=& V^{(1,1)}(\theta,\xi,1)\
{s(i\xi+ \theta) s(i(\xi+ \pi) - \theta) 
\over s(i\xi) s(i(\xi+\pi))} \,, \non  \\
&=&V_{\mathrm{CDD}}(\theta,\xi)\  \,{s( 2\theta-i \pi)\over s(2\theta+i \pi)} \,,
\ee
and
\be
\R{3}{4}{2}{1}{2}(\theta,\xi,1)
&=& V^{(3,4)}(\theta,\xi,1)\
{ s(i\xi+ \theta) s(i(\xi+ 3\pi) - \theta) 
\over s(i\xi) s(i(\xi+3\pi))} \,, \non  \\
&=&V_{\mathrm{CDD}}(\theta,\xi)\ \,{s( 2\theta-i \pi)\over s(2\theta+i \pi)}\,
\prod_{k=1}^{2}{s(i(\xi+ k \pi) -(-1)^{k}\theta)\over 
s(i(\xi+ k \pi) +(-1)^{k}\theta)}\,,
\ee
\be
\R{3}{4}{2}{3}{2}(\theta,\xi,1)
&=& V^{(3,4)}(\theta,\xi,1)\
{s(i\xi- \theta) s(i(\xi+ 3\pi) + \theta) 
\over s(i\xi) s(i(\xi+3\pi))} \,, \non  \\
&=&V_{\mathrm{CDD}}(\theta,\xi)\,{s( 2\theta-i \pi)\over s(2\theta+i \pi)}\,
\prod_{k=0}^{3}{s(i(\xi+ k \pi) -(-1)^{k}\theta)\over 
s(i(\xi+ k \pi) +(-1)^{k}\theta)}\,,
\ee
where we have used the result that $P_{0}(\theta)$ is given by (\ref{Pelementary})
\be
P_{0}(\theta) = V_0(\theta)\,{s( 2\theta+i \pi)\over s(2\theta-i \pi)} \,.
\ee

The amplitudes for the CBC $(1,4)$ are related to those for the CBC
$(1,2)$ described above by the $Z_{2}$ symmetry (\ref{mainZ2}).

\subsubsection*{CBC $(1,3)$}

For the CBC $(1,3)$, the boundary reflection matrix is given by (\ref{calR})
\be
\RR{1}{3}{c}{a}{b}(\theta,\xi)
= \R{2}{1}{c}{a}{b}(\theta,\xi,1) \oplus
\R{2}{4}{c}{a}{b}(\theta,\pi-\xi,1) \,,
\label{m5main13}
\ee
and the corresponding elementary boundary subsets are 
$\calV_{(2,1)}= \{ 2 \}$ and $\calV_{(2,4)}= \{ 3 \}$. The 
nonvanishing amplitudes are self-dual
\be
\R{2}{1}{2}{1}{2}(\theta,\xi,1) &=&\R{2}{4}{3}{4}{3}(\theta,\xi,1)\non\\
&=& V^{(2,1)}(\theta,\xi,1)\
{ s(i\xi- \theta) s(i(\xi+ 2\pi) + \theta) 
\over s(i\xi) s(i(\xi+2\pi))}\non \\
&=&{s( 2\theta-i \pi)\over s(2\theta+i \pi)}\, 
\prod_{k=1}^{2}{s(i(-\xi+ k \pi) -(-1)^{k}\theta)\over 
s(i(-\xi+ k \pi) +(-1)^{k}\theta)}\,,\\[8pt]
\R{2}{1}{2}{3}{2}(\theta,\xi,1)&=&\R{2}{4}{3}{2}{3}(\theta,\xi,1)\non\\
&=& V^{(2,1)}(\theta,\xi,1)\
{s(i\xi+ \theta) s(i(\xi+ 2\pi) - \theta) 
\over s(i\xi) s(i(\xi+2\pi))}  \non  \\
&=&\,{s( 2\theta-i \pi)\over s(2\theta+i \pi)}\,
\prod_{k=2}^{5}{s(i(\xi+ k \pi) -(-1)^{k}\theta)\over 
s(i(\xi+ k \pi) +(-1)^{k}\theta)}\,.
\ee

\subsubsection*{CBC $(2,2)$ and $(2,4)$}

For the CBC $(2,2)$, the boundary reflection matrix is given by (\ref{calR})
\be
\RR{2}{2}{c}{a}{b}(\theta,\xi)
= \R{1}{2}{c}{a}{b}(\theta,\xi,1) \oplus
\R{3}{3}{c}{a}{b}(\theta,\pi-\xi,1) \,,
\label{m5main22}
\ee
and the corresponding elementary boundary subsets are 
$\calV_{(1,2)}= \{ 2 \}$ and $\calV_{(3,3)}= \{ 1, 3 \}$. The 
nonvanishing amplitudes are 
\be
\R{1}{2}{2}{1}{2}(\theta,\xi,1)  
&=& V^{(1,2)}(\theta,\xi,1)\
{ s(i\xi+ \theta) s(i(\xi+\pi) \!-\! \theta) 
\over s(i\xi) s(i(\xi+\pi))} \non  \\
&=&V_{\mathrm{CDD}}(\theta,\xi)\ {s( 2\theta- \pi)\over s(2\theta+i \pi)}\,,
\ee
\be
\R{1}{2}{2}{3}{2}(\theta,\xi,1)  
&=& V^{(1,2)}(\theta,\xi,1)\
{s(i\xi+ \theta) s(i(\xi+ \pi) - \theta) 
\over s(i\xi) s(i(\xi+\pi))} \non  \\
&=&V_{\mathrm{CDD}}(\theta,\xi)\,{s( 2\theta-i \pi)\over s(2\theta+i \pi)}\,,
\ee
and
\be
{\!\!}\R{3}{3}{1}{2}{1}\!\!(\theta,\!\xi,\!1)
\!=\!V^{(3,3)}(\theta,\!\xi,\!1)
{[2]^{-2} s(i\xi\!+\! \theta) s(i(\xi\!+\! 3\pi) \!-\! \theta) 
\!+\! [2]^{-1} s(i\xi\!-\! \theta) s(i(\xi\!+\! 3\pi) \!+\! \theta)
\over s(i\xi) s(i(\xi\!+\!3\pi))}
\ee
\be
{\!\!}\R{3}{3}{3}{2}{3}\!\!(\theta,\!\xi,\!1)
\!=\!V^{(3,3)}(\theta,\!\xi,\!1)
{[2]^{-1} s(i\xi\!+\! \theta) s(i(\xi\!+\! 3\pi)\! -\! \theta) 
\!+\! [2]^{-2} s(i\xi\!-\! \theta) s(i(\xi\!+\! 3\pi) \!+\! \theta)
\over s(i\xi) s(i(\xi\!+\!3\pi))}
\ee
\be
\R{3}{3}{3}{4}{3}(\theta,\xi,1)
&=& V^{(3,3)}(\theta,\xi,1)\,
{s(i\xi- \theta) s(i(\xi+ 3\pi) + \theta) 
\over s(i\xi) s(i(\xi+3\pi))}\non  \\
&=&V_{\mathrm{CDD}}(\theta,\xi)\,{s( 2\theta-i \pi)\over s(2\theta+i \pi)}\, 
\prod_{k=0}^{3}{s(i(\xi\!+\! k \pi) -(-1)^{k}\theta)\over 
s(i(\xi+ k \pi) +(-1)^{k}\theta)} \,.
\ee
where
\be
{\!\!}V^{(3,3)}(\theta,\xi,1)\!=\!V_{\mathrm{CDD}}(\theta,\xi){s( 2\theta\!-\!i \pi)\over s(2\theta\!+\!i \pi)}
{s(i\xi)s(i(\xi\!+\!3\pi)s(i(\xi\!+\!\pi)\!+\!\theta)s(i(\xi\!+\!2\pi)\!-\!\theta)\over
s(i\xi\!+\!\theta)s(i(\xi\!+\!\pi)\!-\!\theta)s(i(\xi\!+\!2\pi)\!+\!\theta)s(i(\xi\!+\!3\pi)\!-\!\theta)}
\ee
Moreover, for this case, there are also ``nondiagonal'' amplitudes
\be
&&\R{3}{3}{1}{2}{3}(\theta,\xi,1)
=\R{3}{3}{3}{2}{1}(\theta,\xi,1)
= V^{(3,3)}(\theta,\xi,1)\
[2]^{-{1\over 2}}{ s(2\theta)
\over s(i\xi) s(i(\xi+3\pi))}\qquad\quad\non\\
&&=V_{\mathrm{CDD}}(\theta,\xi)\,{s( 2\theta-i \pi)\over s(2\theta+i \pi)}\,
{[2]^{-{1\over 2}} s(2\theta)\over 
s(i\xi + \theta) s(i(\xi+ 3 \pi) - \theta)}
\prod_{k=1}^{2}{s(i(\xi+ k \pi) -(-1)^{k}\theta)\over 
s(i(\xi+ k \pi) +(-1)^{k}\theta)} \,.  
\ee

The amplitudes for the CBC $(2,4)$ are related to those for the CBC
$(2,2)$ described above by the $Z_{2}$ symmetry (\ref{mainZ2}).

\subsubsection*{CBC $(2,3)$}

For the CBC $(2,3)$, the boundary reflection matrix is given by (\ref{calR})
\be
\RR{2}{3}{c}{a}{b}(\theta,\xi)
= \R{2}{2}{c}{a}{b}(\theta,\xi,1) \oplus
\R{2}{3}{c}{a}{b}(\theta,\pi-\xi,1) \,,
\label{m5main23}
\ee
and the corresponding elementary boundary subsets are 
$\calV_{(2,2)}= \{ 1, 3 \}$ and $\calV_{(2,3)}= \{ 2, 4 \}$. 
The nonvanishing amplitudes are 
\be
&&\R{2}{2}{1}{2}{1}(\theta,\xi,1)\;=\;\R{2}{3}{4}{3}{4}(\theta,\xi,1)\\
&=& V^{(2,2)}(\theta,\xi,1)\,
{[2]^{-1} s(i\xi+ \theta) s(i(\xi+ 2\pi) - \theta) 
+ [2]^{-2} s(i\xi- \theta) s(i(\xi+ 2\pi) + \theta)
\over s(i\xi) s(i(\xi+2\pi))} \,, \non
\ee
\be
&&\R{2}{2}{3}{2}{3}(\theta,\xi,1)\;=\;\R{2}{3}{2}{3}{2}(\theta,\xi,1)\\
&=& V^{(2,2)}(\theta,\xi,1)\
{[2]^{-2} s(i\xi+ \theta) s(i(\xi+ 2\pi) - \theta) 
+ [2]^{-1} s(i\xi- \theta) s(i(\xi+ 2\pi) + \theta)
\over s(i\xi) s(i(\xi+2\pi))} \,, \non  
\ee
\be
\R{2}{2}{3}{4}{3}(\theta,\xi,1)= \R{2}{3}{2}{1}{2}(\theta,\xi,1)
&=& V^{(2,2)}(\theta,\xi,1)\
{ s(i\xi+ \theta) s(i(\xi+ 2\pi) - \theta) 
\over s(i\xi) s(i(\xi+2\pi))} \non  \\
&=&{s( 2\theta-i \pi)\over s(2\theta+i \pi)}\,
\prod_{k=2}^{5}{s(i(\xi+ k \pi) -(-1)^{k}\theta)\over 
s(i(\xi+ k \pi) +(-1)^{k}\theta)}  
\ee
where
\be
V^{(2,2)}(\theta,\xi,1)={s( 2\theta-i \pi)\over s(2\theta+i \pi)}\,
{s(i\xi)s(i(\xi+2\pi)s(i(\xi+3\pi)+\theta)s(i(\xi+4\pi)-\theta)\over
s(i\xi-\theta)s(i(\xi+2\pi)+\theta)s(i(\xi+3\pi)-\theta)s(i(\xi+4\pi)+\theta)}
\ee
The ``nondiagonal'' amplitudes are self-dual
\be
\R{2}{2}{1}{2}{3}(\theta,\xi,1)
&\!\!\!=\!\!\!&\R{2}{2}{3}{2}{1}(\theta,\xi,1)=\R{2}{3}{2}{3}{4}(\theta,\xi,1) = \R{2}{3}{4}{3}{2}(\theta,\xi,1)\non\\
&\!\!\!=\!\!\!& V^{(2,2)}(\theta,\xi,1)\
[2]^{-{1\over 2}}{ s(2\theta)
\over s(i\xi) s(i(\xi+2\pi))}  \\
&\!\!\!=\!\!\!&\,{s( 2\theta-i \pi)\over s(2\theta+i \pi)}\,
{[2]^{-{1\over 2}} s(2\theta)\over 
s(i \xi - \theta) s(i(\xi+ 2 \pi) + \theta)}\,
\prod_{k=3}^{4}{s(i(\xi+ k \pi) -(-1)^{k}\theta)\over 
s(i(\xi+ k \pi) +(-1)^{k}\theta)} \,. \non
\ee

\section{Conclusion}
\label{sec:Conclude}

We have proposed explicit expressions\,\footnote{The result 
is in terms of paired solutions (\ref{calR}), which 
in turn are given in terms of elementary solutions (\ref{elemgauged}).
The latter are formulated in terms of the reduced elementary solutions
(\ref{elem1}), (\ref{elem2}) and the scalar factor 
(\ref{boundaryscalarfactor}), (\ref{cdd}).}
for the boundary reflection matrices of
the $\phi_{1,3}$-perturbed ${\cal A}_{m}$ unitary minimal models for all
possible Cardy conformal boundary conditions $(r,s)$.  These results
are a generalization of the results for $m=3$ and $m=4$ found by
Ghoshal and Zamolodchikov~\cite{Ghoshal:1993tm} and by Chim~\cite{Chim:1995kf},
respectively.  We have verified that these boundary reflection matrices are consistent with 
the boundary bootstrap and the transformation properties  
under height-reversal and non-invertible symmetries.  We leave to a future investigation the problem of determining the relation between the boundary parameter $\xi$ in the boundary reflection matrices and the parameters $\hat\lambda$ and $h$ in the action (\ref{boundaction}). 
This UV-IR relation could be derived from the analogous sine-Gordon results \cite{Bajnok:2001ug} by quantum group reduction, and would determine the parameters $\Lambda_{\pm}$.
We have also not addressed the problem of determining boundary reflection
matrices for general superpositions of Cardy CBCs.

An alternative approach to computing boundary reflection matrices would be to start from a lattice formulation of the model, and to use the Bethe Ansatz to compute (along lines such as \cite{Fendley:1994cz, Grisaru:1994ru, Doikou:1999xt} for
sine-Gordon/XXZ) the scattering of the physical excitations off the boundary.  Pursuing such alternative approaches would be valuable, since we have made various assumptions in arriving at our results, in particular the boundary flows (\ref{hminus}), (\ref{hplus}) and the
boundary subsets (\ref{symmbsubset}).

There are several interesting questions that one can now hope to address.  Indeed, as noted in \cite{Zamolodchikov:1989rd, LeClair:1989wy, Bernard:1990cw, Reshetikhin:1989qg}, the bulk model (\ref{bulkaction}) has integrals of motion of fractional spin $2/m$; hence, the bulk $S$ matrix (\ref{bulkS}) has a corresponding residual quantum group symmetry.  Moreover, it is known that the bulk model can
be regarded~\cite{LeClair:1989wy, Bernard:1990cw, Reshetikhin:1989qg, Eguchi:1989dq} as a certain restriction of the
sine-Gordon model~\cite{Zamolodchikov:1978xm}.  One would like to understand to what extent these results can be extended to the boundary model
(\ref{boundaction}). In particular, which part of the quantum group respects the integrable $(r,s)$ boundaries \cite{Mezincescu:1997nw, Delius:2001qh} and how they are represented on the reflection factors. Lastly, we expect that the arguments proposed in this paper will extend to all unitary and nonunitary $A$-$D$-$E$ minimal models but, given the length of this paper, these extensions should be considered elsewhere.

\section*{Acknowledgments}

This project originated at the University of Durham in December 2001 from discussions of one of the authors (RN) with P. Dorey and R. Tateo, to whom we are grateful. Further progress resulted from the meeting of two authors (RN and PP) at the APCTP Focus Program ``Finite-size technology in low dimensional quantum field theory'' at POSTECH in December 2003. Decisive progress was made when the authors met at the mathematical research institute MATRIX in Australia in July 2024, and at E\"otv\"os Lor\'and University in July 2025.
We are  indebted to C. Ahn, K. Graham, A. LeClair
and A.B. Zamolodchikov for correspondence and/or discussions.  
This work was supported in part by the NKFIH Grant K134946 (ZB);
the National Science Foundation under
grants PHY-0098088, PHY-0244261 and PHY 2310594 (RN); and by the Australian
Research Council (PP).

\appendix

\definecolor{lightlightblue}{rgb}{.93,.93,1}
\definecolor{lightpurple}{rgb}{.999,.7,.999}
\definecolor{lightyellow}{rgb}{1.,.929,.514}
\def\Wt#1#2#3#4#5{W\Big(\!\begin{array}{cc}#4&#3\\#1&#2\end{array}\!\Big|\,#5\Big)}
\newcommand{\face}[5]{
\psset{unit=0.8cm}
\begin{pspicture}[shift=-.40](0,0)(1,1)
\facegrid{(0,0)}{(1,1)}
\psarc[linewidth=0.5pt,linecolor=red]{-}(0,0){0.16}{0}{90}
\rput(0.,-.1){\spos{tr}{#1}}
\rput(1.,-.1){\spos{tl}{#2}}
\rput(1.,1.1){\spos{bl}{#3}}
\rput(0.,1.1){\spos{br}{#4}}
\rput(.5,.5){\spos{c}{#5}}
\end{pspicture}}
\def\facegrid#1#2{
\psframe[fillstyle=solid,fillcolor=lightlightblue,linewidth=0pt]#1#2
\psgrid[gridlabels=0pt,subgriddiv=1]#1#2}
\def\facegridb#1#2{
\psframe[fillstyle=solid,fillcolor=midblue,linewidth=0pt]#1#2
\psgrid[gridlabels=0pt,subgriddiv=1]#1#2}

\section{Fused Adjacency Matrices and ABF Lattice Models}
\label{app:Adjacency}

Let $G$ denote the $A_{m-1}$ adjacency matrix; i.e., the $(m\!-\!1) 
\times (m\!-\!1)$ matrix with elements $G_{a, b}$ given by (\ref{adjacency}).
The fused adjacency matrices $F^{1} \,, \ldots \,, F^{m}$ are defined 
by the recursion
\be
F^{1} = I \,; \qquad F^{2} = G \,; \qquad F^{r} = G F^{r-1} - 
F^{r-2} \,, \qquad r = 3 \,, 4\,,\ldots \,, m-1
\ee
which truncates with $F^m=0$. The matrix elements $F^{r}_{a b}$ of $F^{r}$ are either 0 or 1 
and are completely symmetric in the three indices.
The $F^r$ are precisely the Verlinde matrices of the $s\ell(2)$ WZW models and 
consequently
there is a Verlinde-type formula for their matrix elements
\be
F^{r}_{a b} = \sum_{j=1}^{m-1} {S_{a j}\, S_{b j}\, S_{r j} \over  
S_{1 j}} \,, \qquad S_{a b} = \sqrt{2\over m} 
\sin \left( {\pi a b \over m} \right) \,.
\ee 

\begin{figure}[htb]
\be
\psset{unit=.9cm}
\begin{pspicture}[shift=-3.9](-2,-.5)(6,4.5)
\multirput(0,0)(1,0){8}{\psline[linewidth=1.pt,linestyle=dashed,dash =2pt 1pt](-1.5,-1.5)(-1.5,5.5)}
\facegrid{(-2,-1)}{(6,5)}
\multirput(0,0)(0,2){3}{\multirput(0,0)(1,0){8}{\rput(-1.5,-.5){\scriptsize $u$}}}
\multirput(0,1)(0,2){3}{\multirput(0,0)(1,0){8}{\rput(-1.5,-.5){\scriptsize $\lambda\!-\!u$}}}
\multirput(0,0)(0,1){6}{\multirput(0,0)(1,0){8}{\psarc[linewidth=0.5pt,linecolor=red]{-}(-2,-1){0.16}{0}{90}}}
\multirput(-2.2,-1)(0,2){4}{\scriptsize $c$}
\multirput(-2.2,0)(0,2){3}{\scriptsize $b$}
\multirput(6.2,-1)(0,2){4}{\scriptsize $c'$}
\multirput(6.2,0)(0,2){3}{\scriptsize $b'$}
\end{pspicture}\nonumber
\ee\\[0pt]
\caption{An $N\times M$ lattice on the cylinder for $(N,M)=(8,6)$ with left and right boundary edges $(b,c)$, $(b'\!,c')$ and $|b-c|=|b'\!-c'|=1$.
The cylinder is built from $M/2$ applications of the double row transfer matrices~\cite{Behrend:1995zj} so the spectral parameters alternate between $u$ and $\lambda-u$ on consecutive rows. The dashed lines indicate that the top and bottom rows are identified.}
\label{cylLattice}
\end{figure}
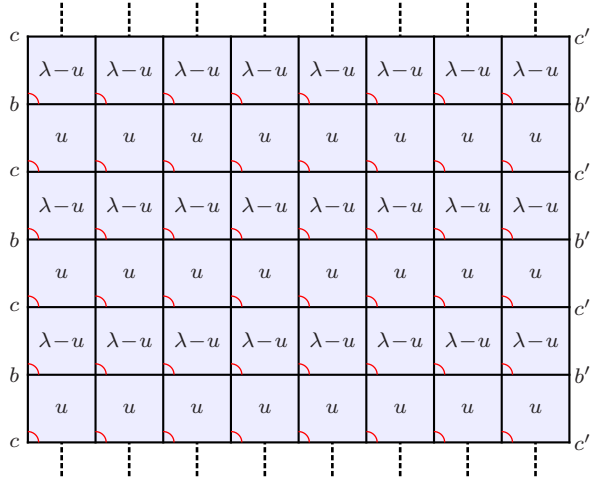

The Boltzmann face weights of the critical $A_{m-1}$ ABF lattice models associated with ``heights'' $a,\!b,\!c,\!d=1,\!2,\ldots,m-1$ are
\be
\face abcdu\,=\!\Wt abcdu\!=\!\sin(\lambda\!-\!u)\delta_{a,c}+\sin u\sqrt{\frac{S_a S_c}{S_b S_d}}\,\delta_{b,d} \,,
\label{faceWeights}
\ee
for $|a\!-\!b|=|b\!-\!c|=|c\!-\!d|=|d\!-\!a|=1$ but vanish otherwise. 
The spectral parameter is $u$, the crossing parameter is $\lambda=\frac{\pi}{m}$ and $S_a=\frac{\sin a\lambda}{\sin\lambda}$. The face weights are invariant under the height reversal symmetry $a\mapsto m-a$. The isotropic point is $u=\lambda/2$. The off-critical face weights involve elliptic functions~\cite{Andrews:1984af} with the elliptic nome playing the role of a temperature-like parameter. An $8\times 6$ lattice on the cylinder is shown in Figure~\ref{cylLattice}.

\section{Lattice Bulk and Boundary Free Energies}
\label{app:LatticeBdyFree}

In this section we solve the lattice inversion relation for the lattice bulk and boundary free energies. 
Consider the $A_{m-1}$ lattice model with crossing parameter $\lambda=\pi/m$ and $m\ge 4$. 
For a strip of width $N$ columns, the inversion relation for the largest eigenvalue of the double row transfer matrix
\be
\kappa(u)=\exp(-2Nf_{\mathrm {bulk}}(u)-f_{\mathrm{bdy}}(u)+o(1)),\qquad\mbox{as $N\to\infty$}
\ee
takes the form
\be
\kappa(u)\kappa(u+\lambda)\!=\![s_1(u)s_1(-u)]^{2N}\! {s_2(2u)s_2(-2u)\over s_2(0)^2 s_1(2u)s_1(-2u)}
{s_0(\xi\!+\!u)s_0(\xi\!-\!u)s_r(\xi\!+\!u)s_r(\xi\!-\!u)\over s_0(\xi)^2s_r(\xi)^2}
\label{latticeinv}
\ee
subject to the crossing symmetry
\be
\kappa(u)=\kappa(\lambda-u)
\ee
where
\be
s_j(u)={\sin(u+j\lambda)\over \sin\lambda}
\ee
This inversion relation comes from keeping only the dominant term in the functional equation satisfied by the double row transfer matrices ((6.57) of \cite{Behrend:1995zj} with $p=q=1$). The boundary condition on the left is the vacuum boundary condition labelled by $(r,s)=(1,1)$ and the boundary condition on the right is the $(r,s)$ boundary condition. This is equivalent to having a $(1,s)$ boundary on the left and an $(r,1)$ boundary on the right. 

The inversion relation factors into a bulk inversion relation of order $2N$ which gives the bulk free energy and two boundary inversion relations of order $1$ which give the boundary free energies. The boundary free energies depend on $r$ but are independent of $s$. Specifically we factor $\kappa(u)$ as
\be
\kappa(u)=\kappa_{\mathrm{bulk}}(u)\kappa_{(r,s)}(u)=\kappa_{\mathrm{bulk}}(u)\kappa_0(u)\kappa_r(u,\xi)
\ee

For the bulk theory, the physical region is $0\le u\le \lambda$. In this region the bulk weights are all nonnegative.
However, in the presence of a boundary, there are two relevant physical regions
\begin{eqnarray}
\mbox{(i)}&&0\le u\le \lambda/2,\qquad \phantom{-3}\lambda/2\le\xi\le3\lambda/2\\
\mbox{(ii)}&&0\le u\le \lambda/2,\qquad -3\lambda/2\le\xi\le-\lambda/2
\end{eqnarray}
In these regions the boundary weights of the $(r,1)$ and $(1,s)$ boundary conditions are all nonnegative (after multiplication by the sign factor $\mbox{sgn}(\xi)$). Moreover, each of the three terms on the right side of (\ref{latticeinv}) is nonnegative.

\subsection*{Bulk free energy}
The inversion relation for the bulk free energy has been solved by Baxter~\cite{Baxter:1982xp}. Since we need similar calculations to obtain the boundary free energies, we summarize the key steps. 
The function $\log\kappa_{\mathrm{bulk}}(u)$ is actually analytic in the analyticity strip
\be
-\lambda/2<\mbox{Re}(u)<3\lambda/2
\ee
and grows as $\exp(\mp iu)$ as $u\to\pm i\infty$. It follows that the second logarithmic derivative of $\kappa_{\mathrm{bulk}}(u)$ can be represented on the full analyticity strip by a Fourier/Laplace integral
\be
{d^2\over du^2}\,\log\kappa_{\mathrm{bulk}}(u)=\int_{-\infty}^{\infty} c(t) e^{2ut}dt
\ee
The inversion and crossing relations with $\kappa_{\mathrm{bulk}}(-u)=\kappa_{\mathrm{bulk}}(u+\lambda)$ are
\be
\log\kappa_{\mathrm{bulk}}(u)+\log\kappa_{\mathrm{bulk}}(u+\lambda)&\!\!=\!\!&\log{\sin(\lambda-u)\sin(\lambda+u)\over \sin^2\lambda},\ \  -\lambda/2<\mbox{Re}(u)<\lambda/2\qquad
\label{bulkinv}\\
\log\kappa_{\mathrm{bulk}}(u)&\!\!=\!\!&\log\kappa_{\mathrm{bulk}}(\lambda-u),\qquad\qquad\ -\lambda/2<\mbox{Re}(u)<3\lambda/2
\ee
From the basic identity
\be
{d^2\over du^2}\,\log{\sin u\over\sin\lambda}=-\int_0^{\infty} {4t\,\cosh(\pi-2u)t\over \sinh \pi t}\,dt,
\qquad 0<\mbox{Re}(u)<\pi
\ee
it follows that
\be
{d^2\over du^2}\,\log{\sin(\lambda-u)\sin(\lambda+u)\over\sin^2\lambda}=
-\int_{-\infty}^{\infty} {4t\,\cosh(\pi-2\lambda)t\over \sinh\pi t}\,e^{2ut}\,dt,
\  \ |\mbox{Re}(u)|<\lambda
\ee
Hence
\be
c(t)=e^{-2\lambda t}c(-t),\qquad\qquad
(1+e^{2\lambda t})c(t)=- {4t\,\cosh(\pi-2\lambda)t\over \sinh\pi t}
\ee
with the solution
\be
c(t)=- {2t\,e^{-\lambda t}\cosh(\pi-2\lambda)t\over \sinh\pi t\,\cosh\lambda t}
\ee
Integrating twice and evaluating the integration constants gives
\be
\log\kappa_{\mathrm{bulk}}(u)&=&-\int_{-\infty}^\infty 
{\cosh(\pi-2\lambda)t\over 2t\,\sinh\pi t\,\cosh\lambda t}\,e^{-(\lambda-2u)t}\,dt+Au+B\nonumber\\
&=&\int_{-\infty}^\infty 
{\cosh(\pi-2\lambda)t \sinh ut\sinh(\lambda-u)t\over t\,\sinh\pi t\,\cosh\lambda t}\,dt,
\quad -\lambda/2<\mbox{Re}(u)<3\lambda/2\qquad\quad
\ee

\subsection*{Boundary free energies}
Following methods introduced in \cite{OBrien:1997}, the $s$-independent boundary free energies $\kappa_{r,s}(u)=\kappa_0(u)\kappa_r(u,\xi)$ are obtained by solving the crossing and inversion relations for the order-1 contributions. 

First, the vacuum contribution $\kappa_0(u)$ satisfies crossing in its analyticity strip $-{\lambda/2}<\mbox{Re}(u)<3{\lambda/2}$ and the inversion relation
\be
\log\kappa_0(u)+\log\kappa_0(u+\lambda)=
\log{\sin^2\lambda\,\sin(2\lambda-2u)\sin(2\lambda+2u)\over 
\sin^2 2\lambda\,\sin(\lambda-2u)\sin(\lambda+2u)},
\ -{\lambda\over 2}<\mbox{Re}(u)<{\lambda\over 2}
\ee
But now in the strip $|\mbox{Re}(u)|<\lambda/2$
\be
{d^2\over du^2}\log{\sin(2\lambda-2u)\sin(2\lambda+2u)\over \sin(\lambda-2u)\sin(\lambda+2u)}=8\!\int_{-\infty}^{\infty}
{t\sinh{(\pi-3\lambda)t\over 2}\sinh{\lambda t\over 2}\over \sinh{\pi t\over 2}}\,e^{2ut}dt
\ee
We conclude that
\be
c(t)=e^{-2\lambda t}c(-t),\qquad\quad 
(1+e^{2\lambda t})c(t)\!\!\!&=&\!\!\!{8\,t\sinh{(\pi-3\lambda)t\over 2}\sinh{\lambda t\over 2}\over 
\sinh{\pi t\over 2}}
\ee
so that the solution is
\be
c(t)= {4\,t\,e^{-\lambda t}\sinh{(\pi-3\lambda)t\over 2}\sinh{\lambda t\over 2}
\over \sinh{\pi t\over 2}\cosh \lambda t}
\ee
Integrating twice and evaluating the integration constants gives
\be
\log\kappa_0(u)&\!\!=\!\!&\int_{-\infty}^\infty 
{\sinh{(\pi-3\lambda)t\over 2}\sinh{\lambda t\over 2}
\over t\,\sinh{\pi t\over 2}\cosh\lambda t}\,e^{-(\lambda-2u)t}\,dt+Au+B\nonumber\\
&\!\!=\!\!&-2\!\int_{-\infty}^\infty 
{\sinh{(\pi-3\lambda)t\over 2}\sinh{\lambda t\over 2}\sinh ut\sinh(\lambda-u)t
\over t\,\sinh{\pi t\over 2}\cosh\lambda t}\,dt,
\  -{\lambda\over 2}<\mbox{Re}(u)<{3\lambda\over 2}\qquad\quad
\ee

The other order-1 contribution  $\kappa_r(u)=\kappa_r(u,\xi)$ depends on $r=1,2,\ldots,m-2$ and $\xi$. It is convenient to allow $r=m-1$ but in this case we restrict $\xi$ to the region $|\mbox{Re}(\xi)|<\lambda$. 
The $r$-dependent contribution $\kappa_r(u)$ satisfies crossing in the analyticity strip $-\lambda/2<\mbox{Re}(u)<3\lambda/2$ and the inversion relation
\be
\log\kappa_r(u)+\log\kappa_r(u+\lambda)=
\log\!\Big[{\sin(\xi+u)\sin(\xi-u)\sin(\xi+r\lambda+u)\sin(\xi+r\lambda-u)\over
\sin^2\xi\sin^2(\xi+r\lambda)}\Big]\quad
\label{rinversion}
\ee
for $u$ in the strip $-\lambda/2<\mbox{Re}(u)<\lambda/2$.
Observing the symmetry
\be
\kappa_r(u,\xi)=\kappa_{m-r}(u,-\xi)
\label{rsymmetry}
\ee
we restrict our attention to the region $\mbox{Re}(\xi)>0$. The solution in the region $\mbox{Re}(\xi)< 0$ is then determined from the solution for the region $\mbox{Re}(\xi)>0$ by applying this symmetry. 

For $\mbox{Re}(\xi)>0$, the relevant analyticity strip for $\xi$ is $0<\mbox{Re}(\xi)<3\lambda/2$. 
To be more precise, the function $\log\kappa_r(u)$ is analytic in the region
\be
-\lambda/2<\mbox{Re}(u)<{3\lambda\over 2},\quad 0<\mbox{Re}(\xi)<{3\lambda\over 2},\quad
\mbox{Re}(\xi)>-\mbox{Re}(u),\quad \mbox{Re}(\xi)>\mbox{Re}(u)-\lambda\quad
\label{rbdyanalyticity}
\ee
which contains the $\xi>0$ physical region $0<u<\lambda/2$ with $\lambda/2<\xi<3\lambda/2$. 
With these restrictions we have
\be
&&\hspace{-.7in}{d^2\over du^2}
\log[\sin(\xi+u)\sin(\xi-u)\sin(\xi+r\lambda+u)\sin(\xi+r\lambda-u)]\non\\
&=&-8\!\int_{-\infty}^{\infty}
{t\cosh{(\pi-2\xi-r\lambda)t}\cosh{r\lambda t}\over \sinh{\pi t}}\,e^{2ut}dt
\ee
We conclude that
\be
c(t)=e^{-2\lambda t}c(-t),\qquad\quad
(1+e^{2\lambda t})c(t)=-{8t\cosh{(\pi-2\xi-r\lambda)t}\cosh{r\lambda t}\over 
\sinh{\pi t}}
\ee
so that the solution is
\be
c(t)=- {4\,t\,e^{-\lambda t}\cosh{(\pi-2\xi-r\lambda)t}\cosh{r\lambda t}
\over \sinh{\pi t}\cosh \lambda t}
\ee
Integrating twice and evaluating the integration constants gives
\be
\log\kappa_r(u,\xi)&\!\!=\!\!&-\int_{-\infty}^\infty 
{\cosh{(\pi-2\xi-r\lambda)t}\cosh{r\lambda t}
\over t\,\sinh{\pi t}\cosh\lambda t}\,e^{-(\lambda-2u)t}\,dt+Au+B\nonumber\\
&\!\!=\!\!&2\!\int_{-\infty}^\infty 
{\cosh{(\pi-2\xi-r\lambda)t}\cosh{r\lambda t}\sinh ut\sinh(\lambda-u)t
\over t\,\sinh{\pi t}\cosh\lambda t}\,dt
\ee
which holds throughout the analyticity region (\ref{rbdyanalyticity}). 

Since we need it, we also note the additional symmetry at $\xi=\lambda/2$
\be
\kappa_r\Big(u,{\lambda\over 2}\Big)=\kappa_{m-r-1}\Big(u,{\lambda\over 2}\Big)
\label{extrasymmetry}
\ee
and the identities
\be
\kappa_1(u,\xi)={s_0(\xi+u)s_1(\xi-u)\over s_0(\xi)s_1(\xi)}\label{Identity1}
\ee
\begin{eqnarray}
&&\log{\kappa_r(u,\xi)\kappa_{\mathrm{bulk}}(u+\xi)\kappa_{\mathrm{bulk}}(u-\xi)\over 
\kappa_1(u,\xi)\kappa_{m-r-1}(u,\lambda-\xi)}
\;=\;-2\int_{-\infty}^\infty {\sinh^2\xi t\,\cosh(m-2)\lambda t\over t\,\sinh\pi t}\,dt\qquad\qquad\nonumber\\
&&\qquad\qquad\;=\;\log\kappa_{\mathrm{bulk}}(\xi)\kappa_{\mathrm{bulk}}(-\xi)
\;=\;\log s_1(\xi)s_1(-\xi)\label{Identity2}
\end{eqnarray}

The boundary free energies are uniquely determined by the inversion relations. We checked the analytic solutions against known solutions for $A_4$ as well as in many cases against boundary free energies obtained numerically. A typical plot of $\kappa_{r,s}(u)$ is shown in Figure~7.
\begin{figure}[htb]
	\centering
	\includegraphics[width=0.6\textwidth]{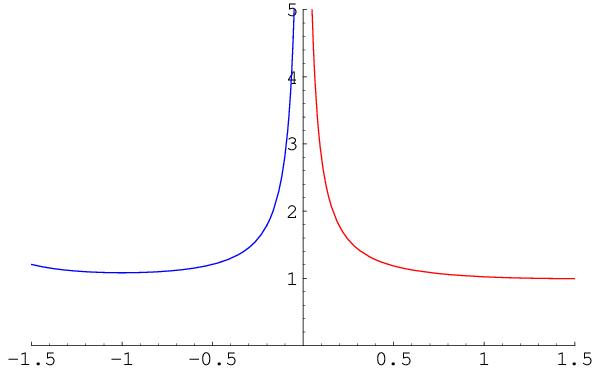}
\caption{Plot of $\kappa_{(r,s)}(u,\xi)$ for $m=5$, $r=2$, any allowed $s$ and $u\to\lambda/2$. The horizontal axis is labelled in units of $\lambda$ for $|\xi|\le 3\lambda/2$. The shape of this graph is typical for any values of $m$, $r$ and $s$ and is not symmetric about $\xi=0$. Notice the divergence at $\xi=0$ separating the $\xi>0$ solution shown in red from the $\xi<0$ solution shown in blue.}
	\label{fig:kapparplot}
\end{figure}
Physically, the $\R{r}{1}{\ \  r}{r+1}{\ \  r}\!\!(u,\xi)$ boundary weights dominate for $\xi>0$ and the $\R{r}{1}{\ \  r}{r-1}{\ \  r}\!\!(u,\xi)$ boundary weights dominate for $\xi<0$. At $u=\pm\xi=\lambda/2$, the boundary triangles can be removed and these precisely correspond to boundary phases where the heights along the right edge of the strip alternate between $r$ and $r+1$ or $r$ and $r-1$ respectively. The two boundary phases coexist at $\xi=0$.

So far we have been considering a strip with left and right boundaries. In QFT however, there is only one boundary, namely, the right boundary. On the lattice it is convenient to think of this situation as moving the left boundary off to infinity.
Notice that crossing symmetry is destroyed by eliminating one of the two boundaries which are related one to the other by crossing. Specifically, this means instead of the crossing symmetric order-1 term $\kappa_0(u)$ which accounts for both boundaries we associate an order-1 non-crossing symmetric term $p_0(u)$ with the right boundary and the crossing symmetric partner $p_0(\lambda-u)$ to the left boundary which is off at infinity. So now there is an ambiguity in how this order-1 term is shared out. Moreover, the analyticity of this extra term is not  determined from simple lattice considerations so it is not uniquely defined. We will see that this is precisely the origin of the CDD factor.

It follows that the non-crossing symmetric order-1 contribution $p_0(u)$ must satisfy the inversion relations
\be
p_0(u)p_0(-u)=1,\qquad p_0(\lambda-u)p_0(\lambda+u)
={\sin(2\lambda-2u)\sin(2\lambda+2u)\over\sin(\lambda-2u)\sin(\lambda+2u)}
\ee
where the second relation follows from (c.\/f. (\ref{elemboundcrossunit}))
\be
p_0(u)={\sin\lambda\,\kappa_{\mathrm{bulk}}(2u)\over\sin(2\lambda-2u)}\,p_0(\lambda-u)
\label{p0crossing}
\ee
after using the bulk inversion relation (\ref{bulkinv}) to eliminate the bulk free energy.
Setting
\be
\log p_0(u)=\int_0^\infty p(t) \sinh 2ut\, dt
\ee
and observing that
\be
\log{\sin(2\lambda-2u)\sin(2\lambda+2u)\over\sin(\lambda-2u)\sin(\lambda+2u)}
=4\int_0^\infty {\sinh{(\pi-3\lambda)t\over 2}\sinh{\lambda t\over 2}\cosh 2ut\over
t\,\sinh{\pi t\over 2}}\,dt
\ee
we find
\be
\log p_0(u)=2\int_0^\infty {\sinh{(\pi-3\lambda)t\over 2}\sinh{\lambda t\over 2}\sinh 2ut\over
t\,\sinh{\pi t\over 2}\sinh2\lambda t}\,dt,\qquad |\mbox{Re}(u)|<3\lambda/2
\ee
But this solution is not unique. Indeed, other solutions are given by 
\be
v_{\mathrm{CDD}}(u,\xi)\,p_0(u),\qquad
v_0(u)\,v_{\mathrm{CDD}}(u,\xi)\,p_0(u)
\ee
where
\be
v_0(u)v_0(-u)=1,&\qquad& v_0(u)=-v_0(\lambda-u)\\
v_{\mathrm{CDD}}(u,\xi)v_{\mathrm{CDD}}(-u,\xi)=1,&\qquad& v_{\mathrm{CDD}}(u,\xi)=v_{\mathrm{CDD}}(\lambda-u,\xi)
\ee
These equations admit many solutions but, for our purposes, the relevant {\it minimal}\/ solutions are
\begin{eqnarray}
v_0(u) &=&\tan\big({mu\over 2}-{\pi\over 4}\big)\\
v_{\mathrm{CDD}}(u, \xi) &=& {\sin(m\xi) + \sin(mu)\over \sin(m\xi) - \sin(mu)}
={\tan({m(\xi+u)\over 2})\over \tan({m(\xi-u)\over 2})}
\label{latticecdd}
\end{eqnarray}
where $v_{\mathrm{CDD}}(u,\xi)$ is the CDD ambiguity factor.

Lastly we point out that, for $m$ odd, all of the integrals of this section can be evaluated explicitly in terms of trigonometric functions as listed in Appendix~D.

\section{Lattice and QFT Scalar Factors for $m$ Odd}
\label{app:LatticeScalar}

\subsection*{Bulk lattice free energy for $m=$ odd}
We find
\be
\log \kappa_{\mathrm{bulk}}(u) = \log\left[ {\prod_{k=1}^{(m-1)/2} 
\sin \left( u+(2k-1)\lambda \right)\over
\sin \lambda \prod_{k=1}^{(m-3)/2} \sin \left( u+ 2k \lambda \right)} \right] 
\ee
so that
\be
\kappa_{\mathrm{bulk}}(\theta) = \left[ {\prod_{k=1}^{(m-1)/2} 
\sinh \left( (\theta+(2k-1)\pi i)/m \right)\over
\sinh (\pi i/m) \prod_{k=1}^{(m-3)/2} \sinh \left( (\theta+ 2k \pi i)/m 
\right)} \right]
\ee
and 
\be
U(\theta) = {\pi\over \kappa_{\mathrm{bulk}}(\theta) \sin(\pi/m)} \,.
\ee 
 
\subsection*{Lattice boundary free energies for $m=$ odd}

{$\bullet\ \kappa_0(u)$:} For $\kappa_0(u)$, the result for $m=5\,,9\,,13\,, \ldots$ is
\be
\kappa_0(u) &=& \tan^2 \!\Big( {mu\over 2}-{\pi\over 4}\Big)\,
{\sin(\lambda) \sin(2u+\lambda) \sin(2u+2\lambda)\over
\sin(2\lambda) \sin^2(2u+(m-1)\lambda)} \non \\
&\times& \prod_{k=0}^{(m-9)/4} {\sin \left( 2u+(5+4k)\lambda \right) 
\sin \left( 2u+(6+4k)\lambda \right)\over
\sin \left( 2u+(3+4k)\lambda \right) \sin \left( 2u+(4+4k)\lambda
\right)}
\ee
while the result for $m=7\,,11\,,15\,,\ldots$ is
\be
\kappa_0(u) = {\sin (\lambda) \over \sin(2\lambda)}
\prod_{k=0}^{(m-7)/4} {\sin \left( 2u+(2+4k)\lambda \right) 
\sin \left( 2u+(3+4k)\lambda \right)\over
\sin \left (2u+(1+4k)\lambda \right) \sin \left( 2u+(4+4k)\lambda
\right)}
\ee

\noindent{$\bullet\ \kappa_r(u,\xi)$,\ \  $r= 1 \,, 2\,, \ldots \,, m-2$:}\\
For $\kappa_r(u,\xi)$ with $\Re\xi>0$, the result for $r$ odd (which holds for $m$ odd and  
$m$ even) is
\be
\kappa_r(u,\xi) ={\prod_{k=0}^{(r-1)/2} 
\sin \left(\xi+u+2k \lambda \right) 
\sin \left(\xi -u+(2k+1) \lambda \right)
\over \sin \xi \sin \left(\xi+ r \lambda \right)
\prod_{k=0}^{(r-3)/2} \sin \left( \xi+u+(2k+1) \lambda\right)
\sin \left( \xi -u+(2k+2) \lambda \right)}
\ee
while the result for $r$ even is
\be
\kappa_r(u,\xi) \!\!\!&=&\!\!\!
{\sin( m u) + \sin ( m \xi)\over \sin( m u) - \sin ( m \xi)}\\
\!\!\!&\times&\!\!\! {\prod_{k=0}^{(m-r-1)/2} 
\sin \left(\xi+u+(r+2k)\lambda\right)
\sin \left(\xi -u+(r+1+2k) \lambda\right)
\over \sin \xi \sin(\xi+ r \lambda)
\prod_{k=0}^{(m-r-3)/2} \sin \left(\xi+u+(r+2k+1) \lambda\right)
\sin \left(\xi -u+(r+2k+2) \lambda \right)}
\non
\ee
Hence,
\be
V_r(\theta,\xi) \!\!\!&=&\!\!\!{1\over \sinh (i \xi/m) \sinh \left( (i \xi+ r i \pi 
)/m\right)}\non \\
\!\!\!&\times&\!\!\!
{\prod_{k=0}^{(r-1)/2} 
\sinh \left( (i \xi+\theta+2k i \pi)/m \right) 
\sinh \left( (i \xi -\theta+(2k+1) i \pi)/m \right)
\over 
\prod_{k=0}^{(r-3)/2} \sinh \left( (i \xi+\theta+(2k\!+\!1) i \pi)/m\right)
\sinh \left( (i \xi -\theta+(2k\!+\!2) i \pi)/m \right)} 
\,, \ \  r \mbox{ odd} \non \\
\!\!\!&=&\!\!\!-\left({\sin \xi - i \sinh \theta \over \sin \xi + i \sinh
\theta}\right)
{1\over \sinh (i \xi/m) \sinh((i \xi+ r i \pi)/m)} 
\label{Vrelementaryodd}  \\
&&\mbox{}\hspace{-1.95cm}\times {\prod_{k=0}^{(m-r-1)/2} 
\sinh \left( (i \xi +\theta+(r\!+\!2k)i \pi)/m\right)
\sinh \left( (i \xi -\theta+(r\!+\!1\!+\!2k) i \pi)/m\right)
\over 
\prod_{k=0}^{(m-r-3)/2} \sinh \left( (i \xi+\theta+(r\!+\!2k\!+\!1) i \pi)/m\right)
\sinh \left( (i \xi -\theta+(r\!+\!2k\!+\!2) i \pi)/m \right)},\ \  r \mbox{ even}.\non
\ee

\noindent{$\bullet\ p_{0}(u)$:} For $p_{0}(u)$, the result for $m=5\,,9\,,13\,, \ldots$ is
\be
p_{0}(u) = -\tan \left( {m u\over 2}-{\pi\over 4}\right) 
\prod_{k=0}^{(m-5)/4} {\sin \left( 2u+(1+4k)\lambda \right) 
\over
\sin \left(-2u+(1+4k)\lambda \right)} \,,
\ee
while the result for $m=7\,,11\,,15\,,\ldots$ is
\be
p_{0}(u) = 
\prod_{k=0}^{(m-7)/4} {\sin \left( 2u+(3+4k)\lambda \right) 
\over
\sin \left(-2u+(3+4k)\lambda \right)} \,,
\ee
and $p_{0}(u) = 1$ for $m=3$.
Hence,
\be
P_{0}(\theta) &=& i\tanh \left( {\theta\over 2}-{i\pi\over 4}\right) 
\prod_{k=0}^{(m-5)/4} {\sinh \left( (2\theta+(1+4k)i \pi)/m \right) 
\over
\sinh \left( (-2\theta+(1+4k)i \pi)/m \right)} \,, \qquad m=5\,,9\,,13\,, 
\ldots \,, \non \\
&=& \prod_{k=0}^{(m-7)/4} {\sinh \left( (2\theta+(3+4k)i \pi)/m \right) 
\over
\sinh \left( (-2\theta+(3+4k)i \pi)/m \right)} 
\,, \qquad m=7\,,11\,,15\,,\ldots \,, \non \\
&=& 1 \,, \qquad m=3 \,.\label{Pelementary}
\ee

\bibliographystyle{utphys}
\bibliography{bdyR}

\end{document}